 \documentclass[preprint,6pt,5p]{elsarticle}

\usepackage{float}
\usepackage[utf8]{inputenc}
\usepackage{graphicx}
\usepackage{subfigure}
\usepackage{epsfig}
\usepackage{amssymb}
\usepackage{amsthm}
\usepackage{url}
\usepackage{array}
\usepackage{amsmath}
\usepackage[titletoc,toc,title]{appendix}
\usepackage{amsfonts}
\usepackage{graphicx}
\usepackage{hyperref}

 \biboptions{comma,square,sort&compress}

\makeatletter
\renewcommand{\fnum@figure}{Fig.\thefigure}

\journal{elsevier}

\begin{document}

\begin{frontmatter}

\title{Causality and Correlations between BSE and NYSE indexes: A Janus Faced Relationship}

%
\author{\large{Neeraj}\corref{cor1}}
\cortext[cor1]{n15rs071@iiserkol.ac.in}
\author{\large{Prasanta K. Panigrahi}\corref{cor2}}
\cortext[cor2]{pprasanta@iiserkol.ac.in}
\address{\large{Indian Institute of Science Education and Research Kolkata, Mohanpur-741 246,West Bengal,India}}
\begin{abstract}
We study  the multi-scale temporal correlations and causality connections between the New York Stock Exchange (NYSE) and Bombay Stock Exchange (BSE) monthly average closing price indexes for a period of 300 months, encompassing the time period of the liberalisation of the Indian economy and its gradual global exposure.
In multi-scale analysis; clearly identifiable 1, 2 and 3 year non-stationary periodic modulations in NYSE and BSE have been observed, with NYSE commensurating changes in BSE at 3 years scale.
Interestingly, at one year time scale, the two exchanges are phase locked only during the turbulent times, while at the scale of three year, in-phase nature is observed for a much longer time frame. 
The two year time period, having characteristics of both one and three year variations, acts as the transition regime. The normalised NYSE's stock value is found to Granger cause those of BSE, with a time lag of 9 months. 
Surprisingly, observed Granger causality of high frequency variations  reveals BSE behaviour getting reflected in the NYSE index fluctuations, after a smaller time lag. This Janus faced relationship, shows that smaller stock exchanges may provide a natural setting  for simulating  market fluctuations of much bigger exchanges. This possibly arises due to the fact that high frequency fluctuations form an universal part of the financial time series, and are expected to exhibit similar characteristics in open market economies.
\end{abstract}


\end{frontmatter}

\section{Introduction}
\label{sec1}
Stock markets can exhibit intricate inter-relationships at different time scales, not easily discernible through traditional methods of analysis ~\cite{ramsey1,ramsey2,book:gencay,crowley}. Unraveling these connections have become even more complex with the advent of globalization. The degree of intricacy can vary from pair to pair, depending on the inter-dependencies arising from the requirements of the two economies, global factors and  possibly also on the size of economies of the two countries.\\ For a relatively insular economy like that of India, the nature of the stock market's correlations with other countries can be subtle. Generally, insularity is expected to protect an economy from short term global trends, while possibly impacting it in the long run. Hence, it is of deep interest to analyse the correlations between open and insular economies to identify their short and long-term behaviour.\\
This will throw light on the nature of interaction between open and protected economies, leading to a better understanding of the global economy as a whole.\\  
The fact that, the Indian economy progressively opened itself post 1991 and it still has not been fully integrated with the open economies, makes it particularly interesting to study its correlation and causal connections with that of US, the largest open economy.  
For this purpose, we carry out a systematic analysis of New York Stock Exchange (NYSE) and Bombay Stock Exchange (BSE) monthly average closing price indexes, for identifying temporal correlations and causality, through a local multi-scale approach. It is assumed that the behaviour of stock markets provides a faithful  indicator for  the state of the  economies.\\
Because of global trends of  liberalisation  in recent years, international diversification of portfolios has caught the attention of investors \cite{rua}. India, being a fast developing economy, has strong potential for attracting international investments. Hence, the present investigation may be useful for  investors in deciding the time scale at which these two markets  are to be considered as a part of their portfolio assets. 
The period of study is chosen from 1986 to 2010, to include various booms and bursts in the American economy, as well as post liberalisation  periods of the Indian economy.
Both of these time series are highly non-stationary, revealing Gaussian random behaviour at certain scales, while possessing well defined periodic modulations at certain other scales 
Keeping these multi-scale variability in mind, we take recourse to discrete wavelets to isolate small-scale high frequency fluctuations from local  smooth trends to unveil their mutual temporal correlations at multiple scales  \cite{ramsey1,ramsey2,book:gencay,book:percival,biswal,schleicher,crowley,parikh2,mani1}. 
Continuous Morlet wavelet is effectively used to identify time varying periodic modulations at longer time scales ~\cite{mallat1,lin}. Recently,  Random Matrix approach has shown linkage between BSE and US market in a smaller time frame, where it was observed that market crisis in the US led to strong correlation amongst the Indian companies in the same period \cite{deo1}. 
Post-liberalisation study of integration of Indian market with the international markets, on the basis of correlations between the BSE and international stock exchanges, has  clearly revealed that reactions of BSE are in tandem with those seen globally \cite{debjiban}. \\
Interestingly, our multi-scale approach exhibits diametrically opposite behaviour at large  and small time scales. In the former case, BSE is found to be strongly driven by NYSE with a small time lag,  with non-stationary periodic variations of  one, two and three year periods.In the one year time scale, the two exchanges are in phase during the market turbulence, whereas, the  three year period exhibits in-phase nature on a much larger time frame, encompassing the crisis periods. The two year period  shows characteristics of both one and three year periods. {\textit{However,  counter intuitively, the observed Granger causality at smaller scales corresponding to high frequency fluctuations, revealed BSE behaviour getting reflected  in the NYSE price index.}
This raises the possibility of smaller stock exchanges providing a natural setting for the simulation of future short term market behaviour of bigger exchanges, as the former being smaller reacts and equilibrates much faster than the latter. This can be attributed to presence of high frequency fluctuations which form the universal part of the financial times series \cite {sen1,uni_non_uni,bouchad2}.\\
The paper  is organised as follows. In the following section,  various statistical characteristics of the fluctuations are investigated, along with their  dynamical behaviour.  Subsequently, multi-scale variations about local linear trends are extracted,  using discrete Daubechies-4 (Db-4) wavelet, and investigated for their probability distributions and correlations. The former revealed  clear distinctions between  the two time series,  bringing out the important role of the outliers. Long term non-stationary periodic variations are extracted through continuous Morlet wavelet. Their phase correlation and coherence  are studied through phase correlations of the dominant wavelet coefficients of the scalogram, cross wavelet spectrum and wavelet coherence.  Multi-scale causality is then investigated, using the Toda-Yamamoto Granger causality test, revealing the clear differences in the causal connections between the high and low frequency variations. We finally conclude with directions for further investigation.
\section{Data}\label{sec:data}
We analyse the monthly average closing price indexes of NYSE and BSE  for a period of 300 months from January 01, 1986 to December 31, 2010 ~\cite{yahoo1, stooq}. As mentioned earlier, this time period includes  several  booms and bursts of both the economies and post liberalisation periods of the Indian economy. The   monthly  averaged price indexes have been scaled by their respective standard deviations for normalisation. These two  time series, as depicted in Fig.\ref{fig:norm_tog}, show global similarities.

\begin{figure}[H]
\centering
\includegraphics[width=10cm,height=5cm]{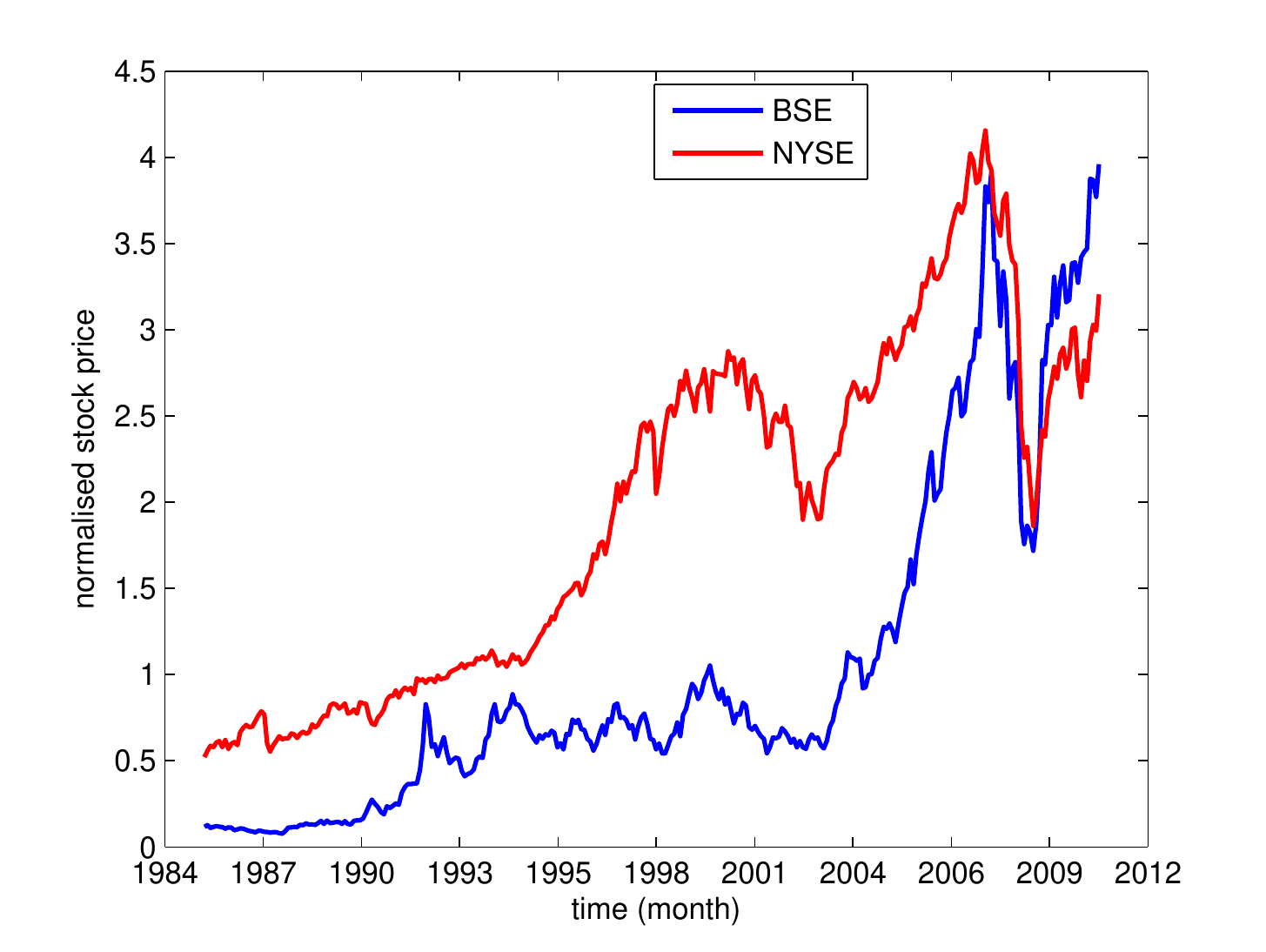}
\caption{ Normalised monthly average closing price indexes of {\small BSE} and {\small NYSE}, exhibiting strikingly similar behaviour }
\label{fig:norm_tog}
\end{figure}
The histograms of the normalised price indexes, shown in Fig.\ref{fig:norm_hist}, reveal characteristic differences, although the global features in Fig.\ref{fig:norm_tog} are similar. {\small NYSE} price index possess bi-modal character, whereas {\small BSE} distribution is uni-modal with a rightly skewed  tail.
\begin{figure}[H]
\hfill
\subfigure[ {\small BSE}]
{\includegraphics[height=4cm,width=4cm]{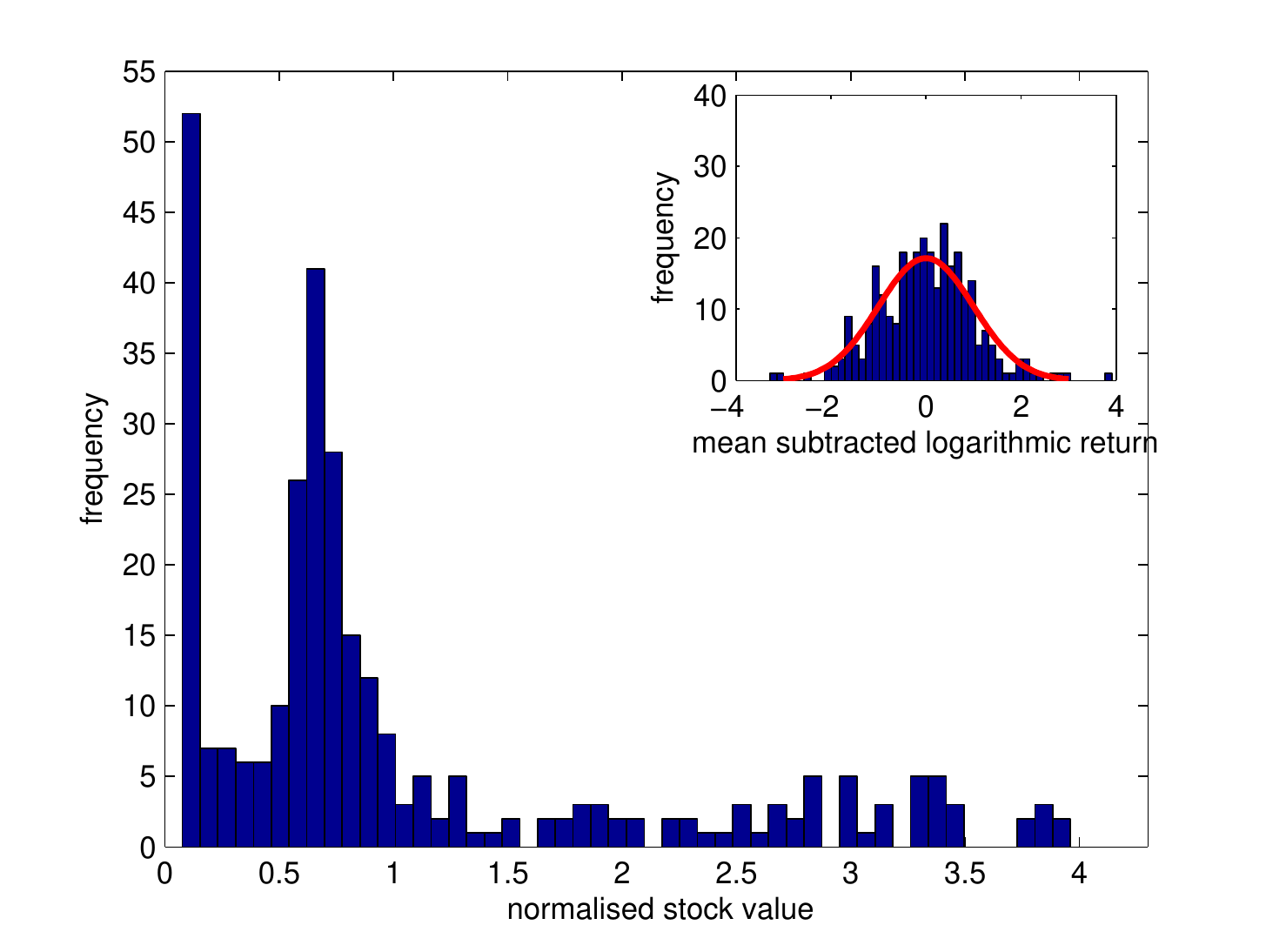}}
\hfill
\subfigure[ {\small NYSE}]
{\includegraphics[height=4cm , width= 4cm]{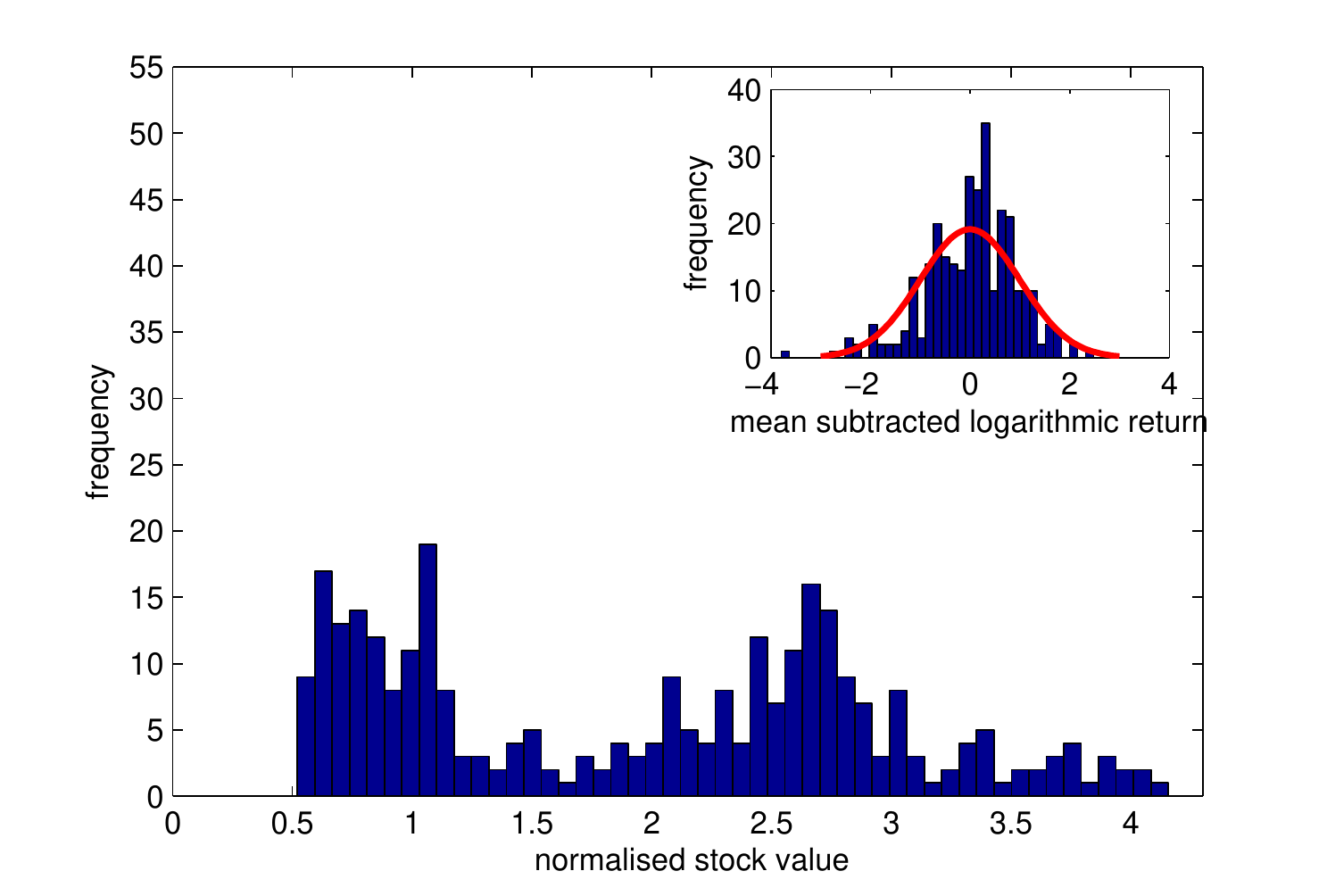}}
\hfill
\caption{Histograms of the normalised {\small BSE} and {\small NYSE} closing price values, revealing significant differences. Inset shows the distribution of mean subtracted normalised logarithmic returns of the two stock exchanges, depicting almost Gaussian behaviour for BSE and larger proportions of smaller returns for NYSE}
\label{fig:norm_hist}
\end{figure}

The distribution of mean subtracted normalised logarithmic returns, 
\begin{equation}
\label{eq:norm_ret}
\hat{R(n)}=\frac{R(n)-\langle R(n) \rangle}{\sqrt{\langle R(n)^2 \rangle - \langle R(n) \rangle^2}}.
\end{equation}
shown in the insets of Fig.\ref{fig:norm_hist}, reveal features different  from the global variations, with 
{\small BSE}  being closer  to normal distribution. In the above, $R(n)=\log x_{n+1}-\log x_n$,  is the logarithmic return and the normalising factor \ \\ $\sqrt{\langle R(n)^2 \rangle - \langle R(n) \rangle^2}$, is known as volatility of returns.
Clustering for small return values, as well as presence of outliers, are evident in case of  {\small NYSE}, which is in sharp contrast with the near Gaussianity  of  {\small BSE}.

\begin{figure}[H]
\centering\includegraphics[height=5cm ,width=10cm]{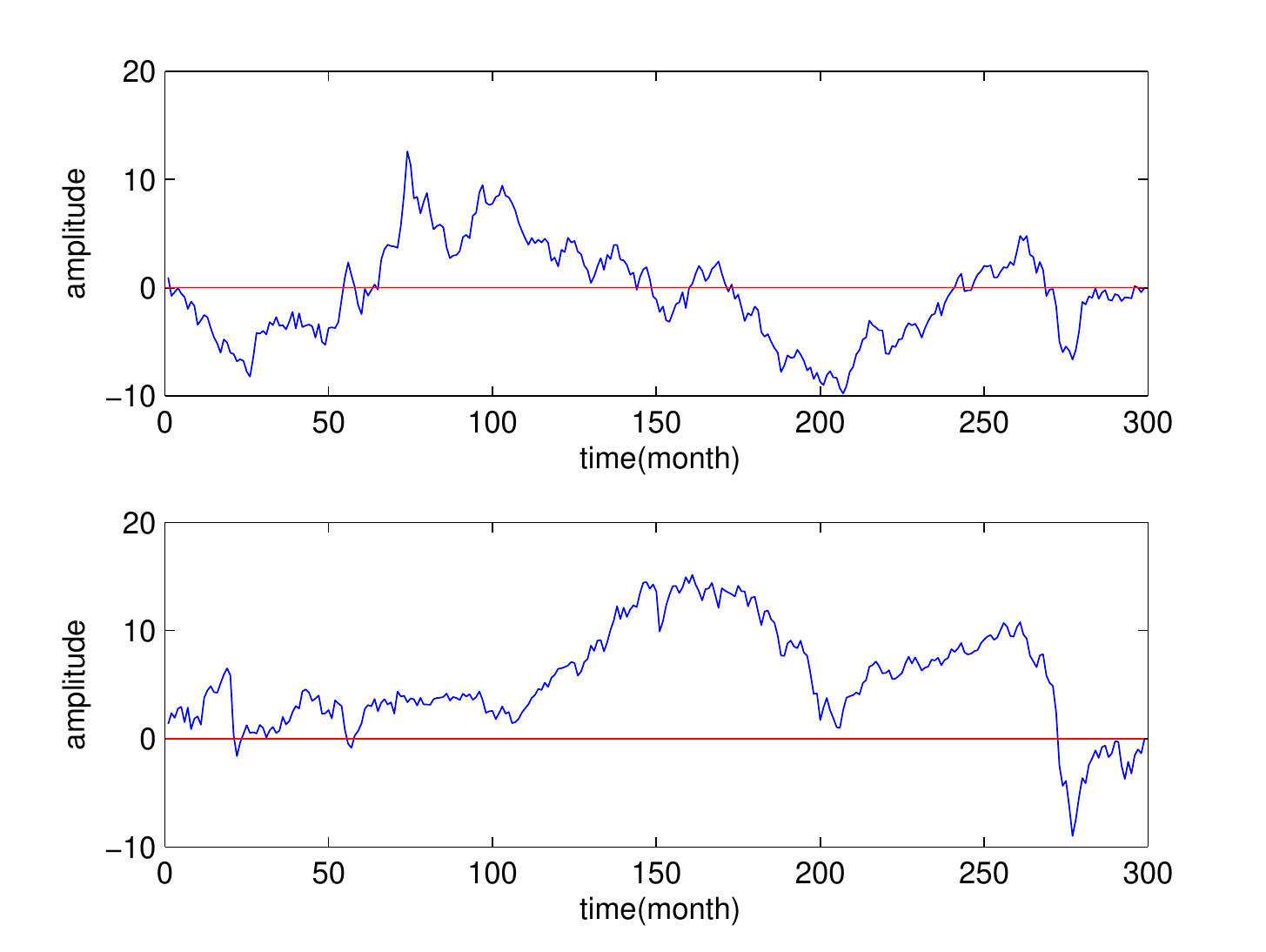}
\caption{Plots of the cumulative sums  of normalised mean subtracted logarithmic price indexes of {\small BSE} (top) and {\small NYSE} (bottom),  showing mean reverting  behaviour}
\label{fig:cumsum}
\end{figure}

From a statistical random walk perspective, the cumulative sums of the normalised mean subtracted logarithmic returns, of both the  stock exchanges, are found to exhibit mean reversion (see supplemental material).  It is evident, as seen in Fig.\ref{fig:cumsum}, that post-liberalisation  mean reversal  for both the time series are similar.\\
For a quantitative characterisation of self-similarities of the time series, the Hurst exponents  are computed \cite{hurst}, revealing anti-persistence nature ~\cite{Alvarez,peter}, with values  0.4261 and 0.4512 for {\small BSE} and  {\small NYSE},  respectively. It is to be noted that, closer the value of the exponent  to zero, stronger is the anti-persistent behaviour and hence the mean reverting nature.
For  {\small BSE}, the mean reversion rate has not changed significantly over the analysis period, while for  {\small NYSE}, it is higher in the first half than the second.
As is well known, stock prices are affected by a  plethora of factors, the dynamics of which do not get easily  revealed in the time series plot of the stock prices \cite{placard}.  A better understanding of the market dynamics is obtained through phase-space analysis, through the study of the returns as a function of the price indexes, akin to the velocity-trajectory plots in particle dynamics \cite{abhinna,bapun}. 
The phase space plots shown in Fig.\ref{fig:phase_space}, clearly reveal periodic and structured variations, both in the short and long time scales. The intersection of  trajectories at some places arises due to the fact that,  the price function has been unfolded in two dimensions, rather than three or more. The stability of a trajectory is dependent on its distance from the horizontal axis,  which   indicates the magnitude of the variations. It is seen in the plots that, {\small BSE} is less stable  than {\small NYSE} for a longer period of time. Interestingly, BSE shows strong positive returns consistently, whereas, NYSE has exhibited negative returns on large occasions.
\begin{figure}[H]
\hfill
\subfigure[BSE]
{\includegraphics[height=4cm,width=4cm]{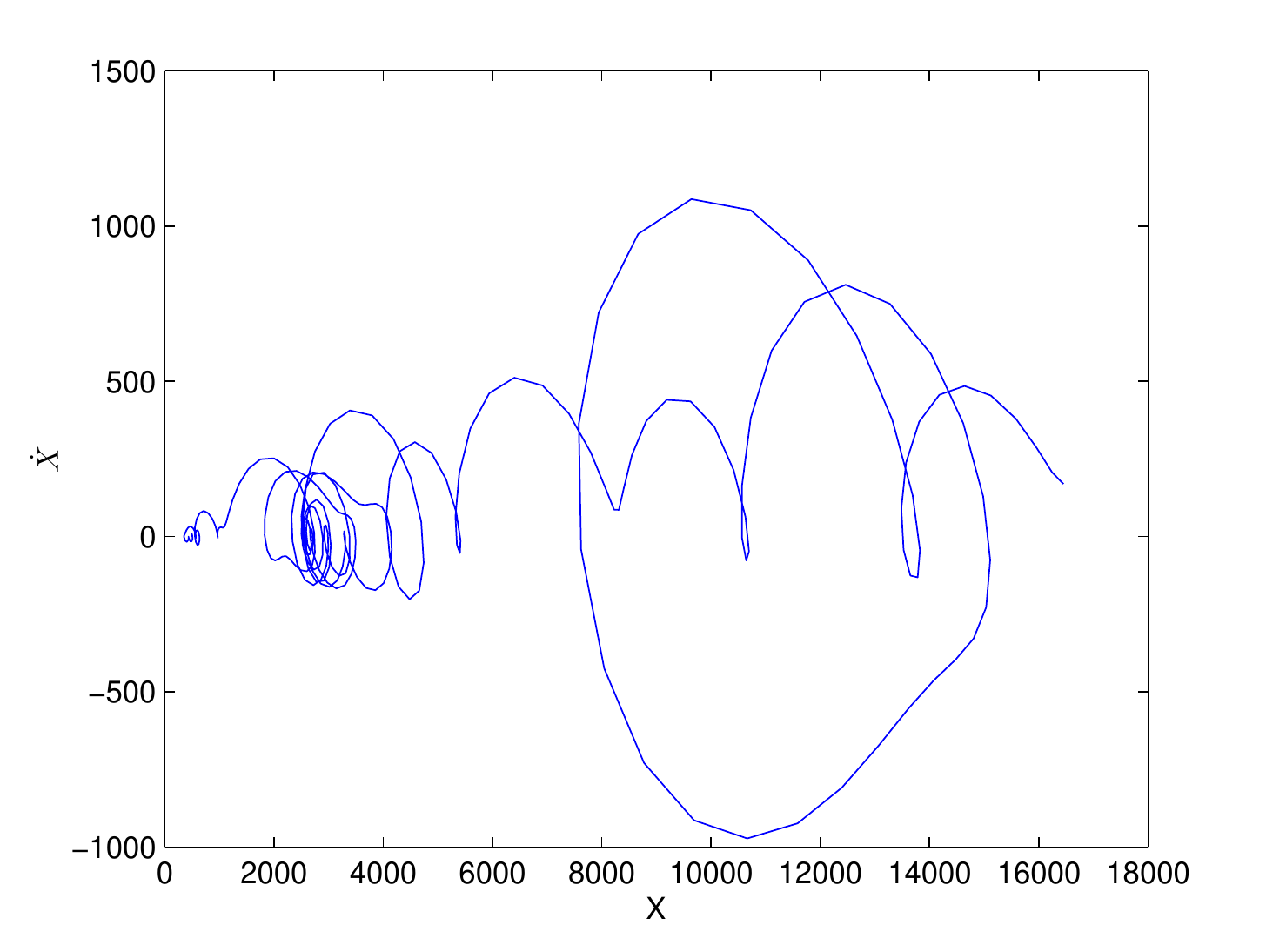}}
\hfill
\subfigure[NYSE]
{\includegraphics[height=4cm,width=4cm]{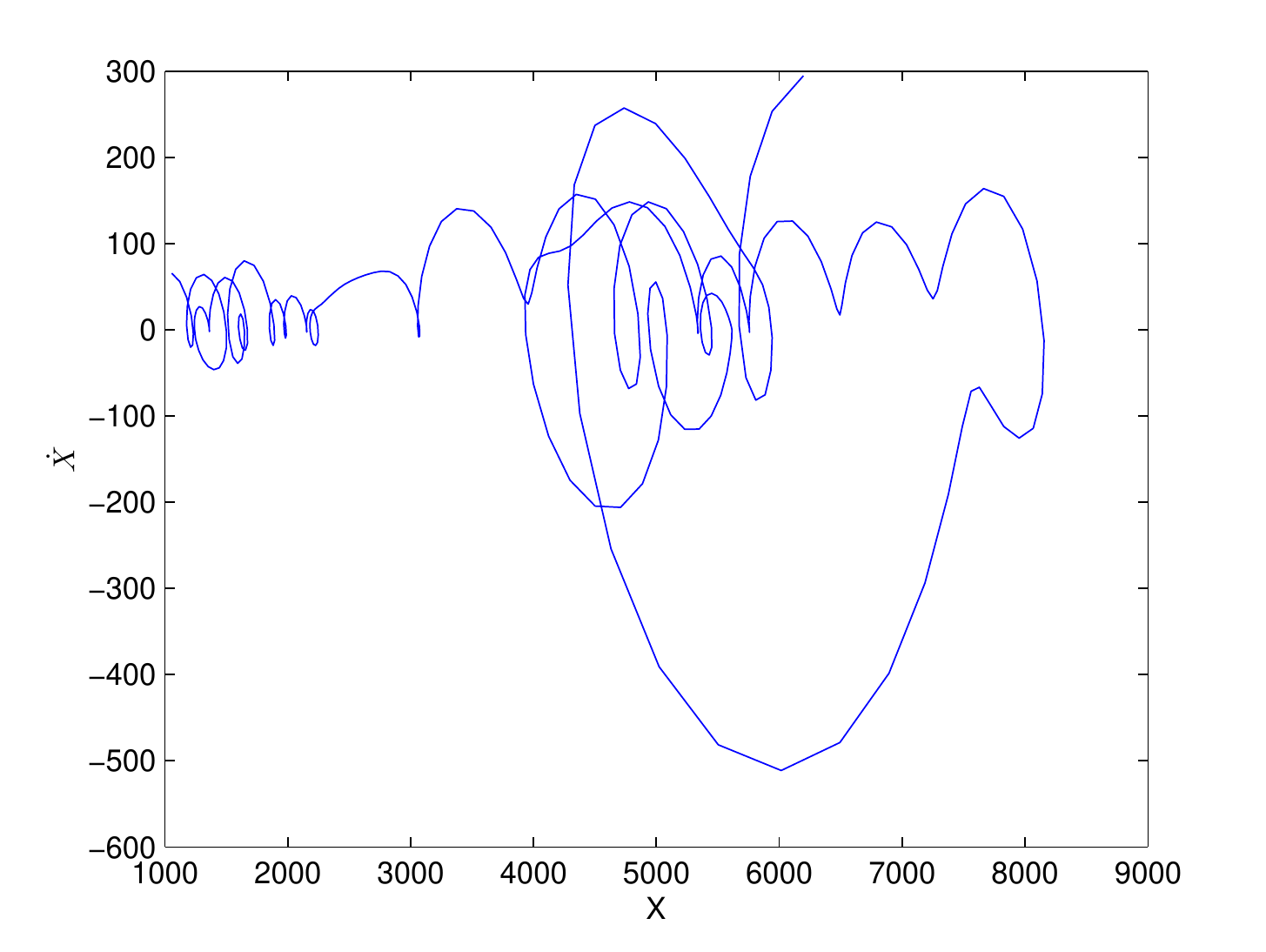}}
\hfill
\caption{Phase-space behaviour of the returns of {\small BSE } and {\small NYSE}, with the abscissa representing the stock prices and Y-axis returns, showing periodic variations}
\label{fig:phase_space}
\end{figure}
For a more systematic multi-scale comparison of the two price indexes,  we study below their time-frequency localisation  and ensuing correlations at various time scales.

\section{Wavelet Transform}\label{sec:wavelet} 
We make use of both discrete and continuous wavelet transforms to study  the local behaviour of the price index fluctuations and non-stationary periodic trends, respectively. The discrete Db-4 wavelet is specifically chosen to extract variations from possible local linear behaviour at different scales~\cite{reso1,reso2,napler}. The periodic modulations present in the data are extracted through the  Morlet wavelet, which uses a Gaussian window, with a sinusoidal sampling function ~\cite{torrence1,lin,pkp_statistical}.

\subsection{Discrete Wavelet Transform}\label{sec:DWT}
Discrete wavelets make use  of two kernels, known as  the father $\phi(t)$ and mother wavelets $\psi(t)$,  satisfying  the following admissibility conditions \cite{daub1,torrence1,mallat1}:\\
\begin{eqnarray}
\int \phi dt < \infty , \int \psi dt &=& 0 , \int \phi^{*} \psi dt =0,\\
\int \vert \phi \vert^2 dt =\int \vert \psi \vert^2 dt &=& 1
\end{eqnarray}
The father, mother and  daughter wavelets  form a complete orthogonal set, the daughter wavelets being the scaled and translated versions of the mother wavelet:
\begin{equation}
\psi_{j,k}(t)=2^{j/2} \psi(2^j t-k), \qquad k \in \mathbb{Z},\quad j \in \mathbb{Z}^+.
\end{equation}
Here, $j$ and $k$ are the scaling and translation parameters~\cite{mallat1}.
\begin{figure}[H]
\centering
\subfigure[Normalised  average monthly values of the {\small BSE}]
{\includegraphics[height=2cm,width=9cm]{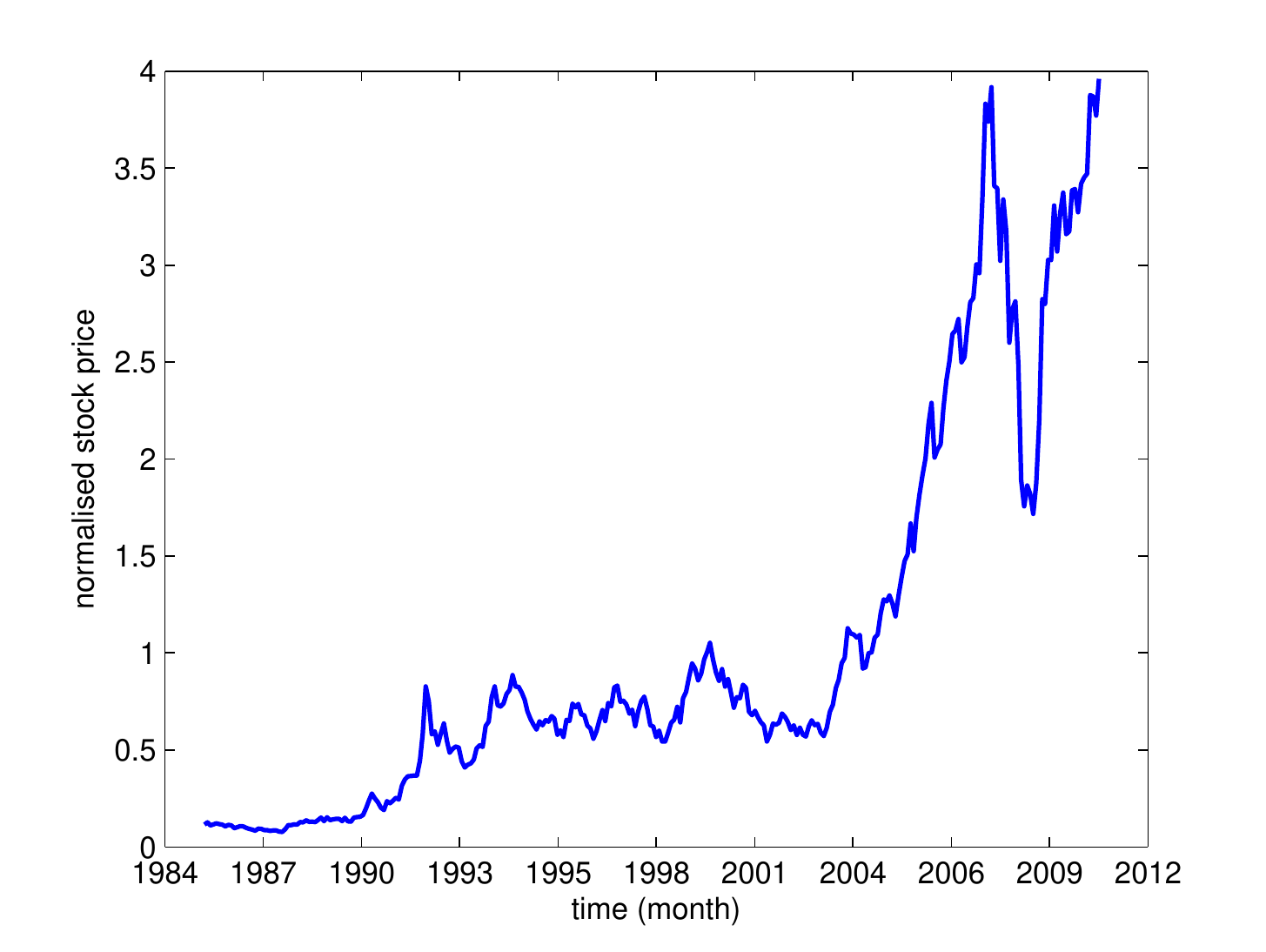}}
\centering
\subfigure[Average behaviour  for  four levels of filtering]
{\includegraphics[height=4cm,width=4cm]{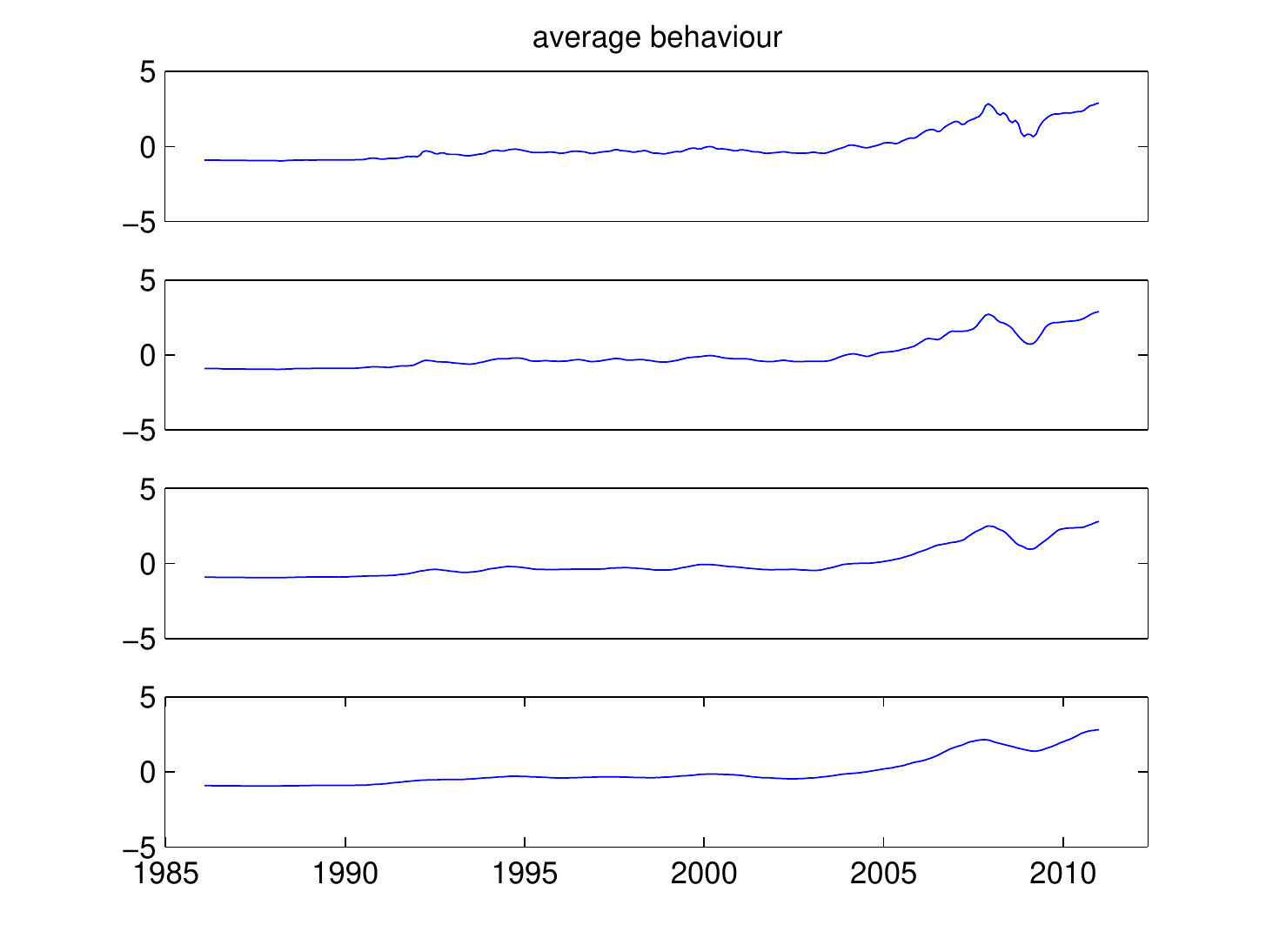}}
\subfigure[Fluctuations  for the first \, \, \, four levels ]
{\includegraphics[height=4cm,width=4cm]{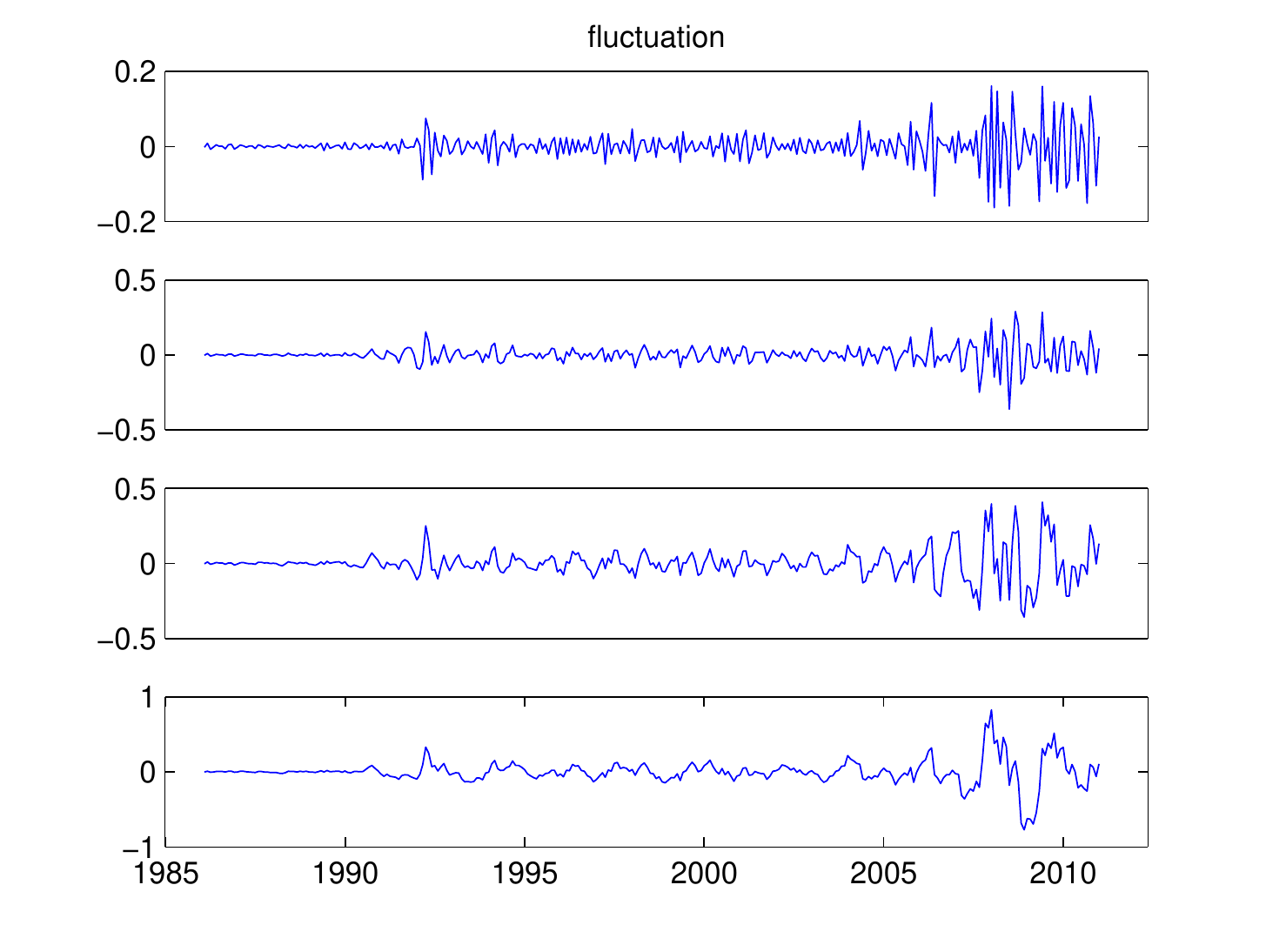}}
\caption{ Plots of (a) normalised mean subtracted {\small BSE}  monthly  average index  and (b)  behaviour  after averaging over progressively bigger temporal domains   (c) corresponding fluctuations  for  the first four levels, exhibiting strong variations in the second half}
\label{fig:low_high_sen_plot}
\end{figure}
\begin{figure}[H]
\centering
\subfigure[ Normalised  monthly average of the  {\small NYSE} ]
{\includegraphics[height=2cm,width=9cm]{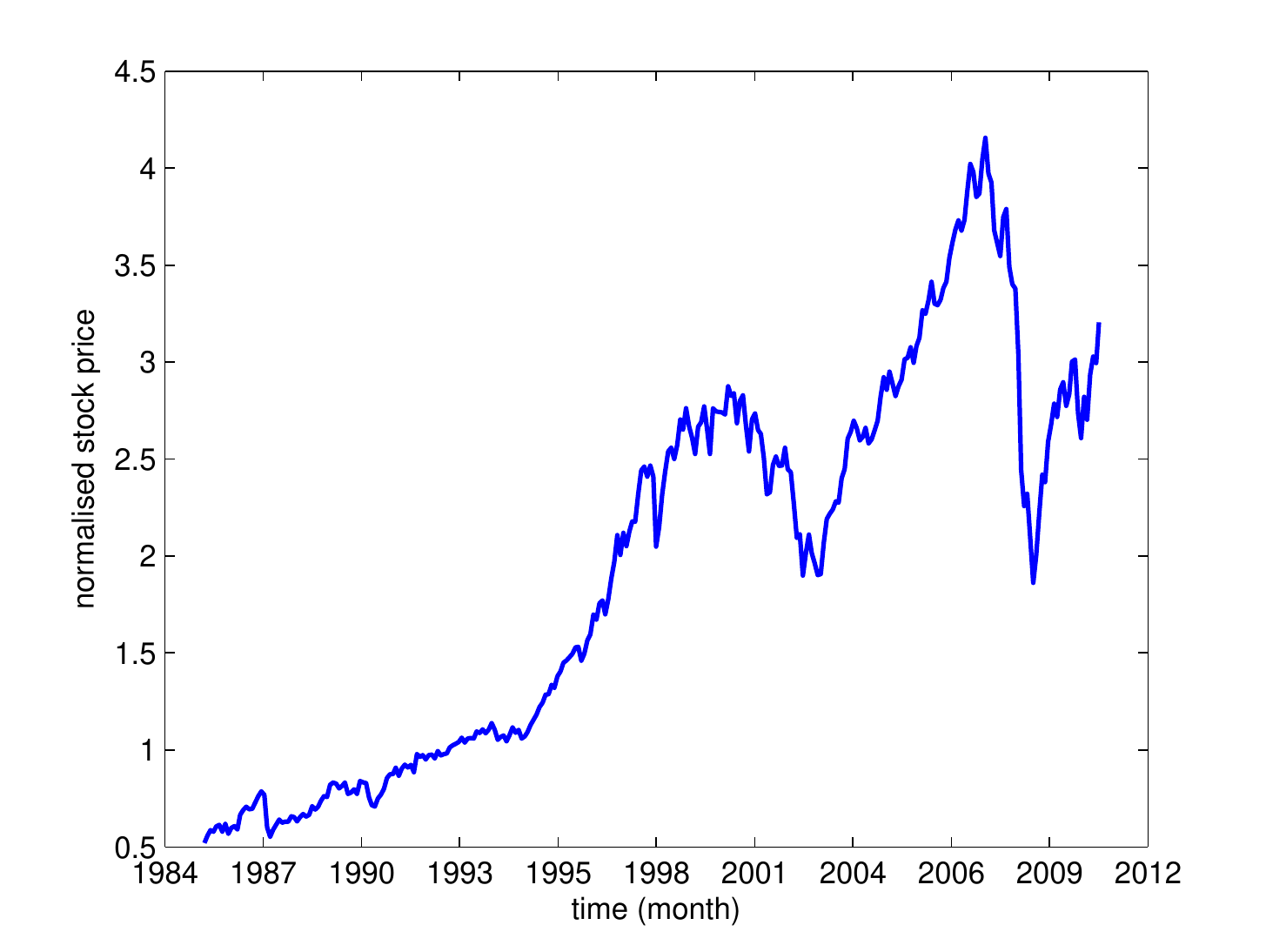}}
\subfigure[Average behaviour for the first  four levels ]
{\includegraphics[height=4cm,width=4cm]{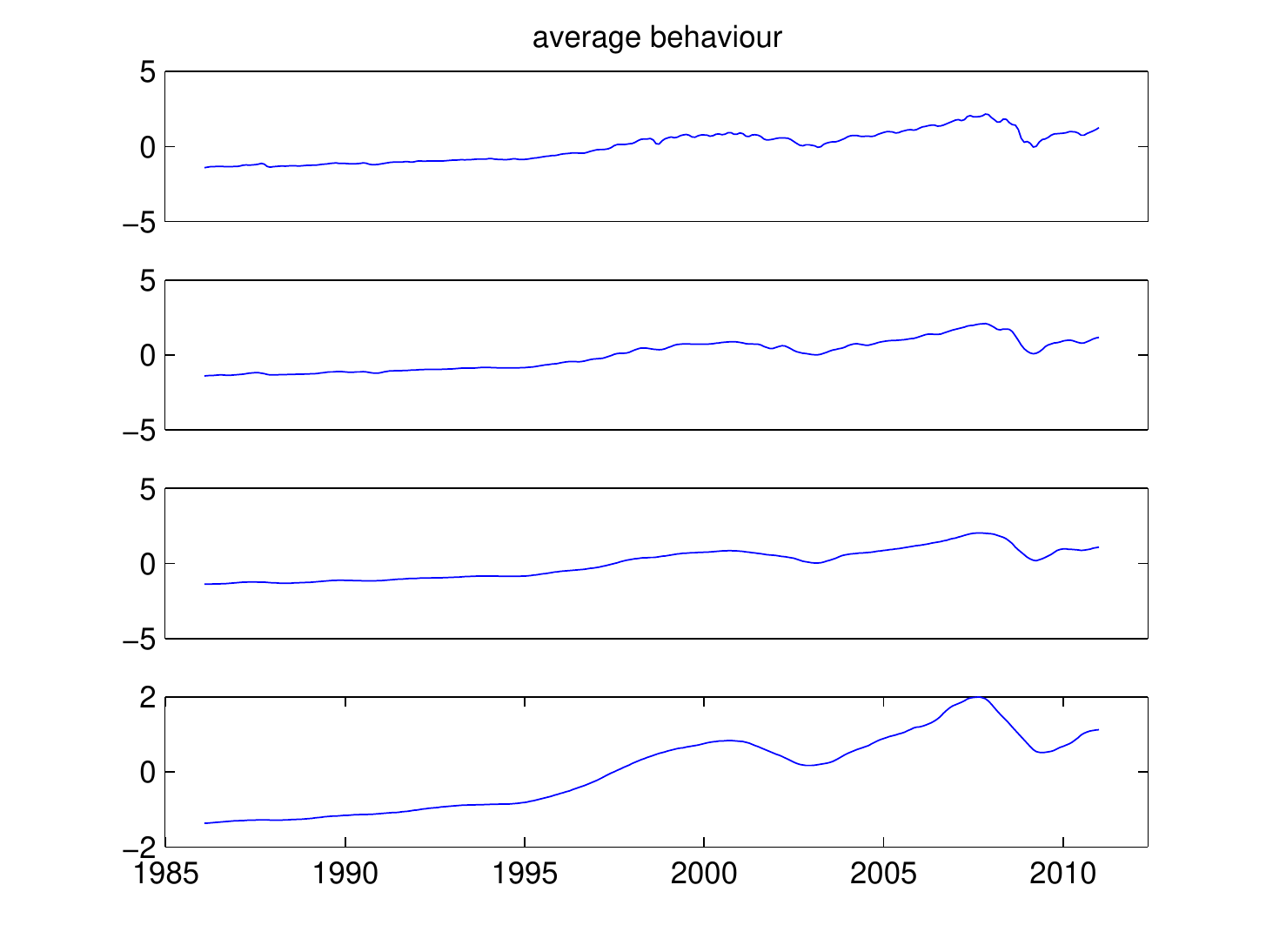}}
\subfigure[Fluctuations for the first \, \, \, four levels ]
{\includegraphics[height=4cm,width=4cm]{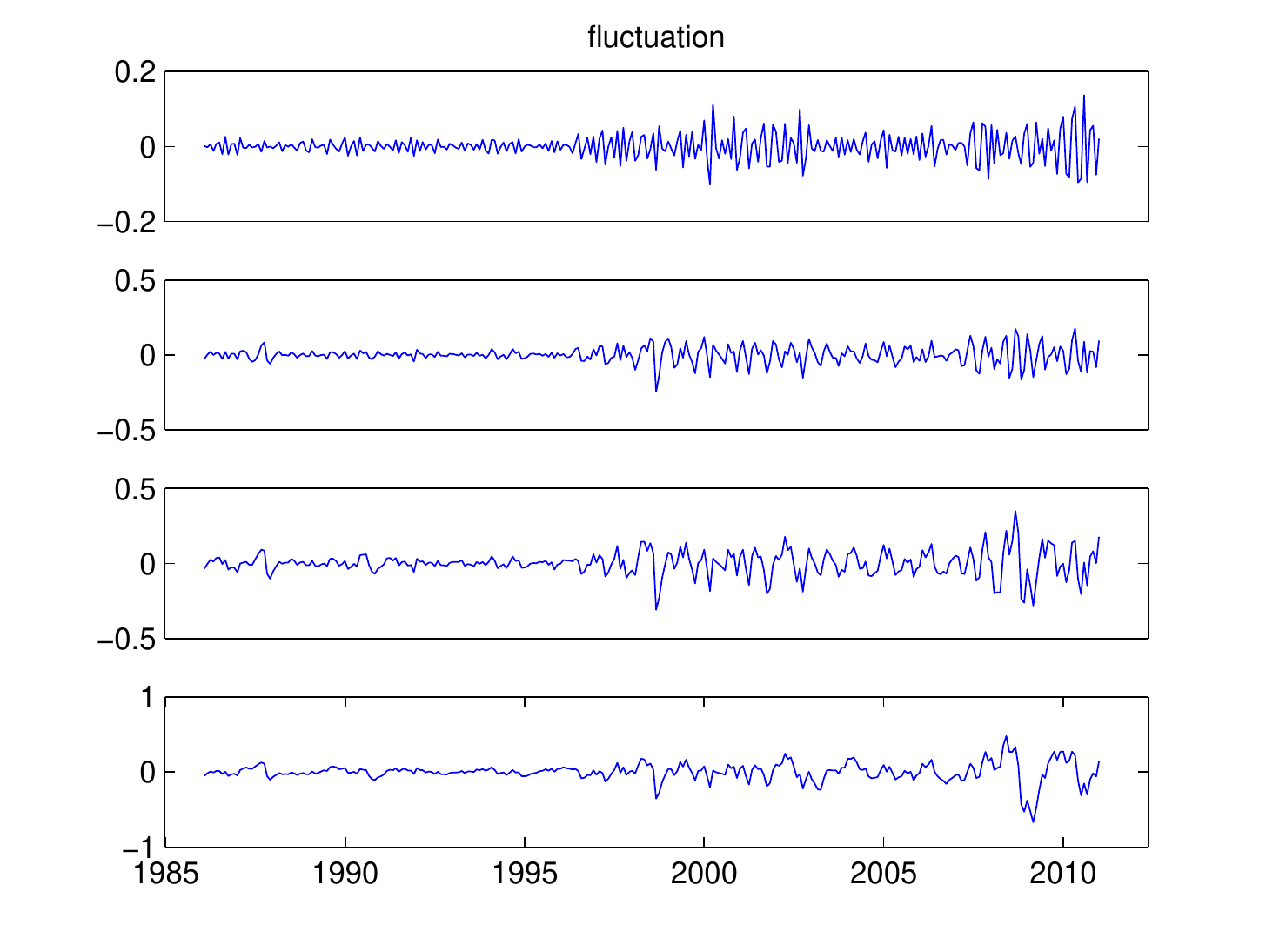}}
\caption{ Plots of (a) normalised {\small NYSE}  monthly average index  and (b) average behaviour  at multiple scales (c) fluctuations for the first four levels, showing similar characteristics as those of BSE}
\label{fig:low_high_nys_plot}
\end{figure}

The average behaviour of the two time series, Figs.\ref{fig:low_high_sen_plot}(a) and  \ref{fig:low_high_nys_plot}(a), over progressively longer data length are  captured in the local trends for the first four levels, as depicted in, Figs.\ref{fig:low_high_sen_plot}(b) and  \ref{fig:low_high_nys_plot}(b)~\cite{mani1,ghosh2011,pkp_statistical}. 
The fact that, the two time series reveal similar global behaviour at different scales, is evident in the average behaviour. 
For computing the variations at different scales, the following procedures are followed \cite{parikh2}.\\
As the low-pass coefficients capture the local linear trends in the data, over a progressively longer domain,  reconstruction of the time series using these low-pass coefficients  extract the  trend, in a corresponding window size. The fluctuations are then obtained at each level by subtracting the reconstructed averaged  time series, from the original data. 
The assymetric nature of the Daubechies wavelet  influences the precision of the resulted values. But, it can be corrected by extricating a new set of fluctuations, i.e., by applying wavelet transform on the reverse profile, followed by averaging the newly obtained fluctuations with the older one.
These variations of BSE and NYSE are depicted in Figs.\ref{fig:low_high_sen_plot}(c) and   \ref{fig:low_high_nys_plot}(c), showing volatile behaviour at small scales and  structured variations at progressively higher scales. The second half of the fluctuations are highly volatile for both the stock exchanges. \\
For testing  normality of the average behaviour and  fluctuations, {\small  Shapiro-Wilk (SW)}  and \\ {\small  Kolmogorov-Smirnov (KS)}  tests are conducted \cite{shapiro1,kolmo3}.
\begin{figure}[H]
\centering
\includegraphics[height=5cm, width=10cm]{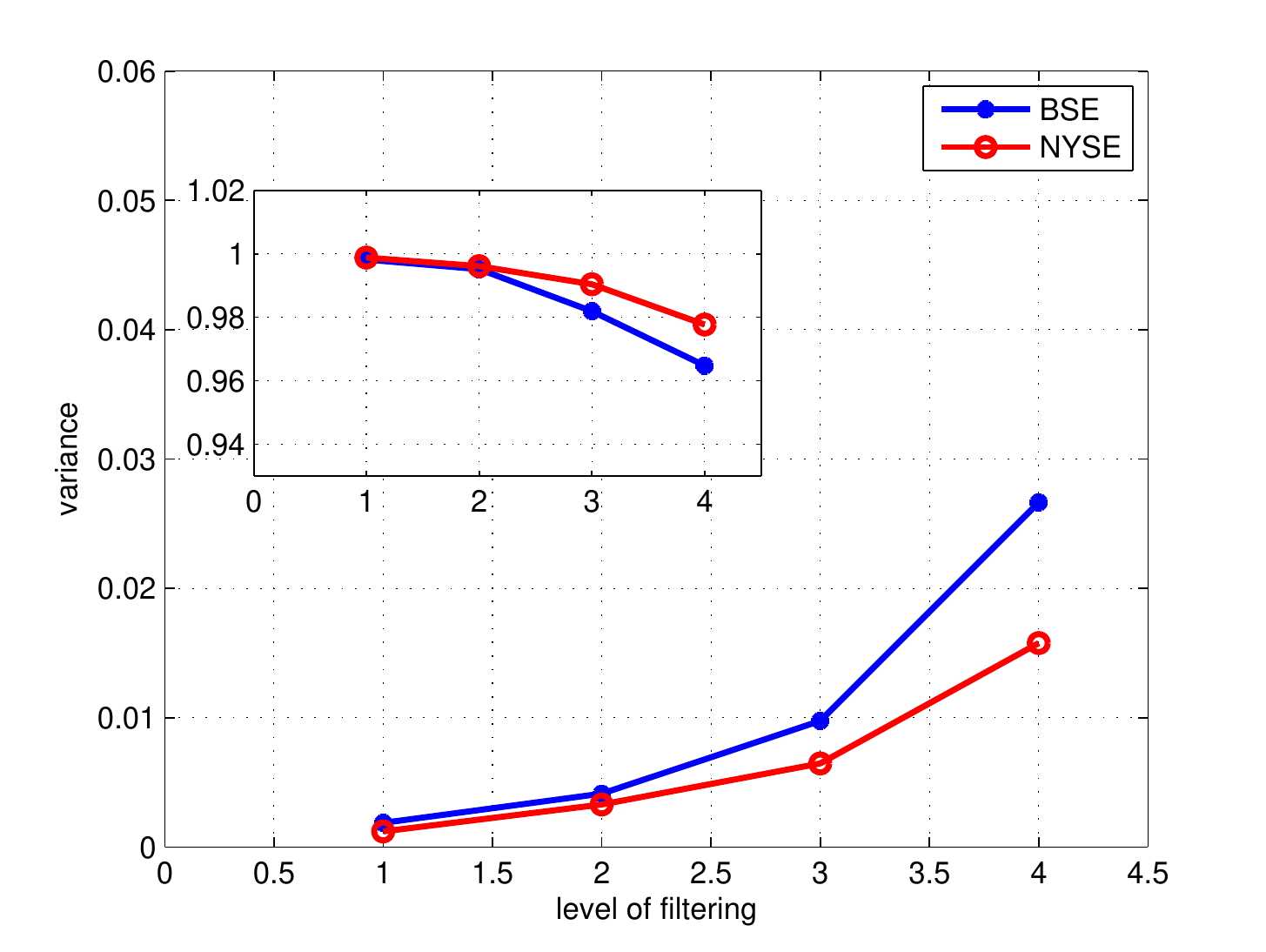}
\caption{The variance of the fluctuations at different levels of filtering, with the  inset showing the  complimentary variance corresponding to the average behaviour of BSE  and NYSE}
\label{fig:variance}
\end{figure}
Null hypothesis is rejected for both the tests. The details of the tests are provided  in the  supplement material, revealing that  fluctuations  do not belong to the normal distribution~\cite{shapiro1}. 
The variance of fluctuations and average behaviour of the two exchanges show broad similarities, at various scales, as shown in Fig.\ref{fig:variance}.

\begin{figure}[H]
\centering
\hfill
\subfigure[level 1]
{\includegraphics[height=4cm,width=4cm]{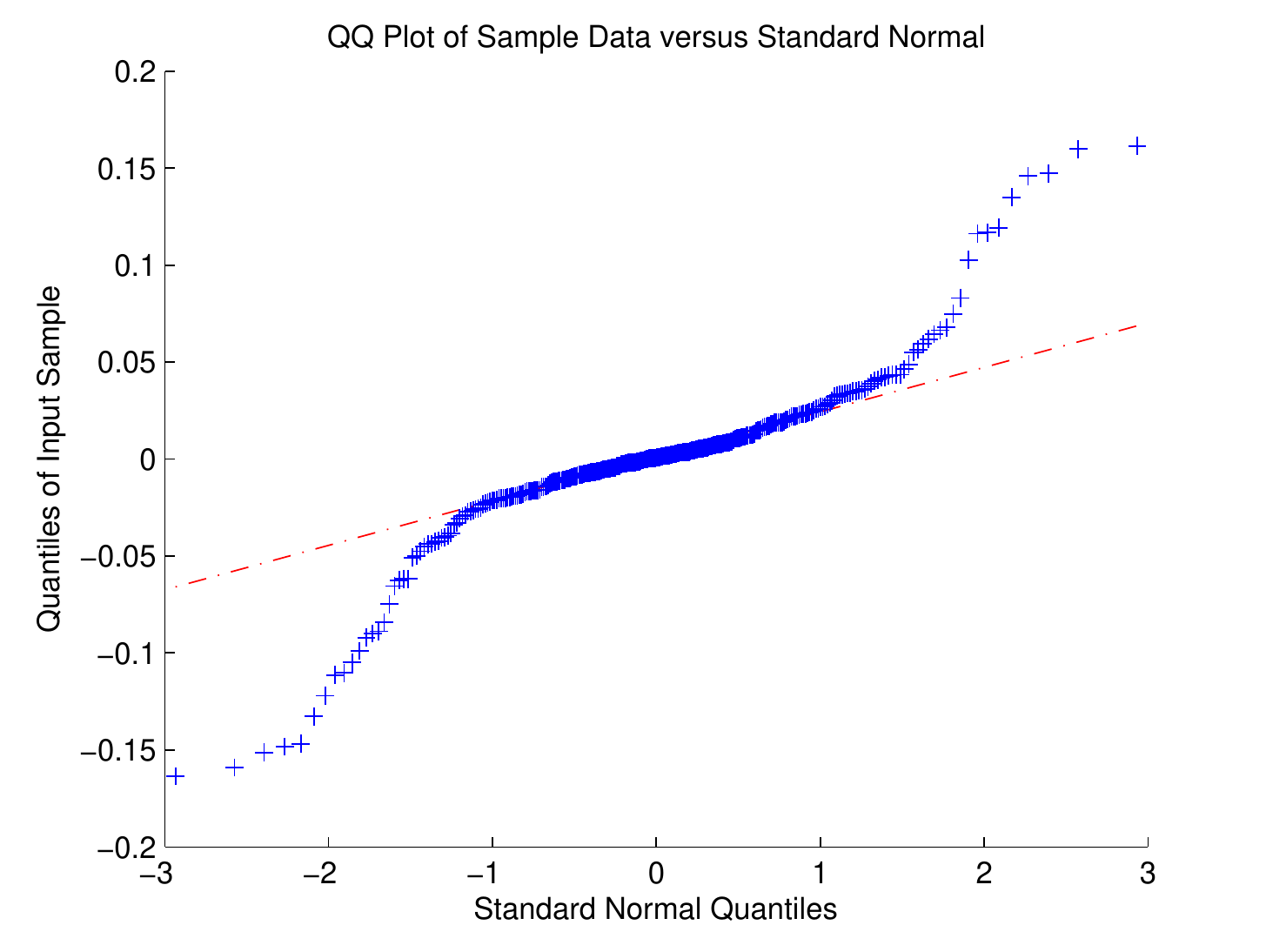}}
\hfill
\subfigure[level 2]
{\includegraphics[height=4cm,width=4cm]{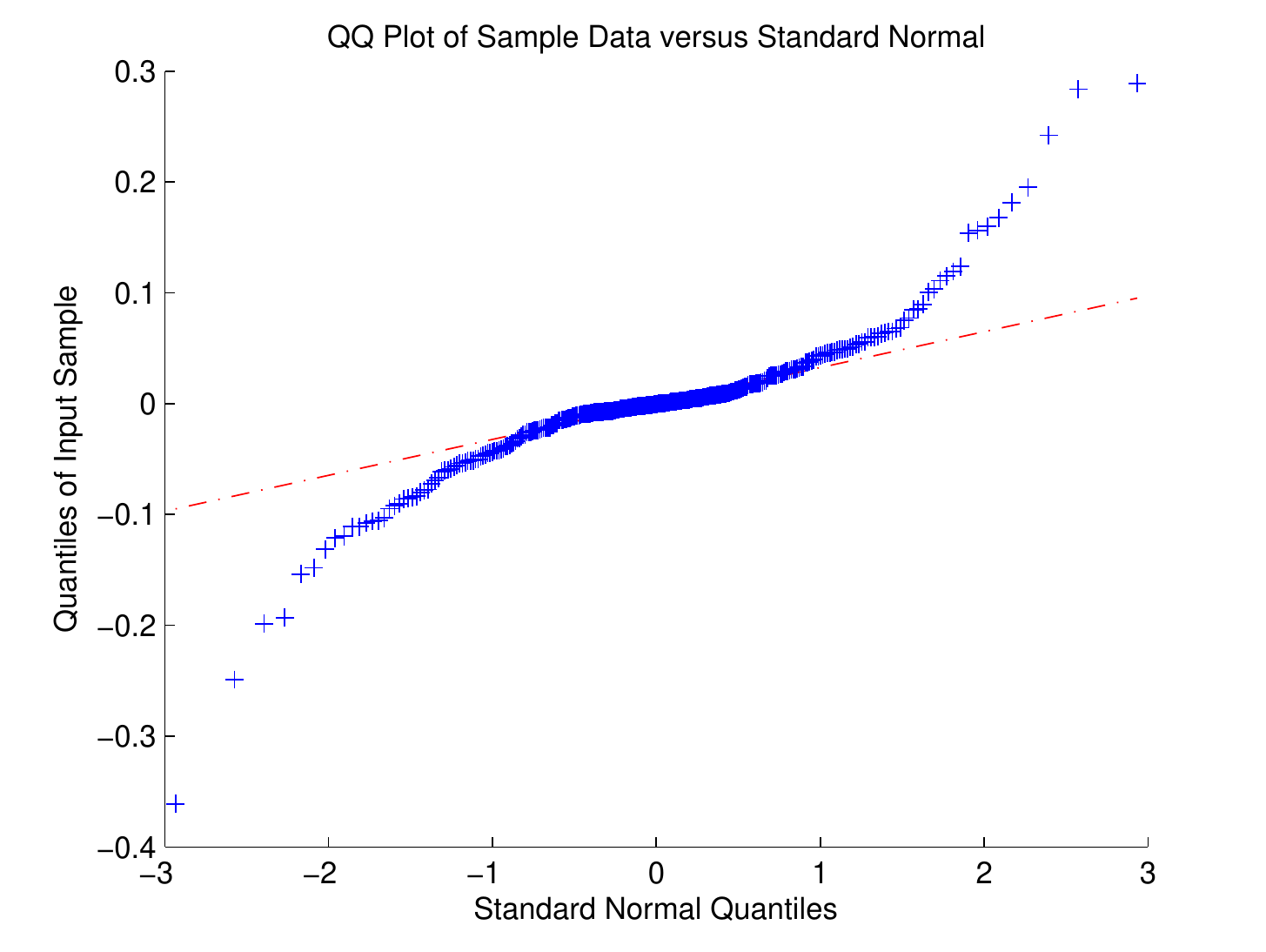}}
\hfill
\subfigure[level 3]
{\includegraphics[height=4cm,width=4cm]{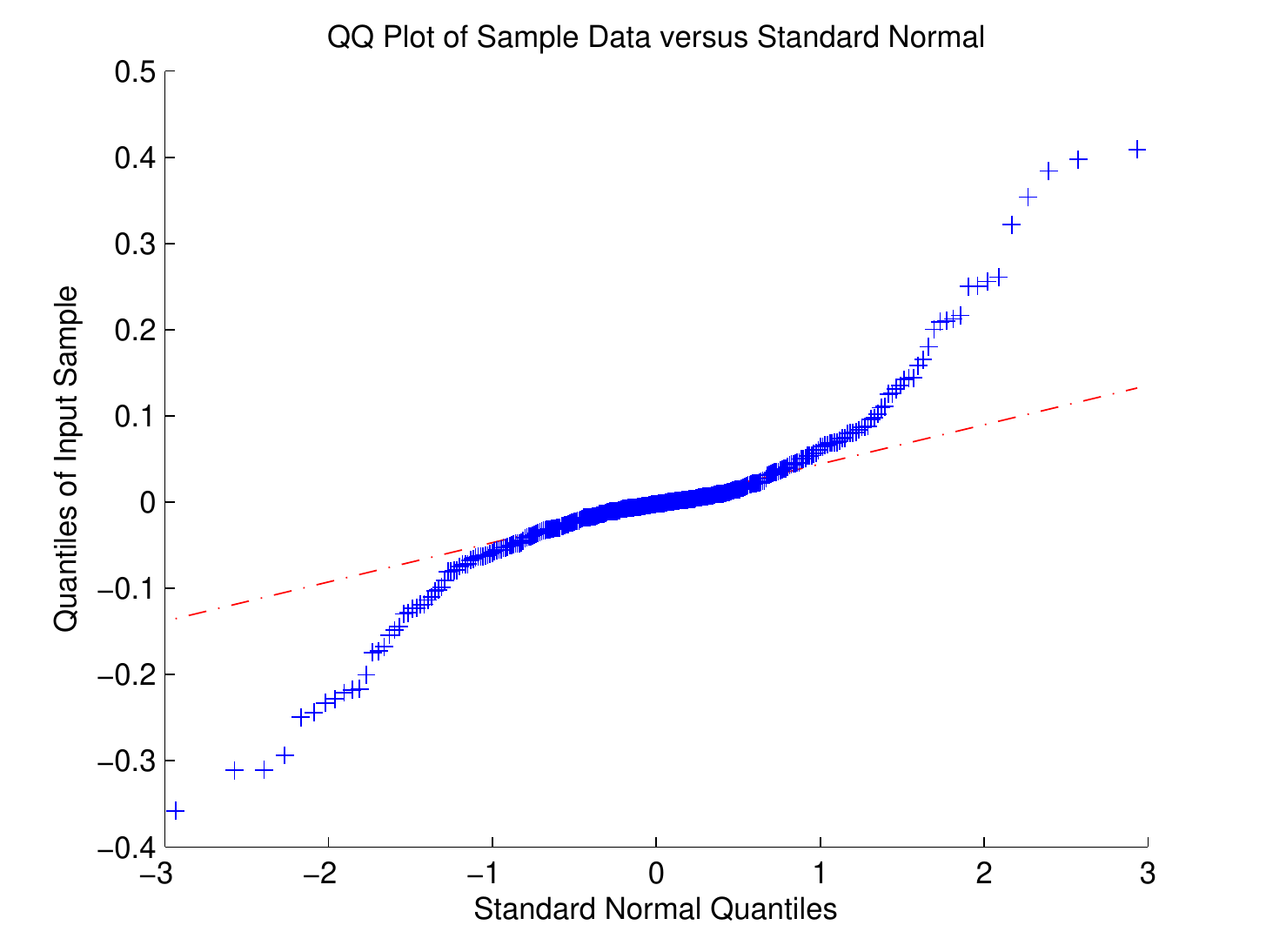}}
\hfill
\subfigure[level 4]
{\includegraphics[height=4cm , width= 4cm]{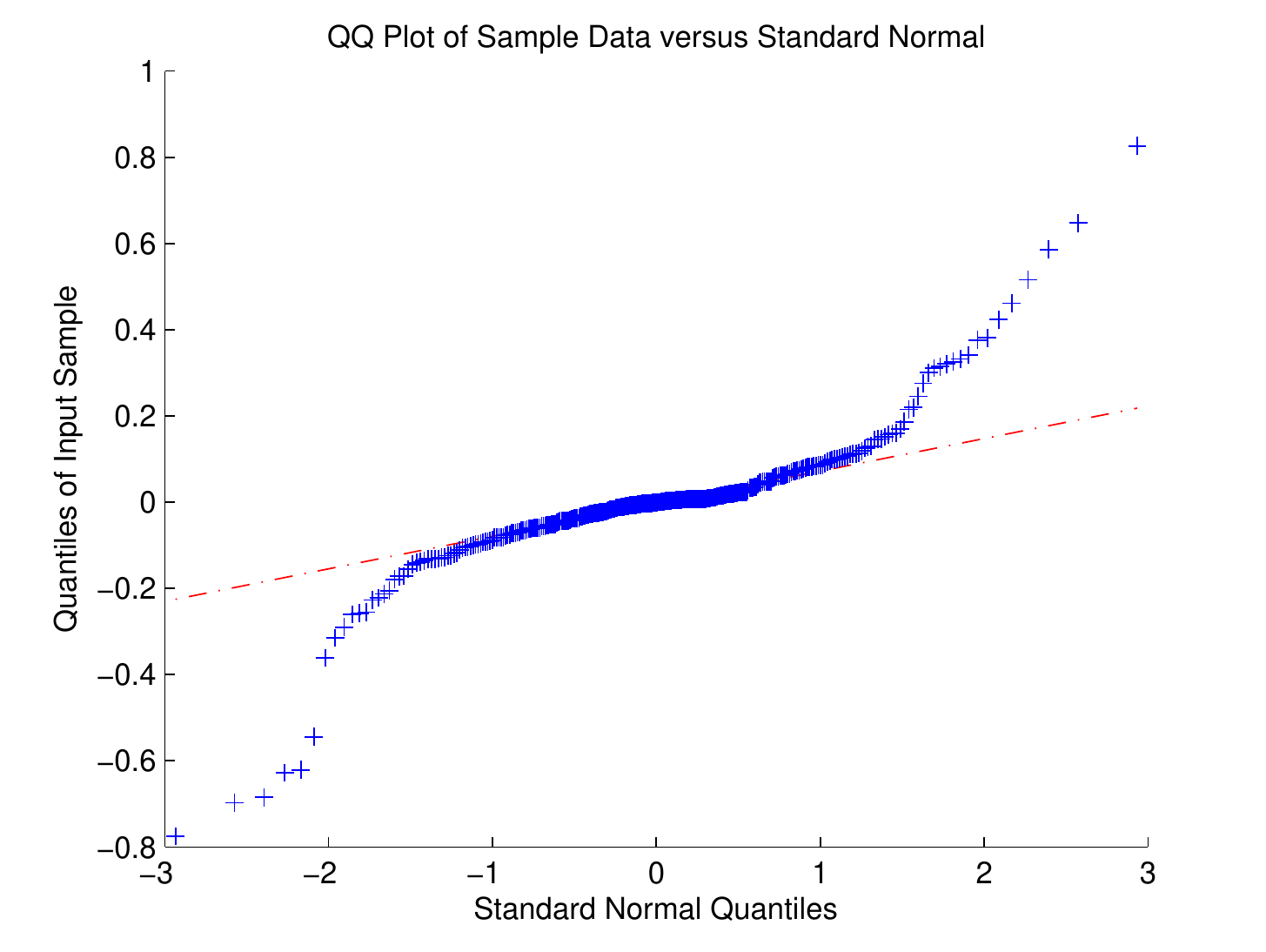}}
\hfill
\caption{Comparing the quantiles  of {\small BSE} fluctuations at different levels of filtering : (a)   1 (b)  2 (c) 3 and (d) 4,  with those of normal distribution, showing deviations at both the ends}
\label{fig:quant_h}
\end{figure}
\begin{figure}[H]
\centering
\hfill
\subfigure[level 1]
{\includegraphics[height=4cm,width=4cm]{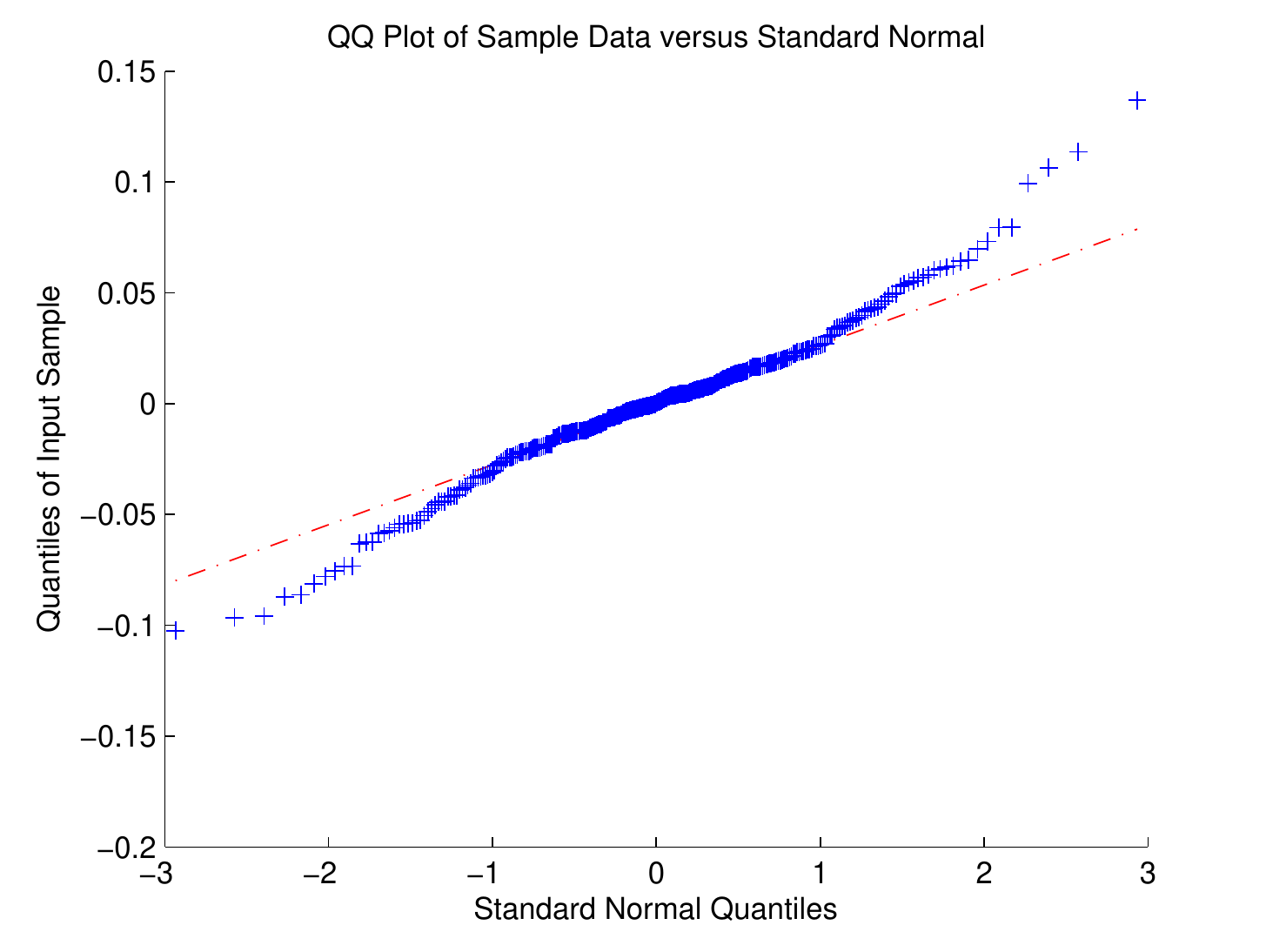}}
\hfill
\subfigure[level 2]
{\includegraphics[height=4cm,width=4cm]{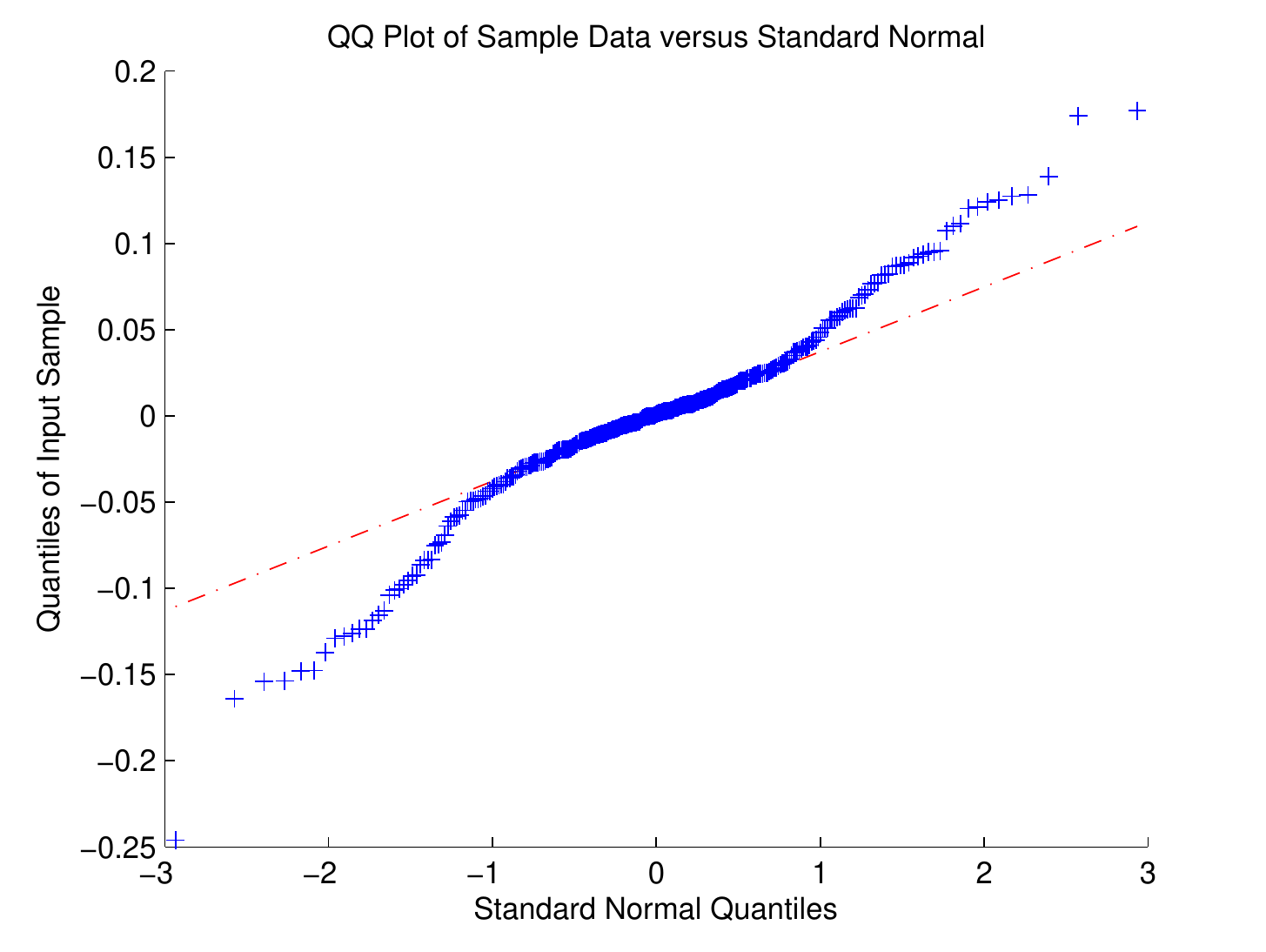}}
\hfill
\subfigure[level 3]
{\includegraphics[height=4cm,width=4cm]{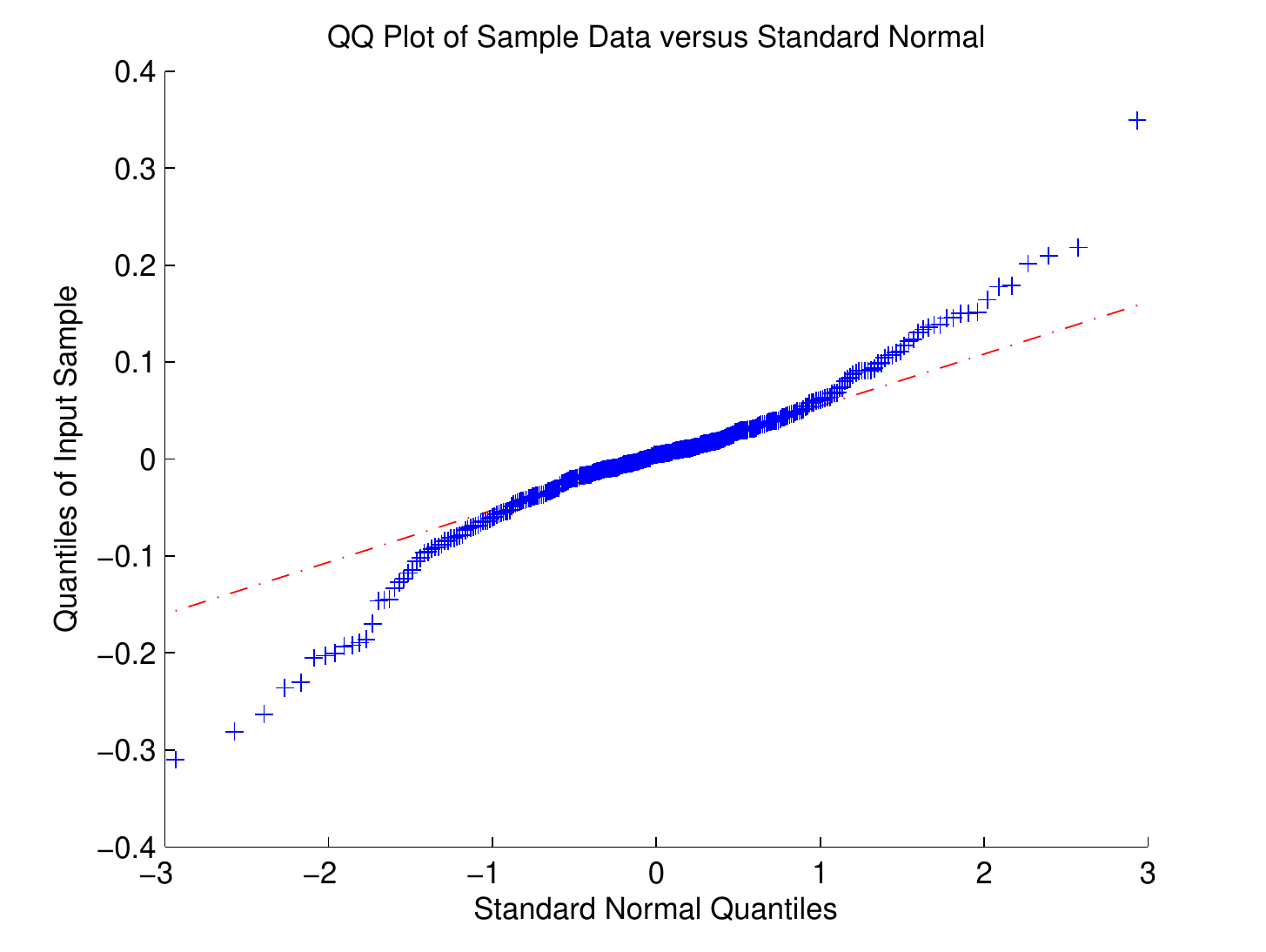}}
\hfill
\subfigure[level 4]
{\includegraphics[height=4cm , width= 4cm]{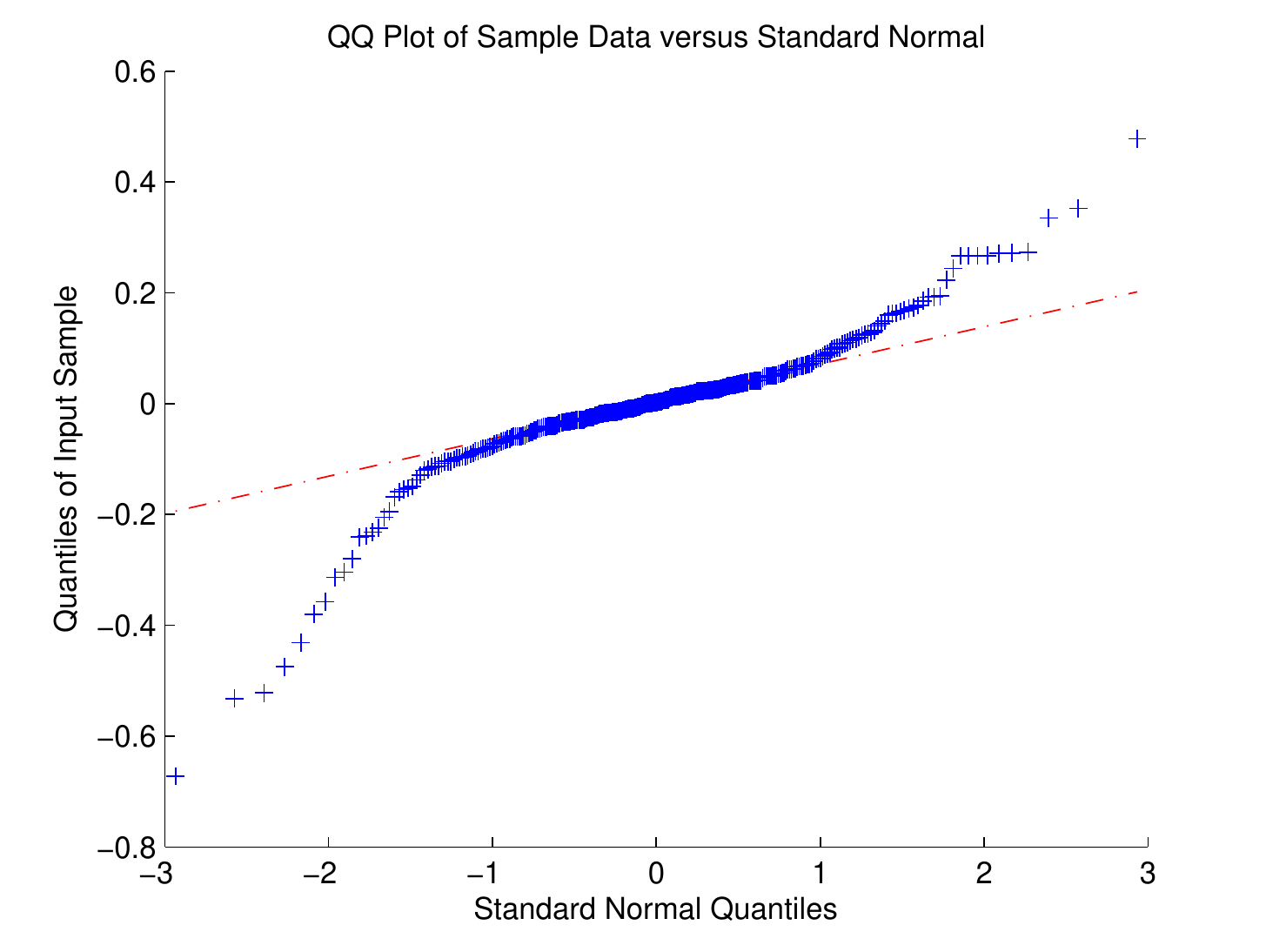}}
\hfill
\caption{Comparison of the quantiles of fluctuations for {\small NYSE} at levels, (a)  1 (b) 2 (c)  3 and (d) 4,  with those of normal distribution, showing deviations at both the ends}
\label{fig:h_quantile_N}
\end{figure}
For a more systematic understanding of the nature of variations at different scales, we  carry out a quantitative estimation of the deviations of the local variations from the normal distribution, taking recourse to Quantile distributions  \cite{quantile1}.
It uses medians, as compared to  the use of mean  in histograms, making  it  more resistant to outliers.\\

The distribution of fluctuations of {\small BSE} and {\small NYSE} for the first four levels, as shown in the Fig.\ref{fig:quant_h},   show a significant  digression from normal behaviour, which are characterised by the dotted lines.
The deviations are significant, both at lower and upper ends of the plots,  indicative of long-tail behaviour on both sides \cite{ghasemi}. Deviation from normality is more in case of {\small BSE} than {\small NYSE}, at all the four levels. The presence of outliers, is clearly revealed in the boxplots of Figs.\ref{fig:boxplot_low} and \ref{fig:boxplot_high},  for average behaviour and fluctuations, respectively.   The  red dotted line shows the  $50th$ percentile, i.e., median ($q_{2}$). If the notches in the box-plot, corresponding to various levels  do not overlap, then it can be concluded, with 95\% confidence, that the true medians differ. Here, the length of the whiskers have been specified as 1.5 times the length of the inter-quartile range, and points beyond that are considered as outliers, denoted  by $'+'$ sign in the box-plots. The  value 1.5 corresponds to approximately +/–2.7$\sigma$ and 99.3 coverage, if the data is normally distributed \cite{potter,naomi}.

\begin{figure}[H]
\centering
\includegraphics[height=8cm, width=10cm]{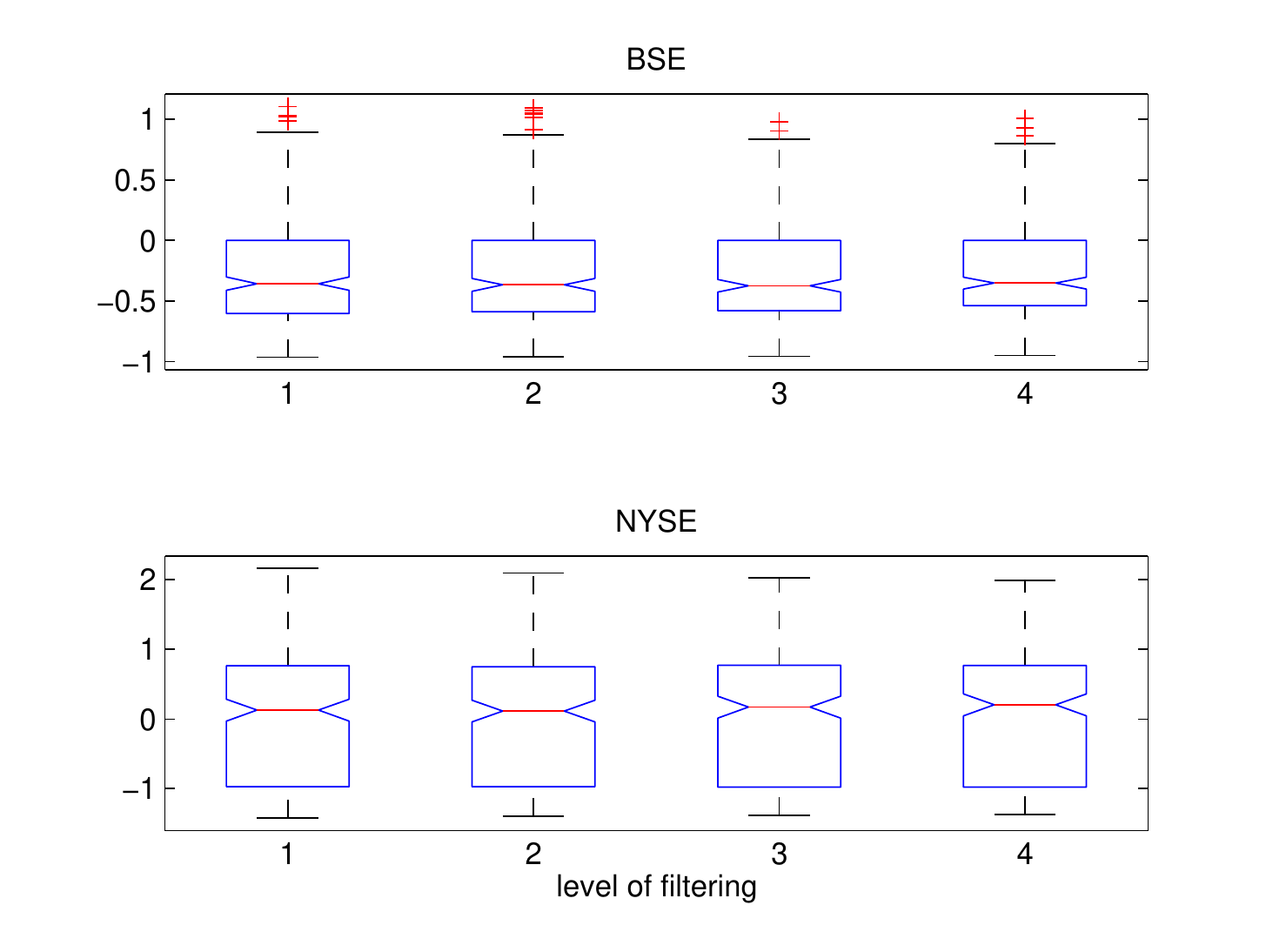}
\caption{Box-plots  showing the skewness and outliers of the average behaviour at different levels, for both the stock exchanges, with lower and upper boundary of the notched boxes representing  $25th$ ($q_{1}$) and $75th$ ($q_{3}$) percentiles, respectively. Notches corresponding to all the four levels overlap. The medians divide the boxes in uneven size, for both the stock  exchanges. In case of BSE, upper half is longer than the lower one, while, it  is vice-versa for NYSE}
\label{fig:boxplot_low}
\end{figure}
\begin{figure}[H]
    \centering
    \includegraphics[height=8cm,width=10cm]{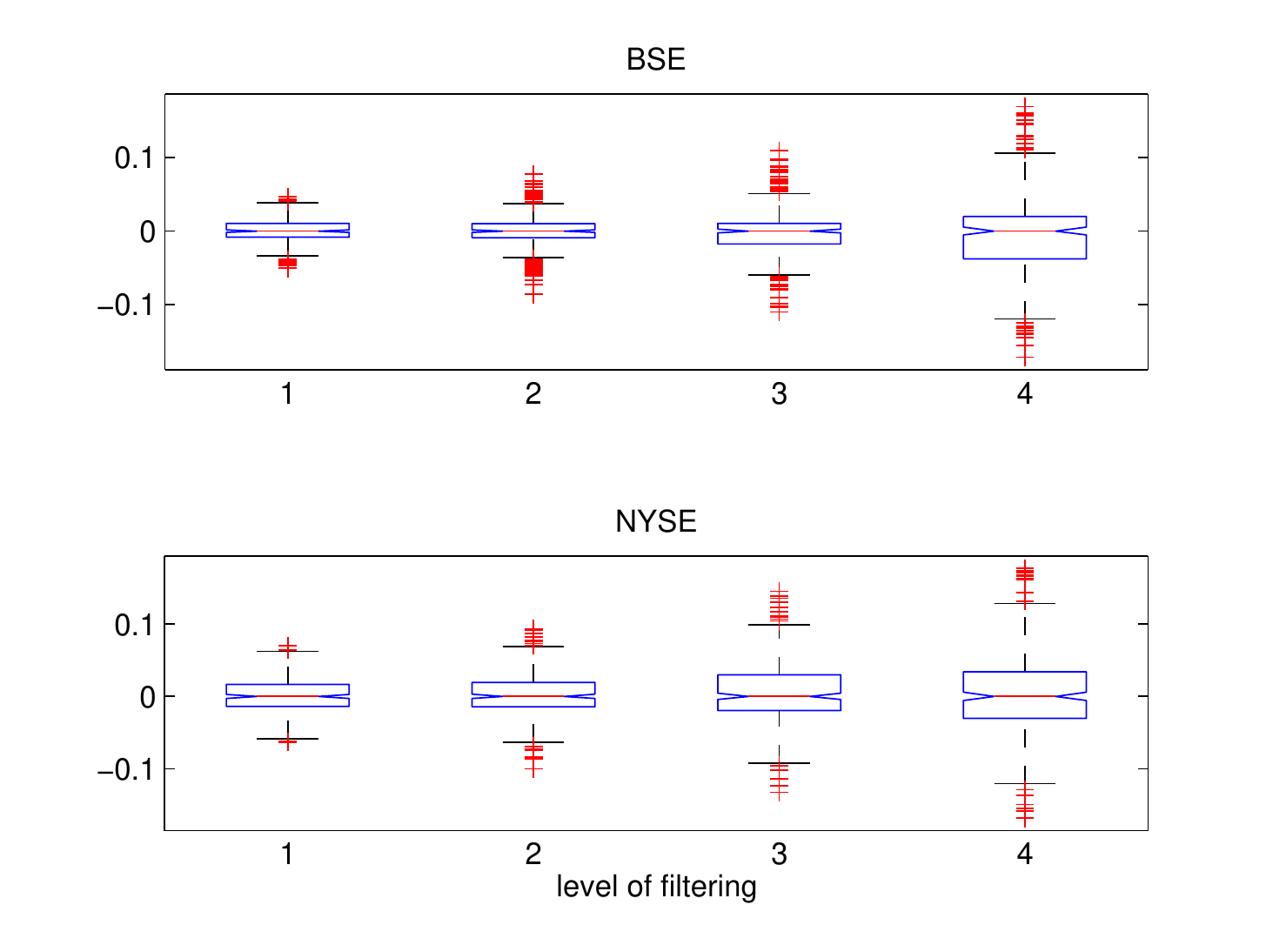}
    \caption{Notched box-plots representation of the fluctuations of  BSE and  NYSE stock exchanges, with similar labelling and qualitative results  as of the  average behaviour. The notches  overlap for both the stock exchanges, signifying that median is resistant to filtering  }
    \label{fig:boxplot_high}
\end{figure}
\begin{figure}[H]
    \centering
    \includegraphics[height=5cm,width=10cm]{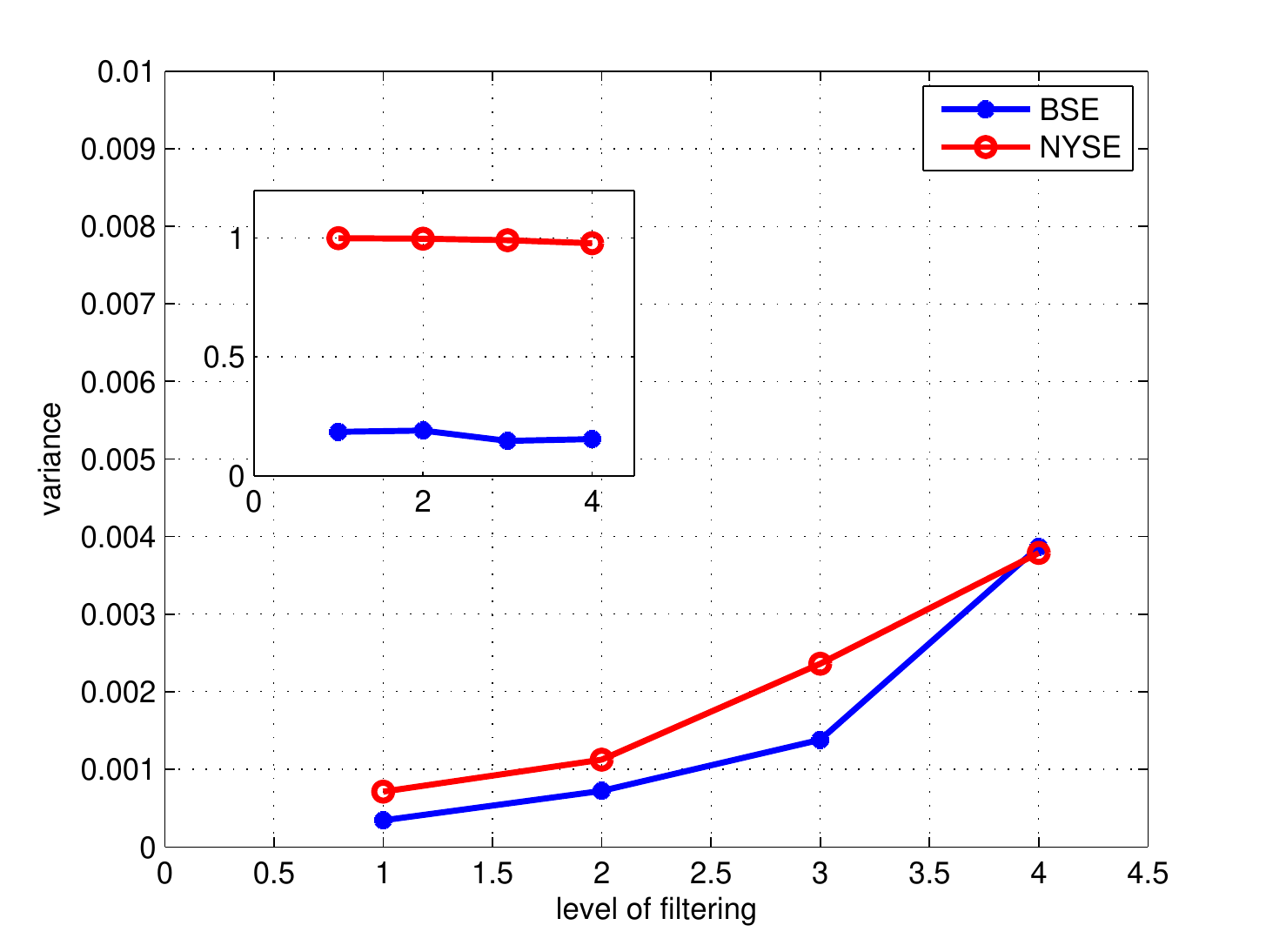}
    \caption{Plots of variance of fluctuations at different levels  with the  inset showing the distribution of variances for average behaviour after removal of the outliers. Significant differences  in the variances of  BSE  and NYSE, are evident  }
    \label{fig:out_variance}
\end{figure}

The average behaviour of BSE possesses outliers for all  the four levels, while in case of  NYSE there are none. The  fluctuations corresponding to both the stock exchanges have outliers, however, they are more in number for BSE, as shown in Figs.\ref{fig:boxplot_low}. However, once the outliers are removed, differences in the variance of the average behaviour of the two, get manifested, as seen in Fig.\ref{fig:out_variance}. The variance of BSE decreased drastically, while that of NYSE remained unchanged, as expected. 
It is physically tenable  that a developing economy like India, has more number of outliers \cite{cohen}. 
But it  also shows that variance in BSE is mainly due to the outliers, while for NYSE, they are not the only reason.  
In case of fluctuations, the two exchanges continued to exhibit similarity even after the removal of the outliers, except at level four, for which the decline is conspicuous.\\

\subsubsection{Pearson and Spearman Correlation}
\label{subsubsec:PS_coef}
\begin{table}[H]
\centering
\caption{Pearson correlation coefficients between BSE and NYSE monthly data for both average behaviour and fluctuations at various levels}
\begin{tabular}{|c|c|c|}
\hline 
 & Pearson Coefficients & \tabularnewline
\hline 
\hline 
 & Average behaviour  & Fluctuation \tabularnewline
\hline 
Level   & Correlation coefficients & Correlation  coefficients\tabularnewline
\hline 
1 & 0.7769 & 0.4532\tabularnewline
\hline 
2 & 0.7777 & 0.5025\tabularnewline
\hline 
3 & 0.7775 & 0.6974\tabularnewline
\hline 
4 & 0.7807 & 0.7341\tabularnewline
\hline 
\end{tabular}
\label{table:pearson}
\end{table}
\begin{table}[H]
\centering
\caption{Spearman coefficients between {\small BSE} and {\small NYSE} monthly data for both average behaviour and fluctuations for various levels}
 \begin{tabular}{|c|c|c|}
\hline 
 & Spearman Coefficients & \tabularnewline
\hline 
\hline 
 & Average behaviour  & Fluctuation \tabularnewline
\hline 
Level  & Correlation coefficients & Correlation  coefficients\tabularnewline
\hline 
 &  & \tabularnewline
\hline 
1 & 0.8969 & 0.4092\tabularnewline
\hline 
2 & 0.9013 & 0.3648\tabularnewline
\hline 
3 & 0.9073 & 0.4473\tabularnewline
\hline 
4 & 0.9121 & 0.4061\tabularnewline
\hline 
\end{tabular}

\label{table:spearman}
\end{table}

For estimations of the multi-scale linear and  monotonic correlations,  between the two stock exchanges, we compute both the Pearson  and Spearman correlations  of their  average behaviour and fluctuations.
Pearson and Spearman coefficients  corresponding to all the four levels are tabulated in Tables.\ref{table:pearson} and \ref{table:spearman}, respectively. The two stock exchanges exhibit monotonic relationship for all the four levels, as  Spearman correlation coefficients are higher than Pearson, revealing that throughout the two stock exchanges are moving concurrently, though the increment or decrement in the stock values for the two may not be equal.
In the case of fluctuations, Pearson coefficients are higher than Spearman, indicating that linear relationship is stronger than a monotonic one.

\subsubsection{Multi-scale Probability Density }\label{subsec:pd} 
Probability density estimation of the average behaviour and fluctuations are carried out, with the help of  Kernel smoothing  (KS) operations. As is known, Kernel is a special type of Probability Density Function (PDF) with the  property of being non-negative, real valued, even and normalised, over its support value:
\begin{equation}
\ \int_{-\infty}^{\infty} K(u) du=1\;
\label{eq:ker1}
\end{equation}

\begin{equation}
\ K(-u)=K(u)\, 
\label{eq:ker2}
\end{equation}
%
Kernel smoothing   is a non-parametric method of estimating PDF, as it does not assume any underlying distribution for a variable. At every data point, a Kernel is created with the data point at the centre to ensure that the Kernel is symmetric about it. The PDF is then estimated by adding all these Kernel functions and dividing by the number of data.\\ 

\begin{figure}
\centering
\hfill
\subfigure[level 1 ]
{\includegraphics[height=5cm,width=4.5cm]{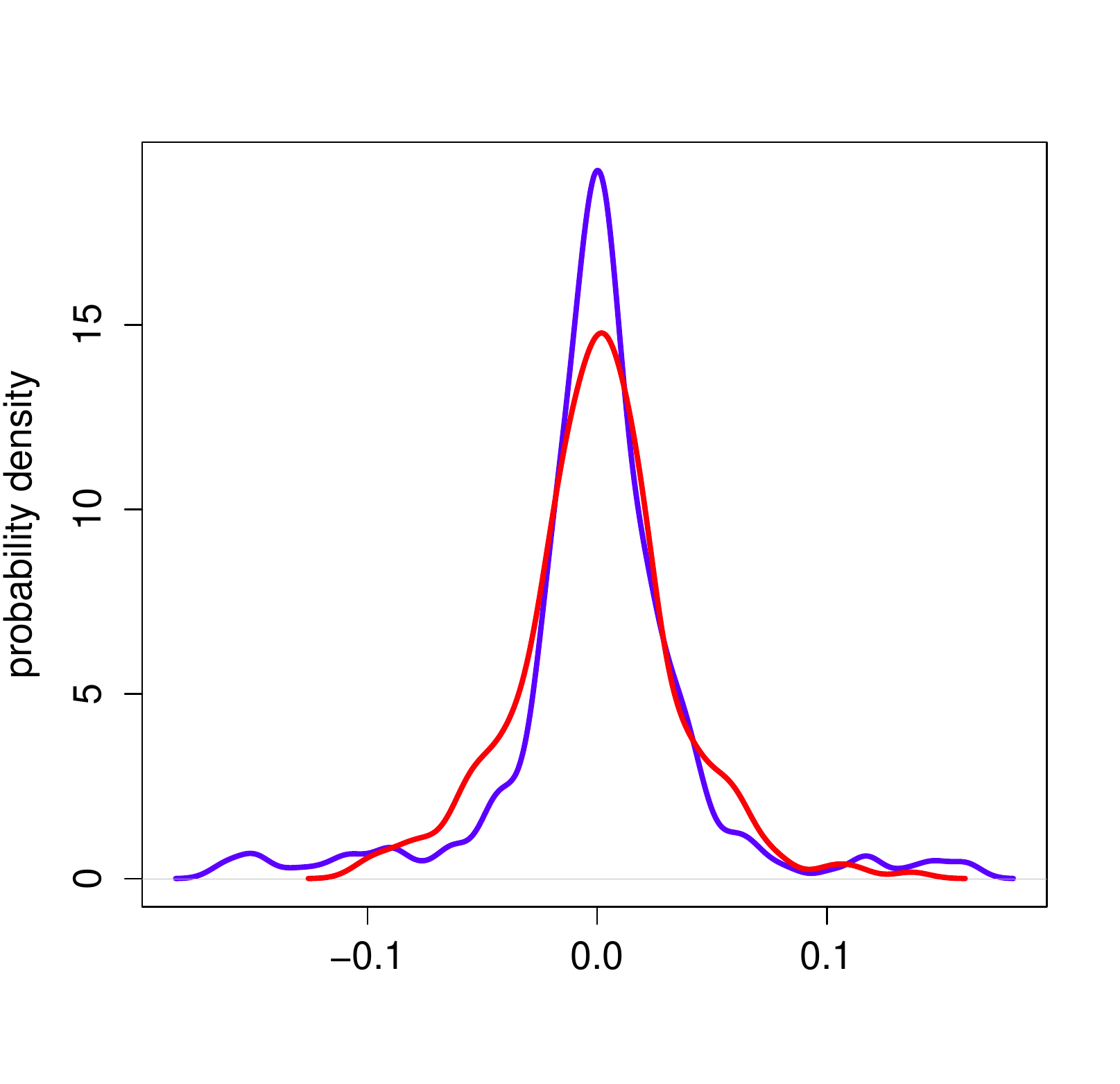}}
\hfill
\subfigure[level 2 ]
{\includegraphics[height=5cm,width=4cm]{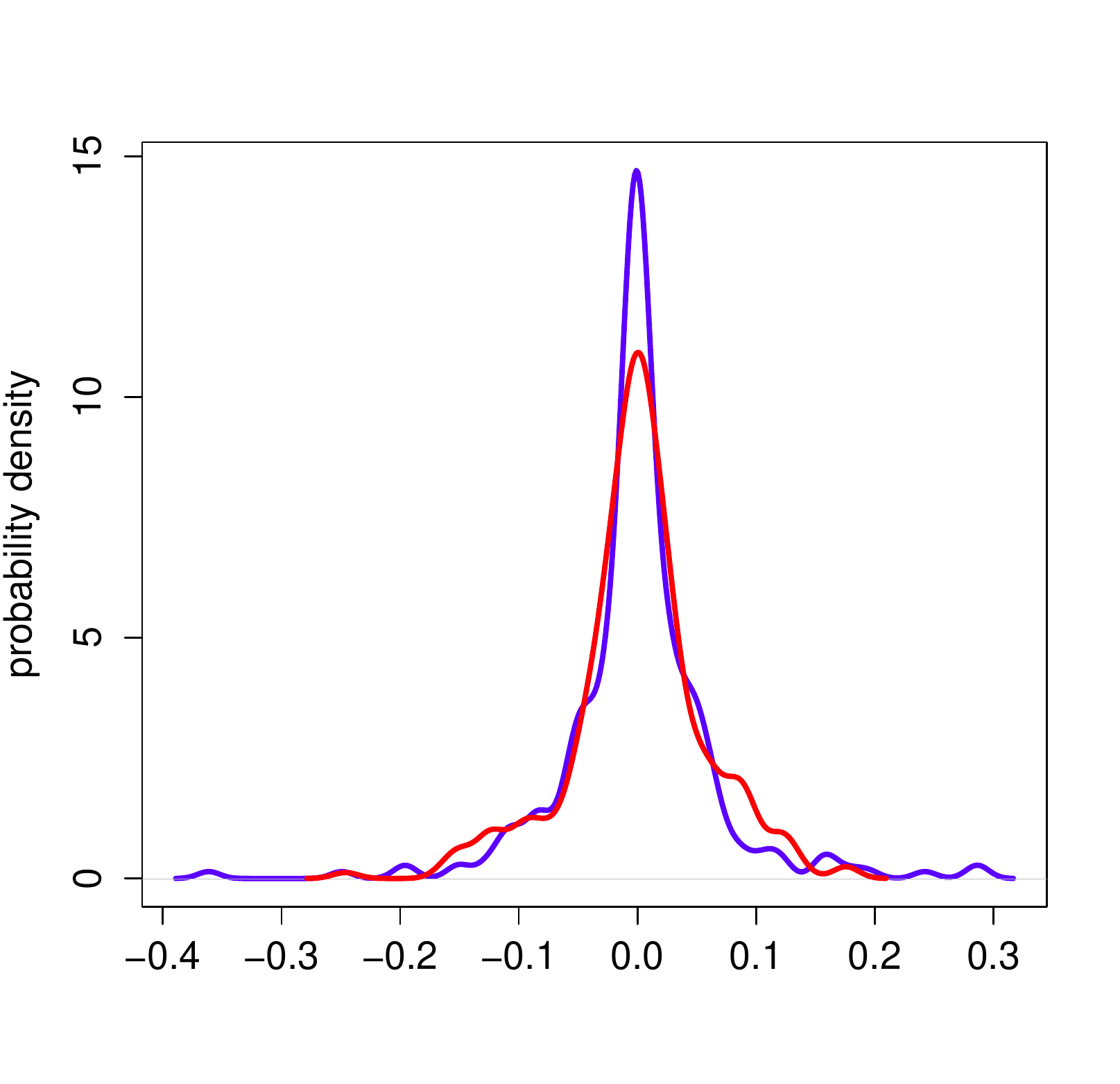}}
\hfill
\subfigure[level 3 ]
{\includegraphics[height=5cm,width=4cm]{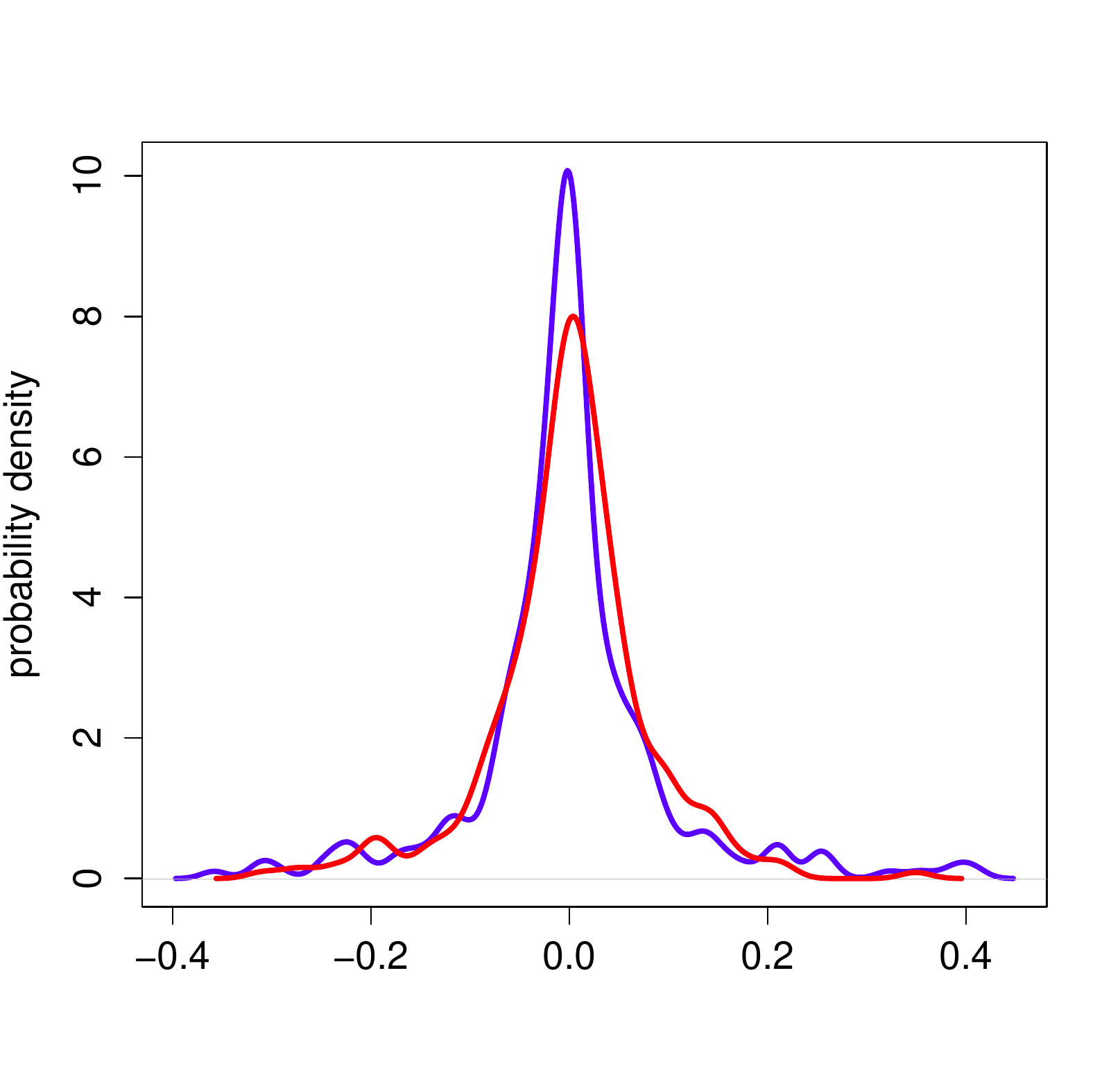}}
\hfill
\subfigure[level 4 ]
{\includegraphics[height=5cm,width=4cm]{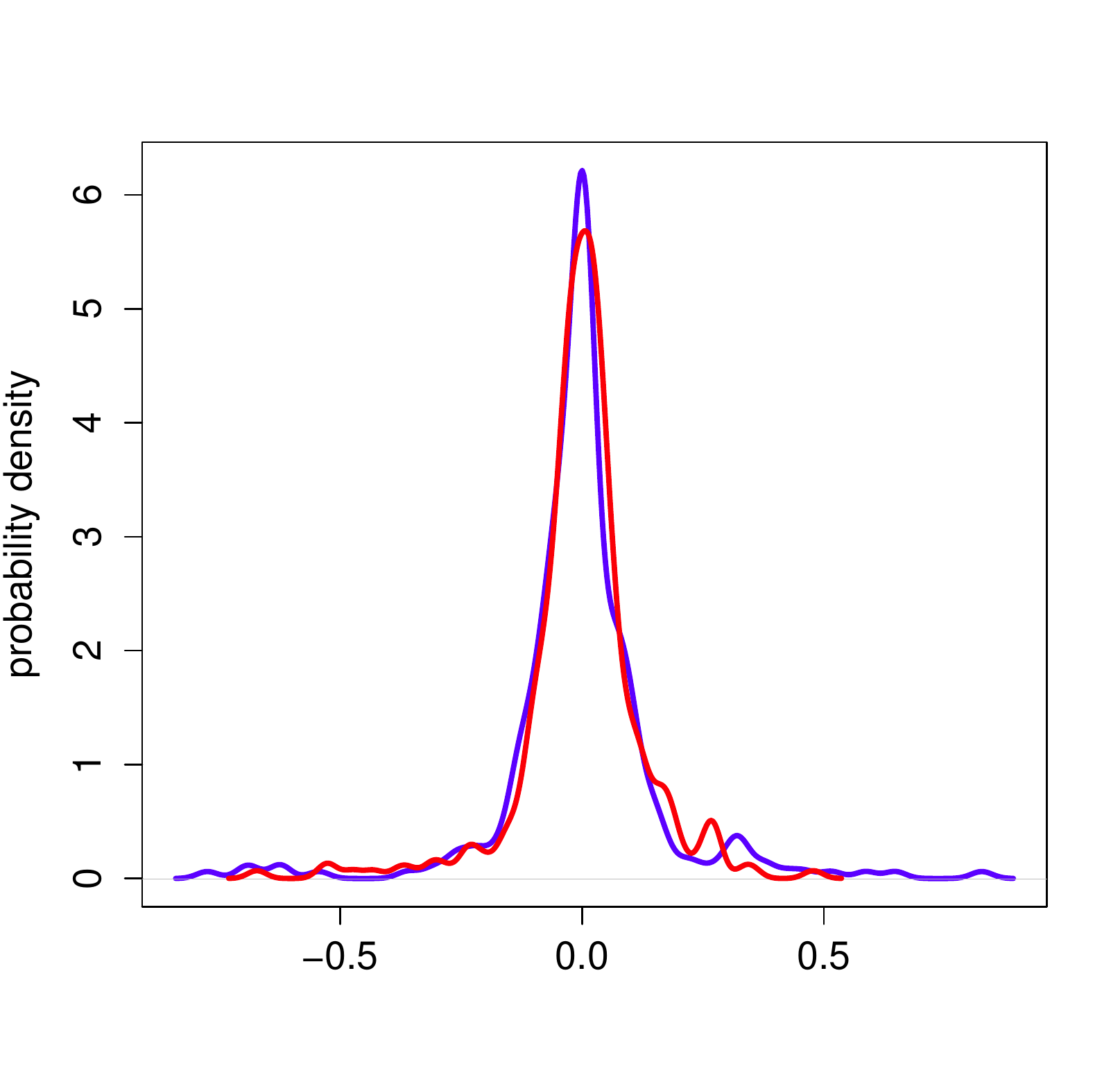}}
\hfill

\caption{Juxtapositions of  the probability density of fluctuations  of {\scriptsize BSE} (blue) and {\scriptsize NYSE } (red) for levels  (a) 1 (b) 2 (c) 3 and (d) 4, showing strong leptokurtic nature,  more pronounced for BSE}
\label{fig:h_pdf}
\end{figure}
Average behaviour of both the stock exchanges exhibit bi-modal nature, it is symmetric for NYSE and  asymmetric with a pronounced rightly skewed tail for BSE, as seen in Fig.\ref{fig:lpdf}. 
The  distributions are rightly skewed and leptokurtic, and  symmetrical and platykurtic, at various scales for BSE and NYSE, respectively. The risk of having extreme events due to outliers are low in NYSE, as skewness value is high for BSE and low for  NYSE.
The tails in probability density plots, at various levels, got truncated on the removal of outliers for BSE, as seen in Fig.\ref{fig:wo_lpdf}.
Skewness and kurtosis coefficients have been tabulated in Tables.\ref{tab:high_sk} and \ref{tab:ave_sk}, 
which show that, in the case of fluctuations, the probability density is highly leptokurtic for BSE in comparison to that of NYSE, as seen in Fig.\ref{fig:h_pdf}. The kurtosis values are unaffected even after the advent of structured variations, revealing that occurrence of extreme events are not limited to certain scales, rather they are  ubiquitous across all the scales. Such a behaviour is more pronounced in BSE.
Stronger leptokurtic behaviour  bestowed with fat tails and thus extreme fluctuations, on both sides, are more probable in BSE,  making  it more susceptible  to risks. As seen earlier, it is also observed  in the boxplots of Fig.\ref{fig:boxplot_high}. The distributions are symmetrical for both the stock exchanges, as the skewness values are close to zero. 
%
\begin{table}
\centering
\caption{Skewness and kurtosis corresponding to the fluctuations of the two stock exchanges.}
\label{tab:high_sk}
\begin{tabular}{|c|c|c|c|c|c|}
\hline 
 &  &             Fluctuation &  &  & \tabularnewline
\hline 
\hline 
 &  &  &  &  & \tabularnewline
\hline 
 &  &   Skewness &  &              Kurtosis & \tabularnewline
\hline 
Level & BSE & NYSE &  & BSE & NYSE\tabularnewline
\hline 
1 & -0.23088 & 0.14554 &  & 7.5069 & 4.4489\tabularnewline
\hline 
2 & -0.013232 & -0.32914 &  & 10.041 & 4.7613\tabularnewline
\hline 
3 & 0.49524 & -0.3104 &  & 7.2145 & 5.746\tabularnewline
\hline 
4 & -0.17078 & -1.0385 &  & 10.977 & 8.9887\tabularnewline
\hline 
\end{tabular}
\end{table}
\begin{figure}[H]
\centering
\hfill
\subfigure[level 1 ]
{\includegraphics[height=6cm,width=4cm]{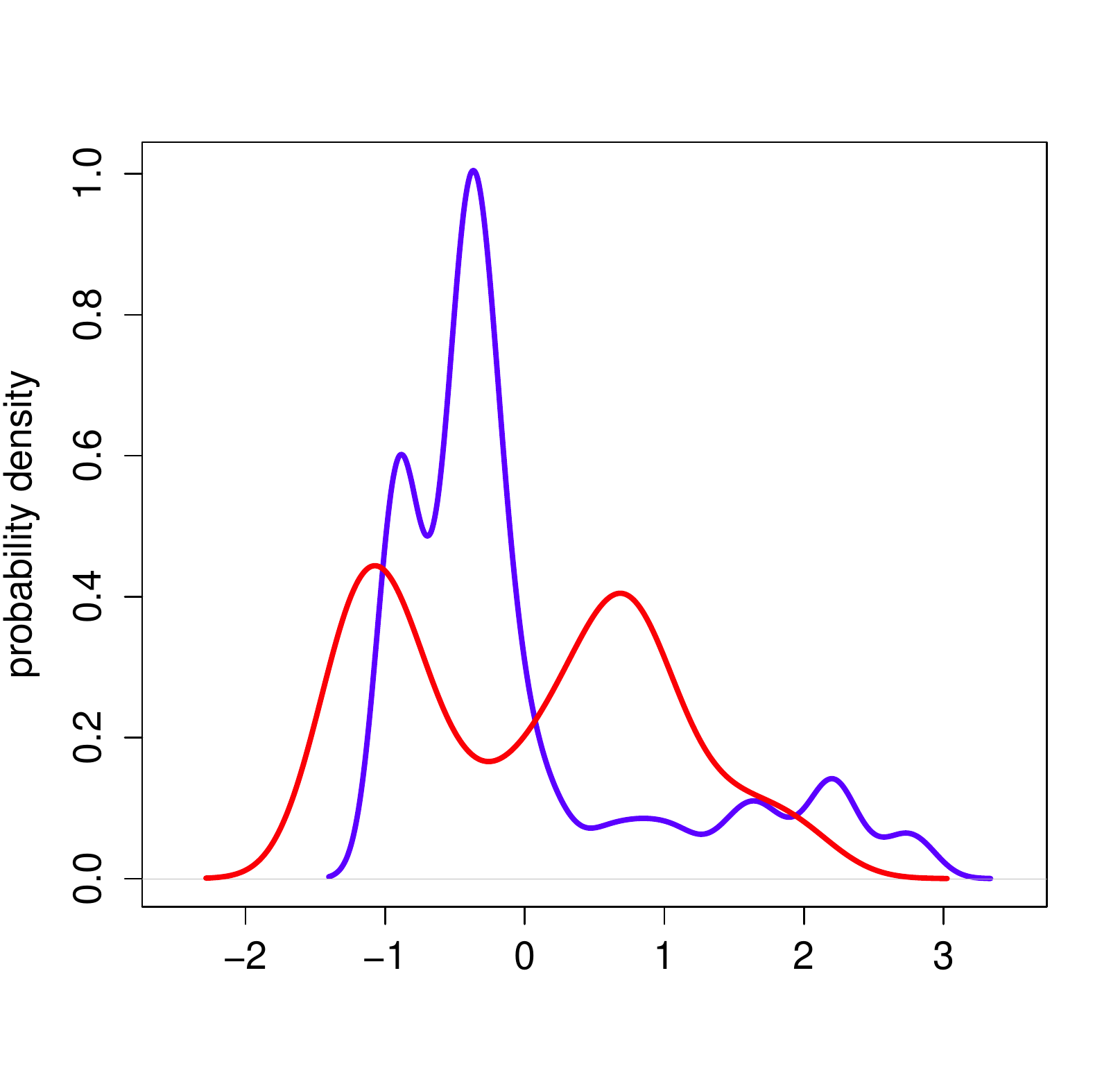}}
\hfill
\subfigure[level 2 ]
{\includegraphics[height=6cm,width=4cm]{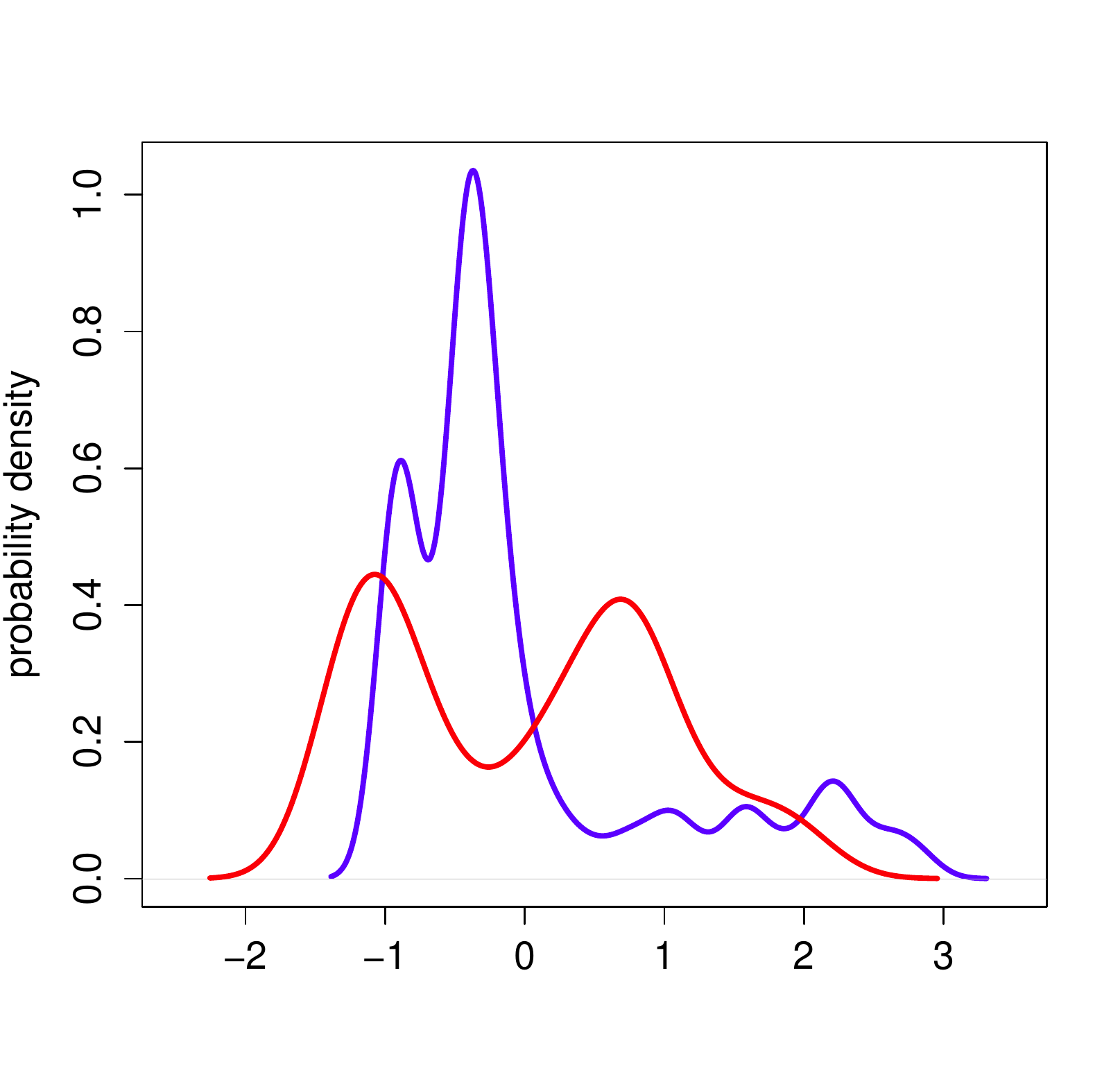}}
\hfill
\subfigure[level 3 ]
{\includegraphics[height=6cm,width=4cm]{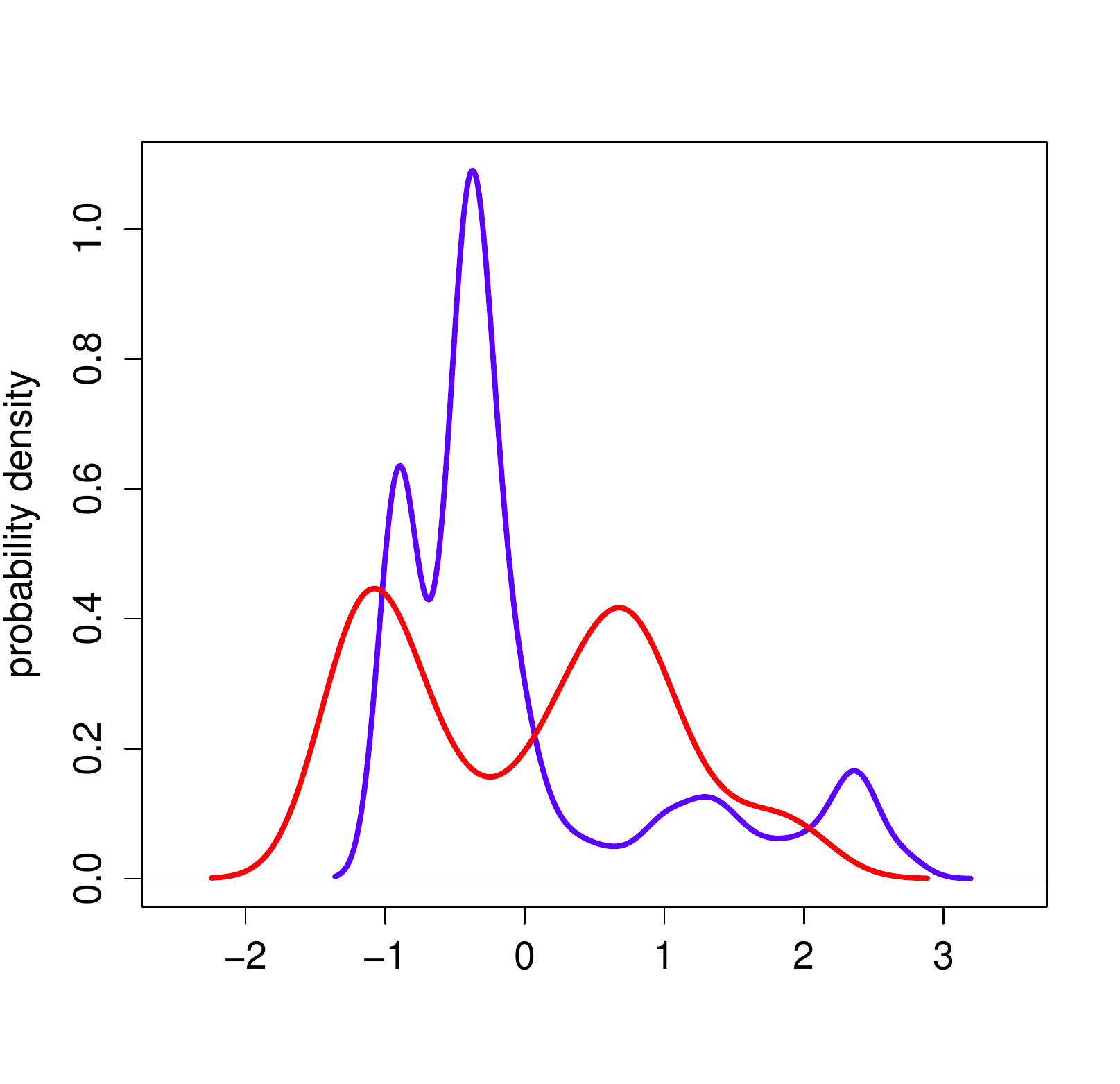}}
\hfill
\subfigure[level 4 ]
{\includegraphics[height=6cm,width=4cm]{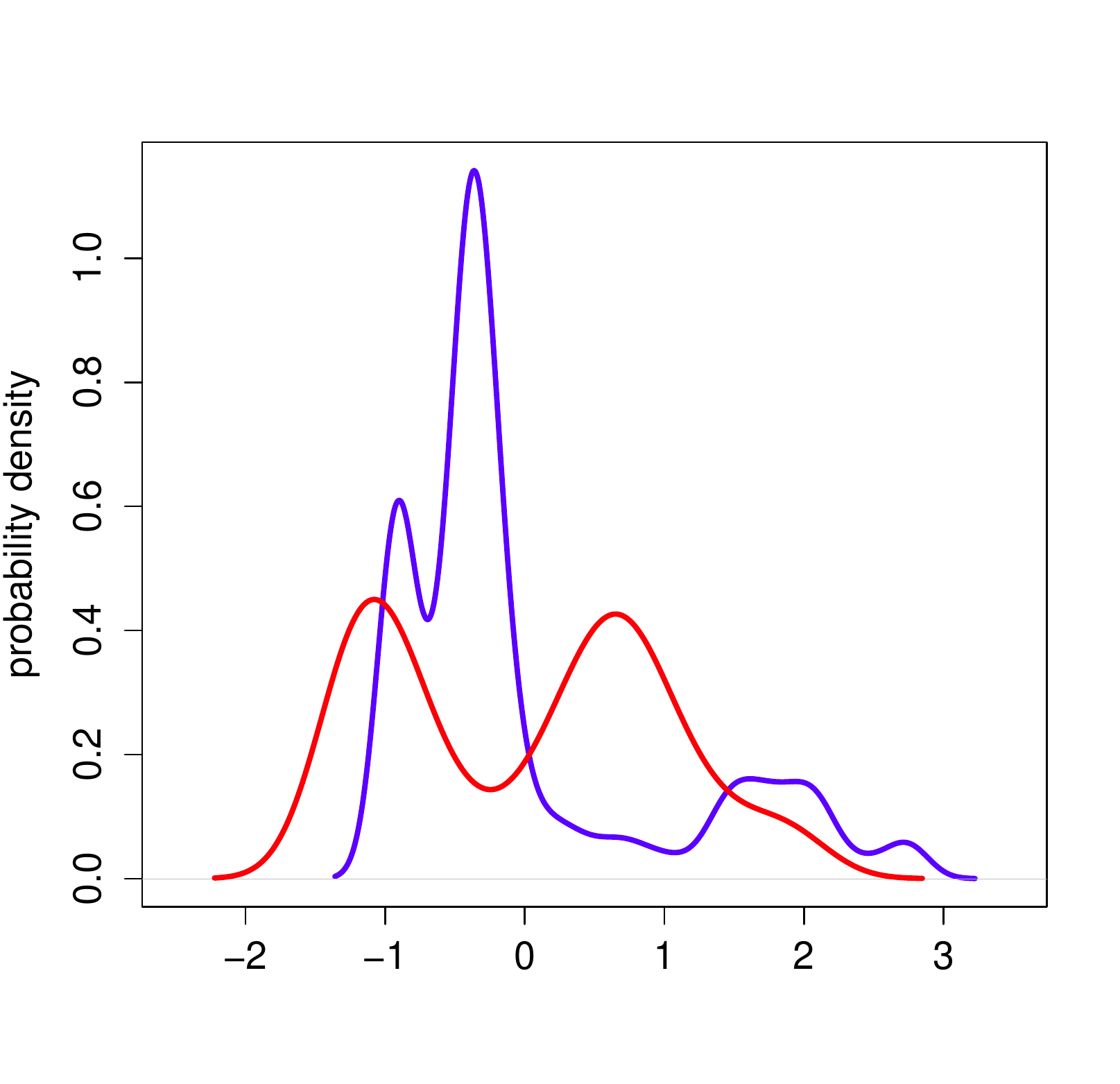}}
\hfill

\caption{Juxtapositions of  the bi-modal probability densities of average behaviour  of  {\scriptsize BSE} (blue) and  {\scriptsize NYSE } (red) for levels  (a)  1 (b)  2   (c)  3  and (d)  4, showing significantly different behaviour}
\label{fig:lpdf}
\end{figure}
\begin{table}
\caption{  Skewness and kurtosis  values, corresponding to the average behaviour, for  four levels of filtering.}
\label{tab:ave_sk}
\begin{tabular}{|c|c|p{1.8cm}|c|p{1.2cm}|}
\hline 
 &     &          Average behaviour  &  &  \tabularnewline
\hline 
\hline
 &  &  &  &   \tabularnewline
\hline 
 &         Skewness   &  &   Kurtosis & \tabularnewline
\hline 
Level & BSE & NYSE &   BSE & NYSE\tabularnewline
\hline 
1 & 1.4119 & 0.18182 &  3.8667 & 1.8049\tabularnewline
\hline 
2 & 1.4076 & 0.17737 &  3.8454 & 1.8001\tabularnewline
\hline 
3 & 1.3822 & 0.1692 &   3.745 & 1.8003\tabularnewline
\hline 
4 & 1.3351 & 0.13364 &  3.5765 & 1.7515\tabularnewline
\hline 
\end{tabular}
\end{table}
%

\begin{figure}
\centering
\hfill
\subfigure[level 1 ]
{\includegraphics[height=5cm,width=4cm]{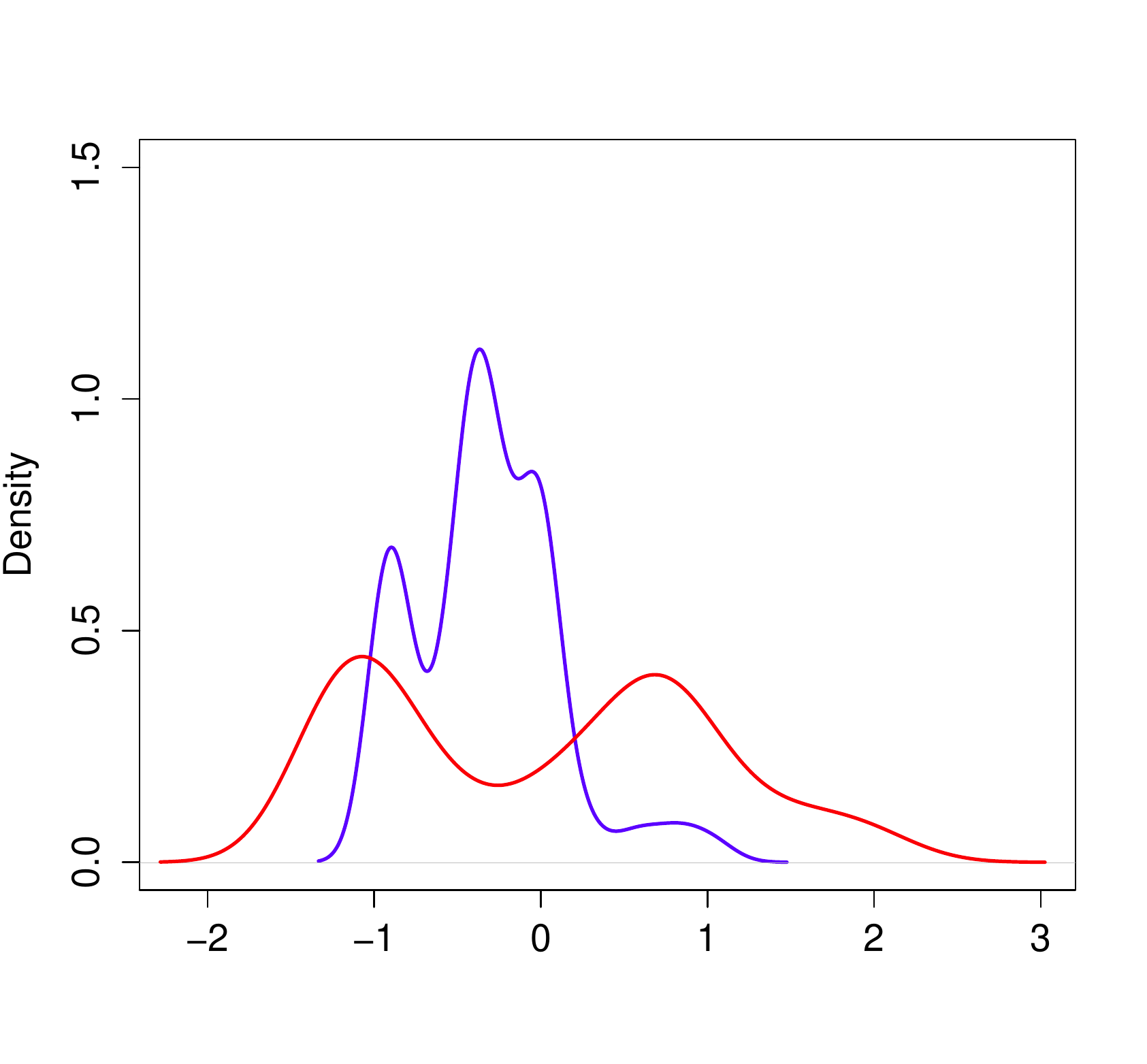}}
\hfill
\subfigure[level 2 ]
{\includegraphics[height=5cm,width=4cm]{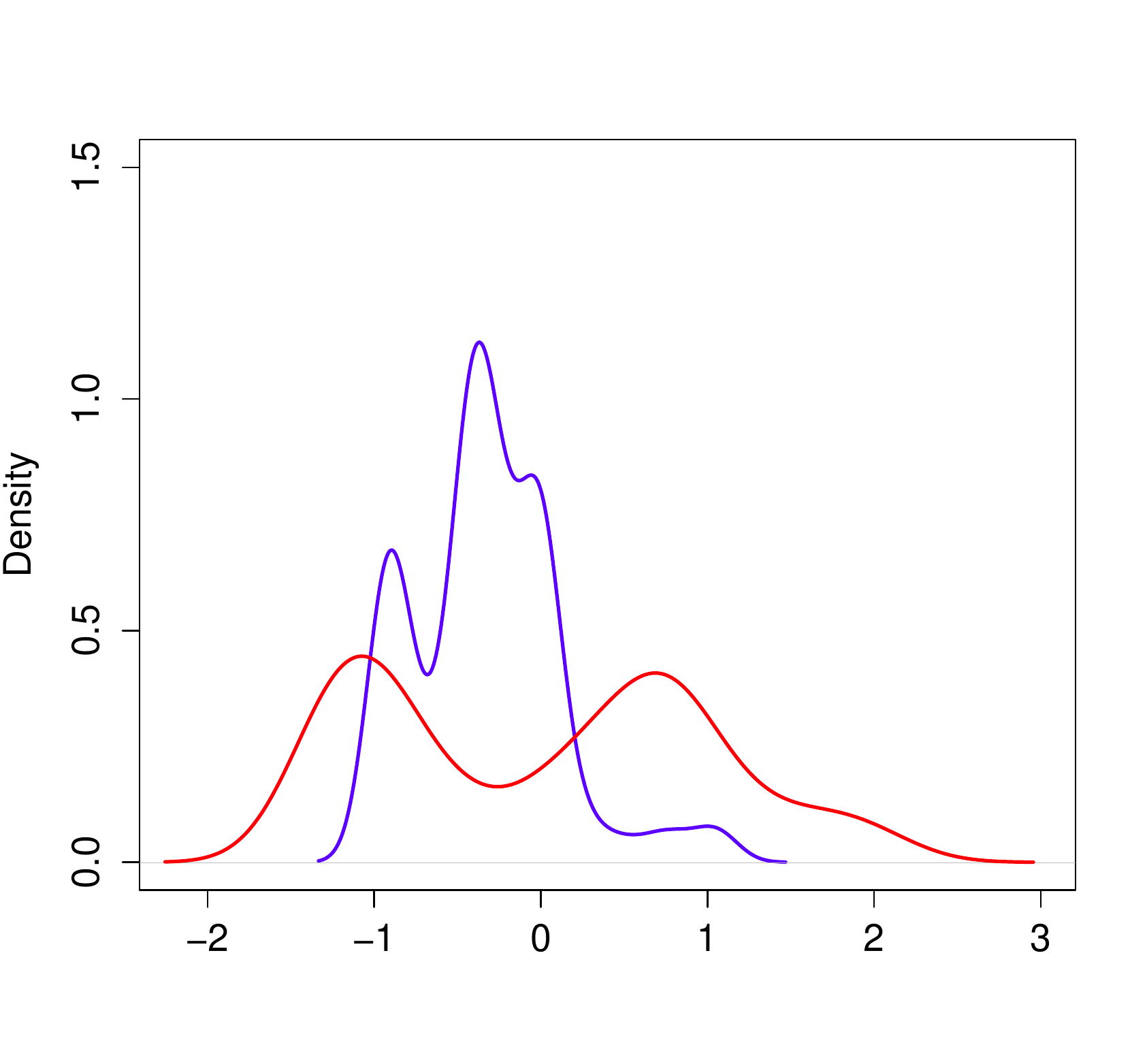}}
\hfill
\subfigure[level 3 ]
{\includegraphics[height=5cm,width=4cm]{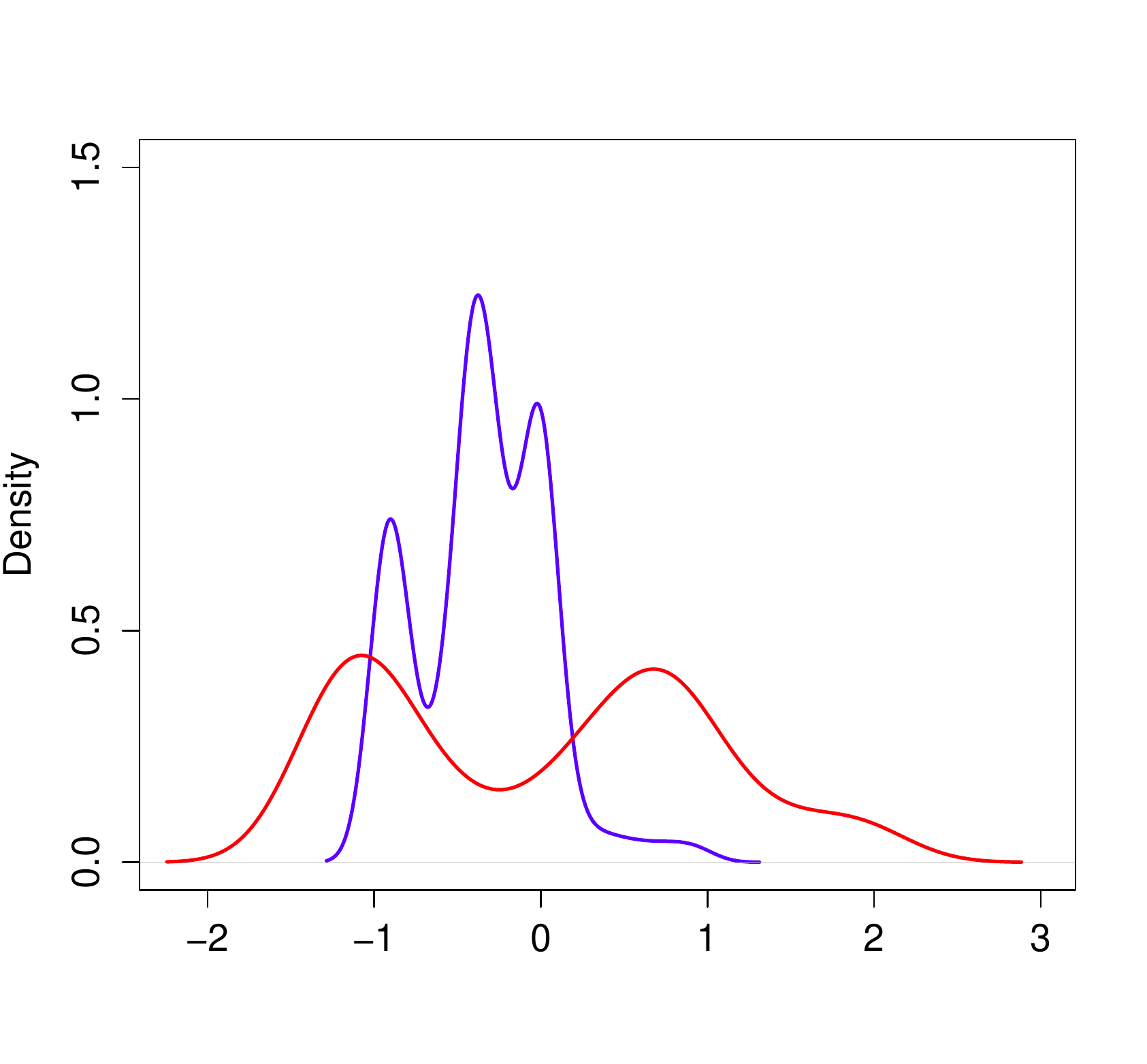}}
\hfill
\subfigure[level 4 ]
{\includegraphics[height=5cm,width=4cm]{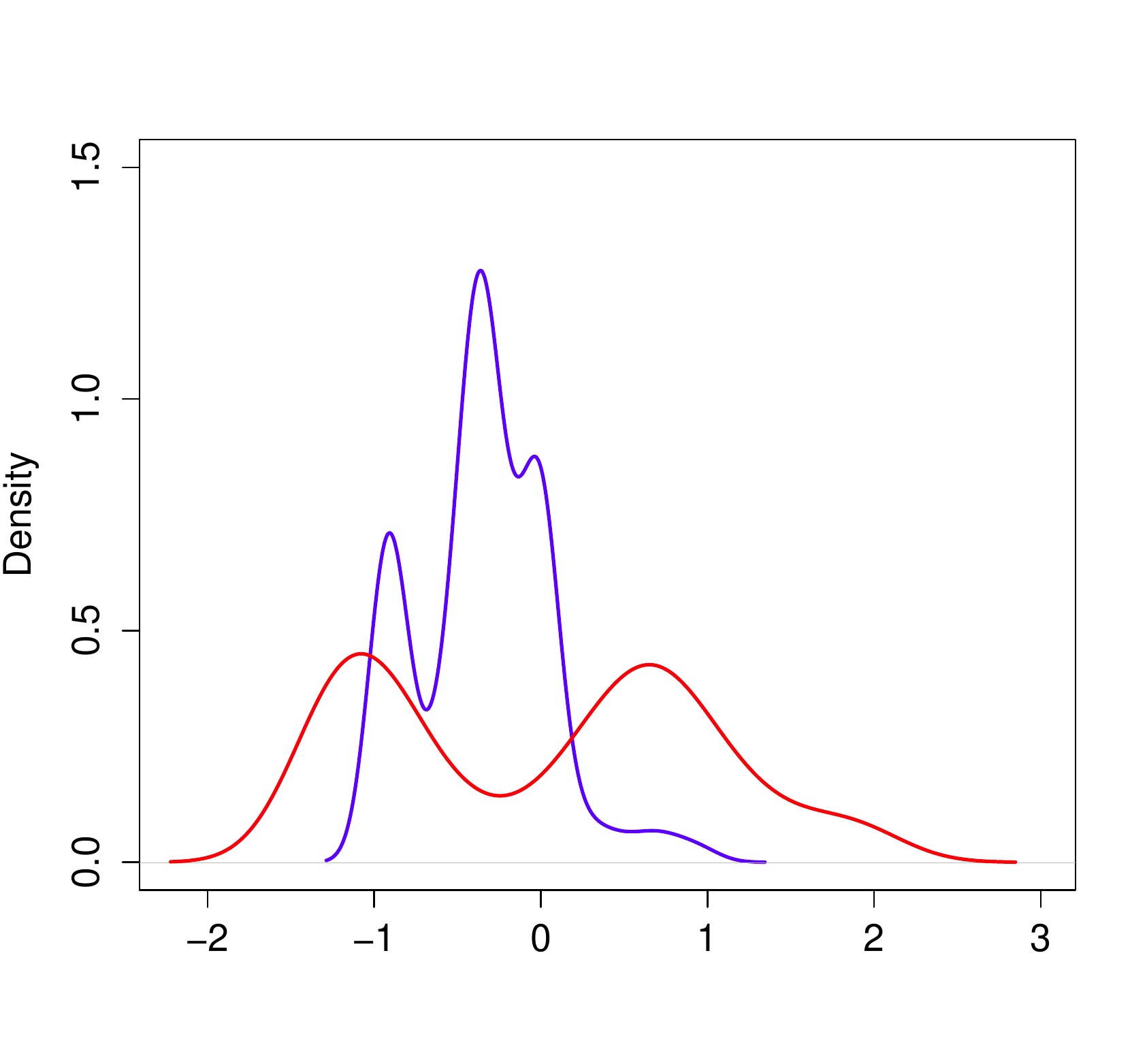}}
\hfill
\caption{ Probability densities of average behaviour  of  {\scriptsize BSE} (blue) and  {\scriptsize NYSE } (red), after removal of outliers,  showing glaring differences  in the tails of the two exchanges. Fat tail of the BSE, for  each of the four levels, got truncated after the removal of the outliers  }
\label{fig:wo_lpdf}
\end{figure}
It needs to be emphasised that the density plots of fluctuations are similar for both the indexes. For the average behaviour, bimodal structure of NYSE is distinctly different from that of BSE, which has a single dominant peak. Bimodal structure of the average behaviour of the nirmalised stock values reveal two stable points around which the NYSE price indexes fluctuates. This corroborates the observed behaviour in the phase space plots, where two stable points are clearly visible.;
For quantifying the periodic modulations in these non-stationary time series, we now proceed to continuous wavelet transform, using Morlet wavelet. 

\subsection{Continuous Wavelet Transform (CWT)}\label{sec:CWT}
CWT brings out the structured variations and their phase and temporal correlations ~\cite{mallat1,stephane1,abhinna}. 

An integrable, well localized (in both the physical and Fourier domains), zero mean function, the mother wavelet $\psi(n)$, is used as the analyzing function. Given a discrete data set, $X = \{x_n\colon n\in \mathbb{Z}^+\}$, the wavelet coefficients are calculated by convolving the data, with the scaled and translated $\psi(n)$:
\begin{equation}
W_n(s)=\sum_{n'=0}^{N-1} x_{n'} \psi^{*}\left(\frac{n-n'}{s} \right),
\end{equation}
where $s$ is the scale.
The admissibility conditions are,
\begin{eqnarray}
\label{eq:eq2}
\int_{\mathbb{R}^r} \vert\hat{\psi}(\vec{k})\vert^2 \frac{d^r \vec{k}}{\vert \vec{k} \vert^r} &<& \infty, \\
\label{eq:eq3}
\mbox{where}, \hat{\psi}(\vec{k}) &=& \frac{1}{(2\pi)^r} \int_{\mathbb{R}^r} \psi(\vec{x})e^{-\imath \vec{k}\cdot\vec{x}}d^r \vec{x}\\
\label{eq:eq4}
\mbox{and }\int_{\mathbb{R}^r}\psi(\vec{x})d^r \vec{x}&=&0.
\end{eqnarray}
Here, $r$ is  the number of spatial dimensions.
The complex Morlet wavelet is given by,
\begin{equation}
\label{eq:eq5}
\psi(n)=\pi^{-1/4}e^{\imath \omega_0 n}e^{-n^2 /2},
\end{equation}
where, n is the localized time index. $\psi(n)$ is a marginally admissible function, it is made admissible by taking $\omega_0=6$. The Fourier wavelength of $\psi(n)$, $\lambda_F$, is given by,
\begin{equation}
\label{eq:lamb_morl}
\lambda_F=\frac{4\pi s}{\omega_0+\sqrt{2+\omega_0^2}} \sim 1.03s
\end{equation}

The cone of influence (COI) is the region beyond which convolution errors make the wavelet coefficients unreliable for analysis~\cite{torrence1,stephane1}. 
The scalograms in Fig.\ref{fig:scalogram}, clearly demonstrate existence of well defined periodic modulations at different scales.
\begin{figure}[H]
\centering
\subfigure[ 3D scalogram  plot of {\small BSE}]{\includegraphics[width=10cm,height=4cm]{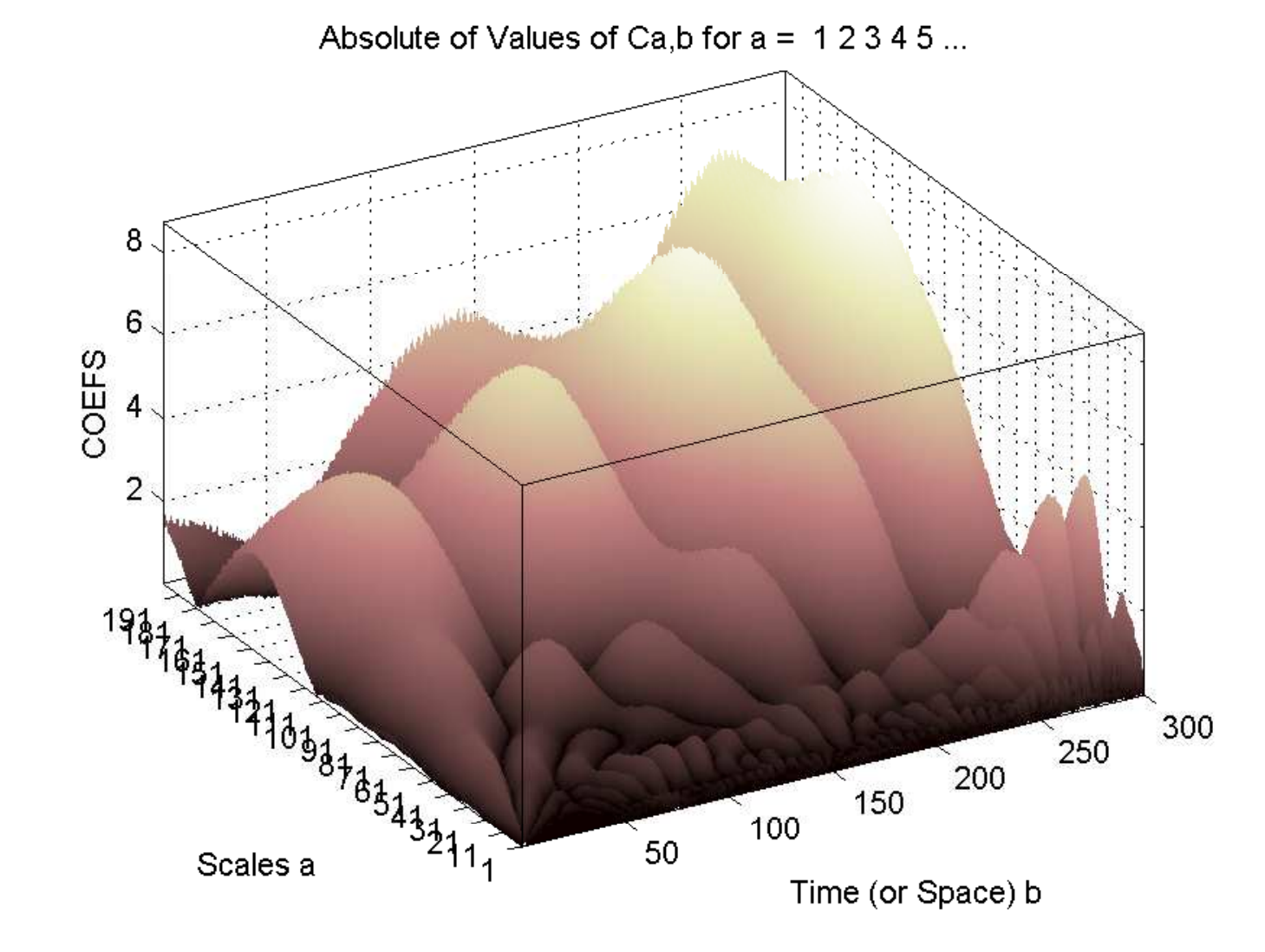}}
\label{fig:scalo_sen}

\centering
\subfigure[ 3D-scalogram  plot of {\small NYSE}]{\includegraphics[width=10cm,height=4cm]{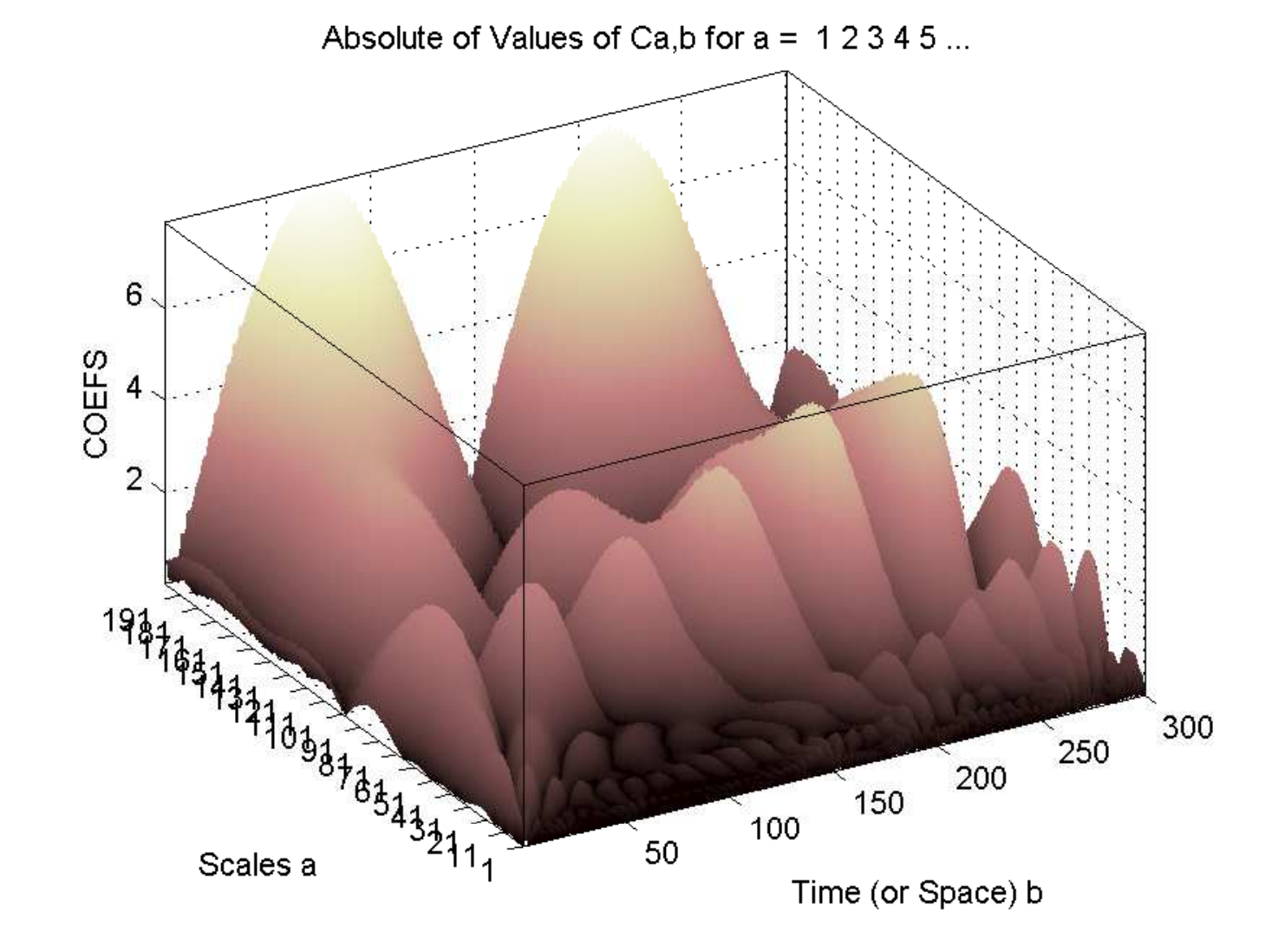}}
\label{fig:scalo_nys}
\caption{3D scalograms of the CWT coefficients, periodic modulations of the two indexes at certain scales are evident}
\label{fig:scalogram}
\end{figure}

Semi-log plots in Figs.\ref{fig:semilog_sen} and  \ref{fig:semilog_nys}, showing the periodic modulations present in the parameters.
\begin{figure}[H]
\centering
\includegraphics[width=8cm,height=4cm]{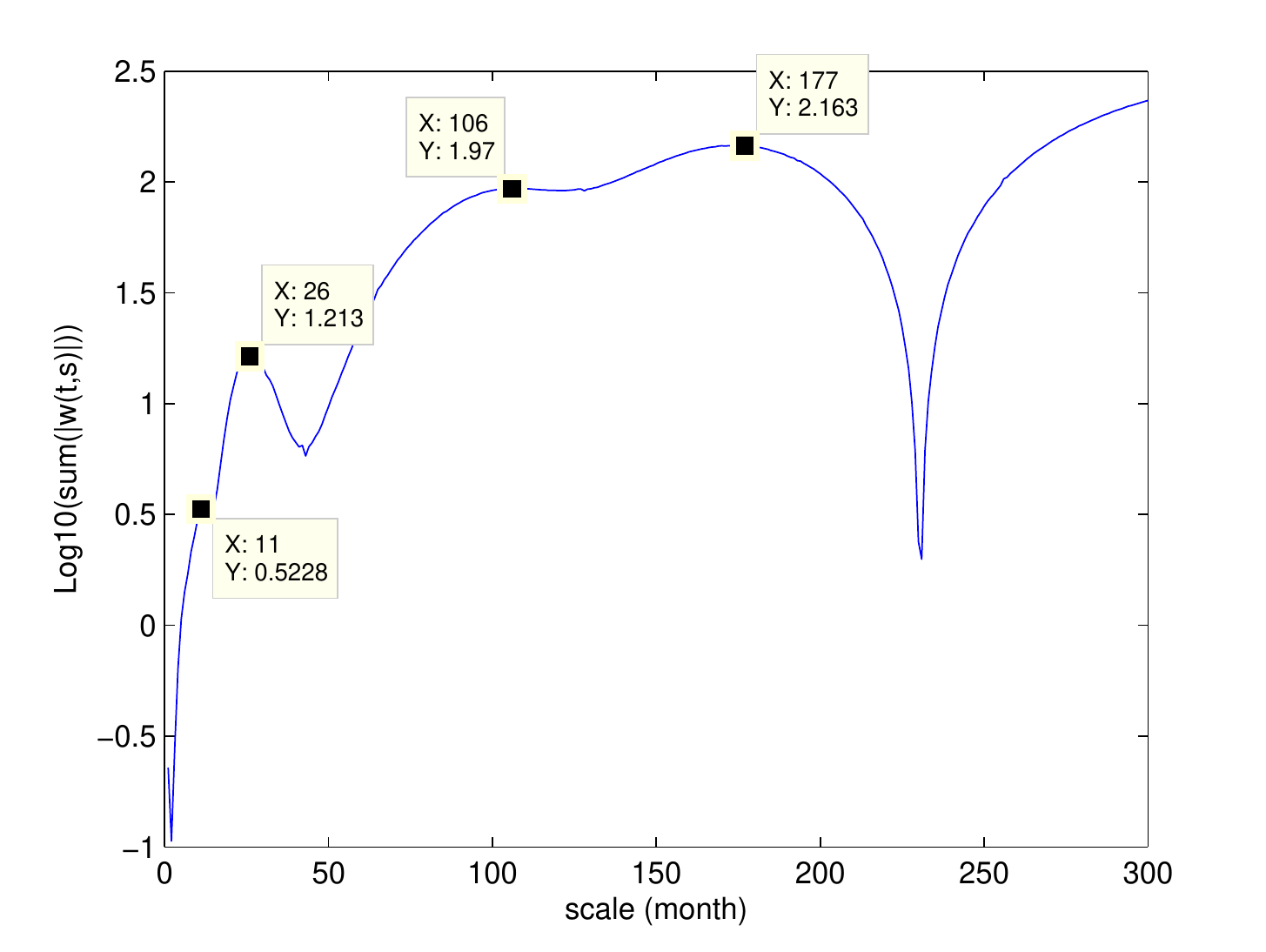}
\caption{Semi-log plot of the wavelet power for BSE, summed over all time, at different scales, unveiling  periodicities of approximately one and two years}
\label{fig:semilog_sen}
\end{figure}
\begin{figure}[H]
\centering
\includegraphics[width=8cm,height=4cm]{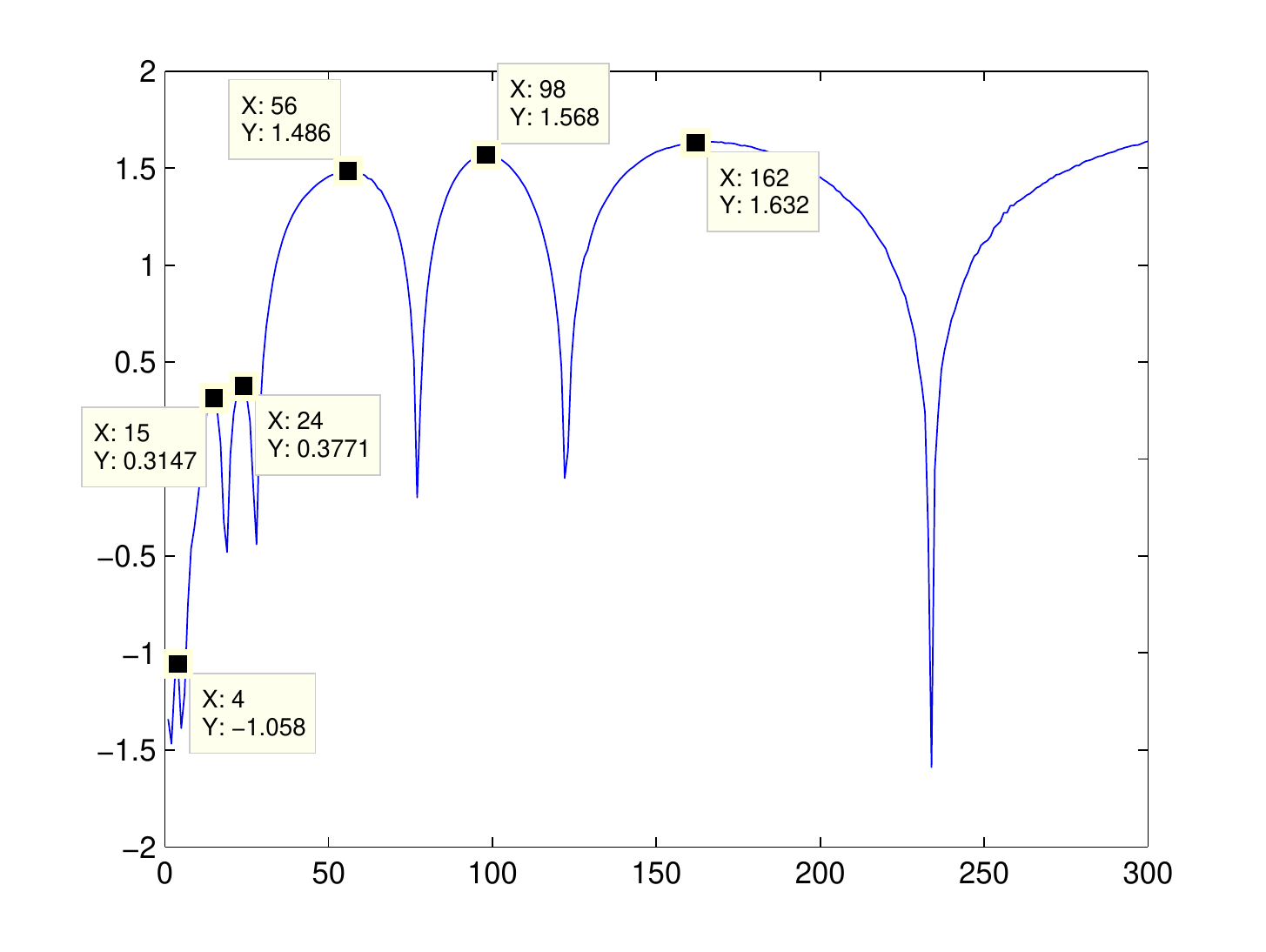}
\caption{Semi-log plot of the wavelet power for NYSE data,  exhibiting the dominant  periodicities of approximately one, two and five  years }
\label{fig:semilog_nys}
\end{figure}
Peaks in the semi-log plots  pin point the dominant scales; these are approximately 1, 2 and 9 years for {\small BSE} and 1, 2 and 5  years for {\small NYSE}. The nine year period in BSE, spans for approximately 1/3 of the analysis time frame, deeming it unfit for further analysis. The void due to its exclusion get filled by 3 year period, though this scale is not exhibited in the semi-log plots, but  regarding wavelet power, it lags behind the two year time period only by  a small margin, and supersedes  the same in case of NYSE. The CWT coefficients at this scale sport strong correlation  with the monthly averaged normalised  time series. Another reason, why it should be looked into, is because the power at this scale is on a curve with a negative slope for BSE, but positive one in case of NYSE. These factors make it an interesting time scale to explore and unveil its characteristics.  

\begin{figure}[ht]
\centering
\includegraphics[width=10cm,height=6cm]{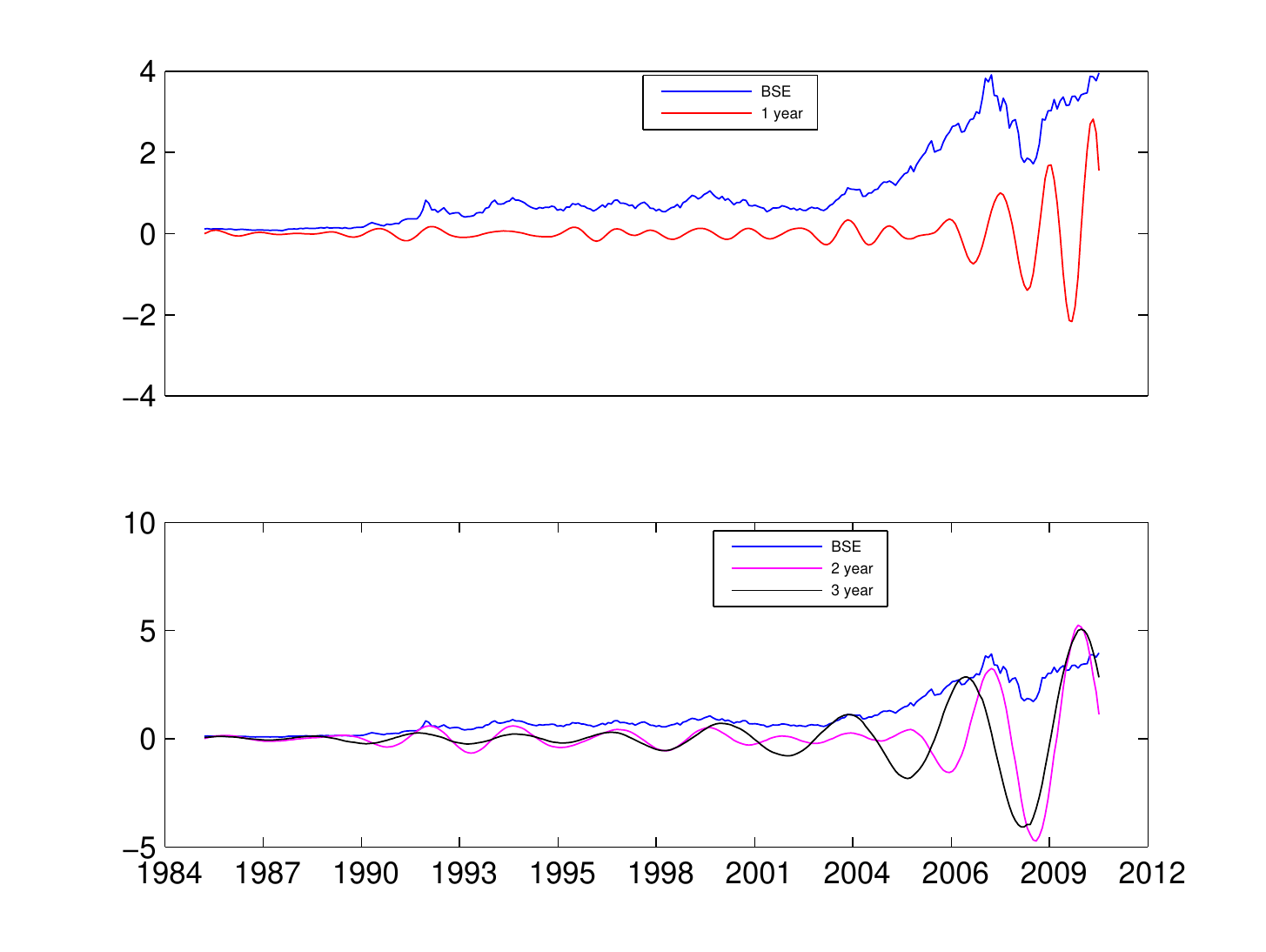}
\caption{Plots of  normalised stock value of {\small BSE} and   its CWT coefficients, at scale of 1, 2 and 3 years,  revealing phase correlations}
\label{fig:coeff_sen}
\end{figure}
\begin{figure}[ht]
\centering
\includegraphics[width=10cm,height=6cm]{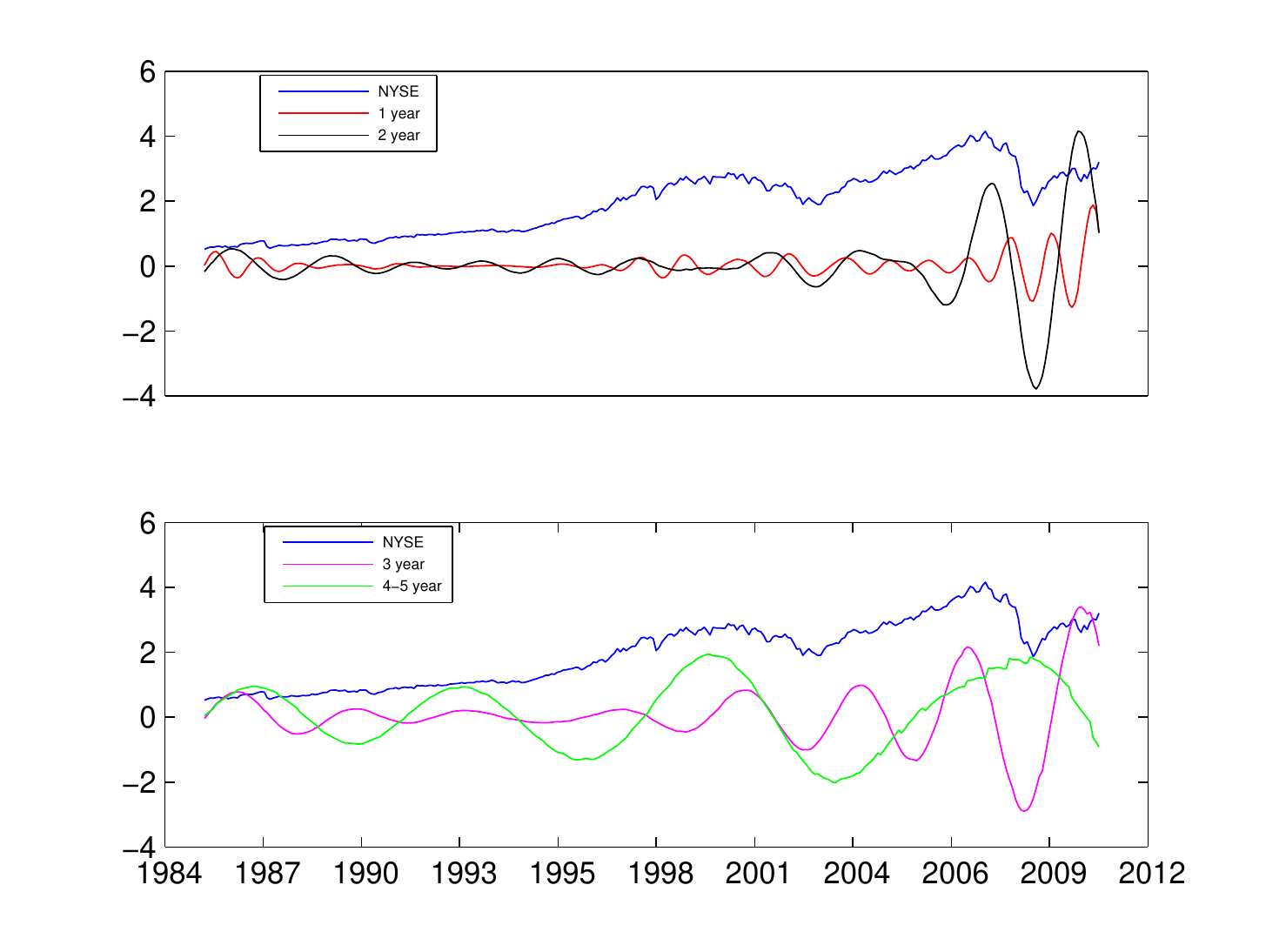}
\caption{Plots of normalised stock value of {\small NYSE} and its CWT coefficients, at scale of approximately 1, 2, 3 and 4-5 years, showing different phase correlations at different time scales}
\label{fig:coeff_nys}
\end{figure}

The {\small CWT} coefficients juxtaposed with the mean subtracted normalised indexes, as shown in Figs.\ref{fig:coeff_sen} and \ref{fig:coeff_nys}, reveal the degree of similarities between the coefficients, corresponding to the periods  shown in semi-log plots. 
\subsubsection{Comparing the coefficients of {\small BSE} and {\small NYSE}}
\label{subsec:coefficient}                           
A careful study of {\small CWT} coefficients, as shown in Fig.\ref{fig:coeff}(a), (b) and (c), reveal that  {\small BSE} and {\small NYSE} are moving in tandem,  except when one of them is non-periodic, revealing a transition period leading to an unstable regime \cite{abhinna,pkp_aerosol}.\\
The two stock exchanges moved  in an un-synchronised fashion throughout, except between 2008 to 2010, 1987 to 2006 and 1987 to 1997 at 1, 2 and 3 year scales, respectively. The  synchronised behaviour,  between 2008-2010, at one year time scale and  between 2006-2010 at  a scale of two year, shows the crisis centric in-phase movement of the two exchanges.
At three year time scale, the synchronised behaviour appears post 1997, with a lag of 5-6 months and slowly the lag get marginalised with time, as seen in Fig.\ref{fig:coeff}(c), exhibiting the degree of integration of BSE. 
Multi-level synchronisation between coefficients of two stock exchanges is possible only if the driving forces of both the stock exchanges are of similar nature or one of them is driving the other. It is clear that NYSE is the driving force with which  BSE got aligned, with a time lag.

\begin{figure}[H]
\centering
{\includegraphics[height=10cm,width=10cm]{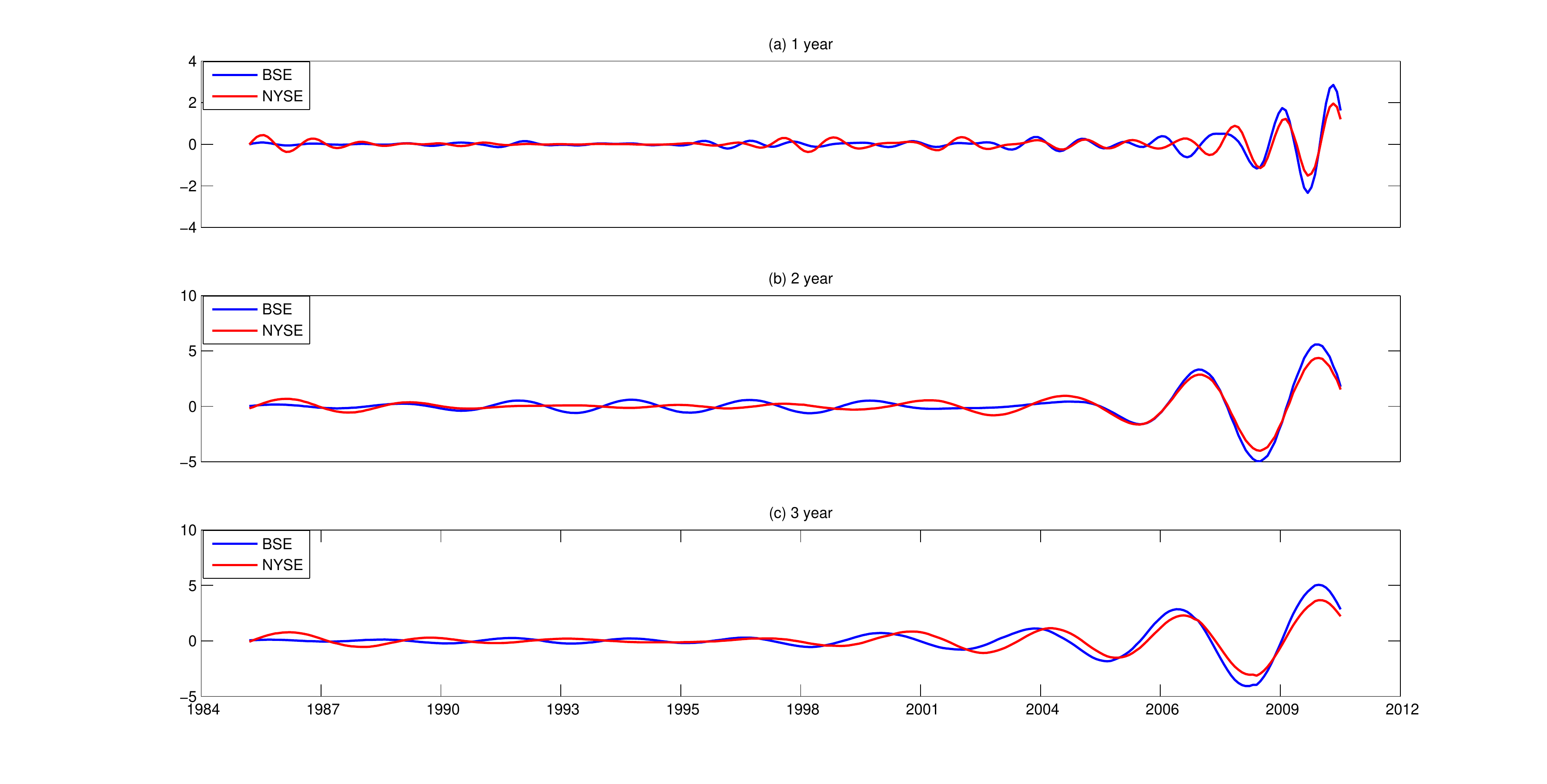}}

\caption{ CWT coefficients of BSE (blue) and NYSE (red),  depicting  in and out of phase periodic behaviour,  of two stock exchanges approximately at scale of (a) 1 (b) 2 and (c) 3 years, respectively, and in-phase nature at higher scales are evident}
\label{fig:coeff}
\end{figure}
\subsubsection{Phase Plot}\label{subsec:phase}

We now investigate more systematically  the phase structure of periodic modulations, and the manner in which  the trends at different scales are in synchronization with each other ~\cite{torrence1,torrence2,abhinna,pkp_aerosol}. 
Figs.\ref{fig:CWT_coeff} (a), (b) and (c), depict the evolution of phase angle of the average behaviour of {\small BSE} and {\small NYSE}  at different  scales. 
In-phase nature, between the coefficients of CWT of the two stock exchanges, is observed between 2002-2011 corresponding  to one year time scale, from 2002-2007 at two year scale and between 1997-2011 at a scale of three years.\\
They are in anti-phase mode between 1987-2002 at one year scale, from 1997-2002 and again from 2007-2011 at two year scale and during pre-liberalisation period at three year scale. The anti-phase nature at one year scale vindicates the finding that it is a crisis centric scale. 
At three year scale, it is seen that in-phase nature starts to emerge after 1996, though with a lag of 3-4 months which subsequently get marginalised and the two  became phase locked, even during the crisis periods.  \\ 
In-phase nature in the second half can be attributed to the opening up of the Indian market, thus paving  way for synchronised behaviour.

\begin{figure}
\centering
\subfigure[scale 12]
{\includegraphics[width=10cm,height=10cm]{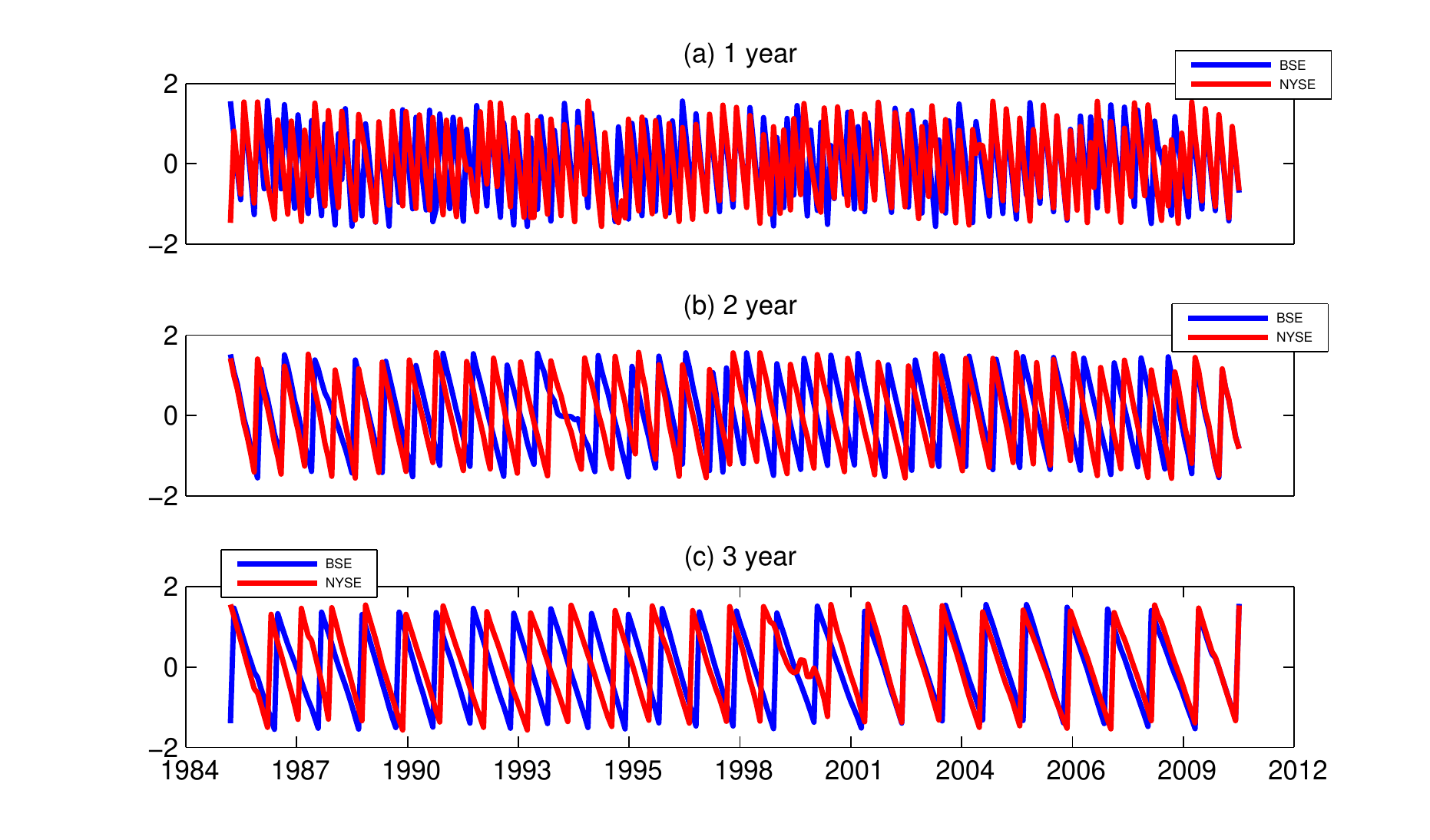}}
\caption{Phase plots of CWT coefficients of average behaviour data at level 1 {\small BSE } (blue) and {\small NYSE} (red) at a time scale of approximately, (a) 1  (b) 2 and (c) 3 year, respectively}
\label{fig:CWT_coeff}
\end{figure}

\subsubsection{Cross Wavelet Spectrum }\label{subsec:CWT}

We proceed and identify  regions in time frequency domain, where two time series share maximum power. This is done through the Cross wavelet spectrum. The arrows reveal information about phase angles between the two time series ~\cite{torrence1,torrence2,grinsted1}.\\
In Fig.\ref{fig:h_xwt}, common power region lies  in the second half of the plots. In case of fluctuations, it is maximum at around 2, 4, and 8 month scales corresponding to levels 1, 2 and 3, respectively.

\begin{figure}[H]
\centering
\hfill
\subfigure[level 1 ]
{\includegraphics[height=3cm,width=4cm]{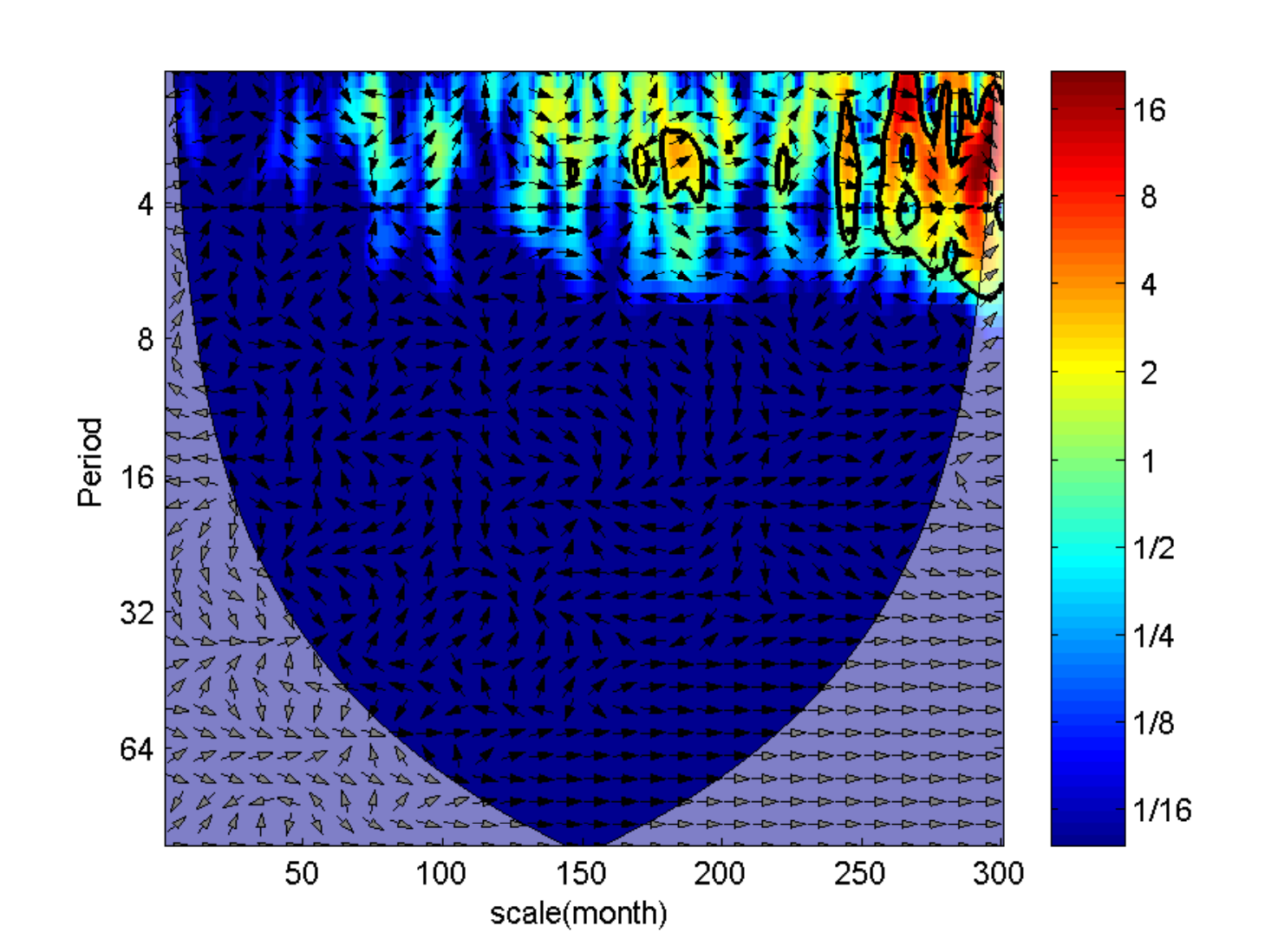}}
\hfill
\subfigure[level 2 ]
{\includegraphics[height=3cm,width=4cm]{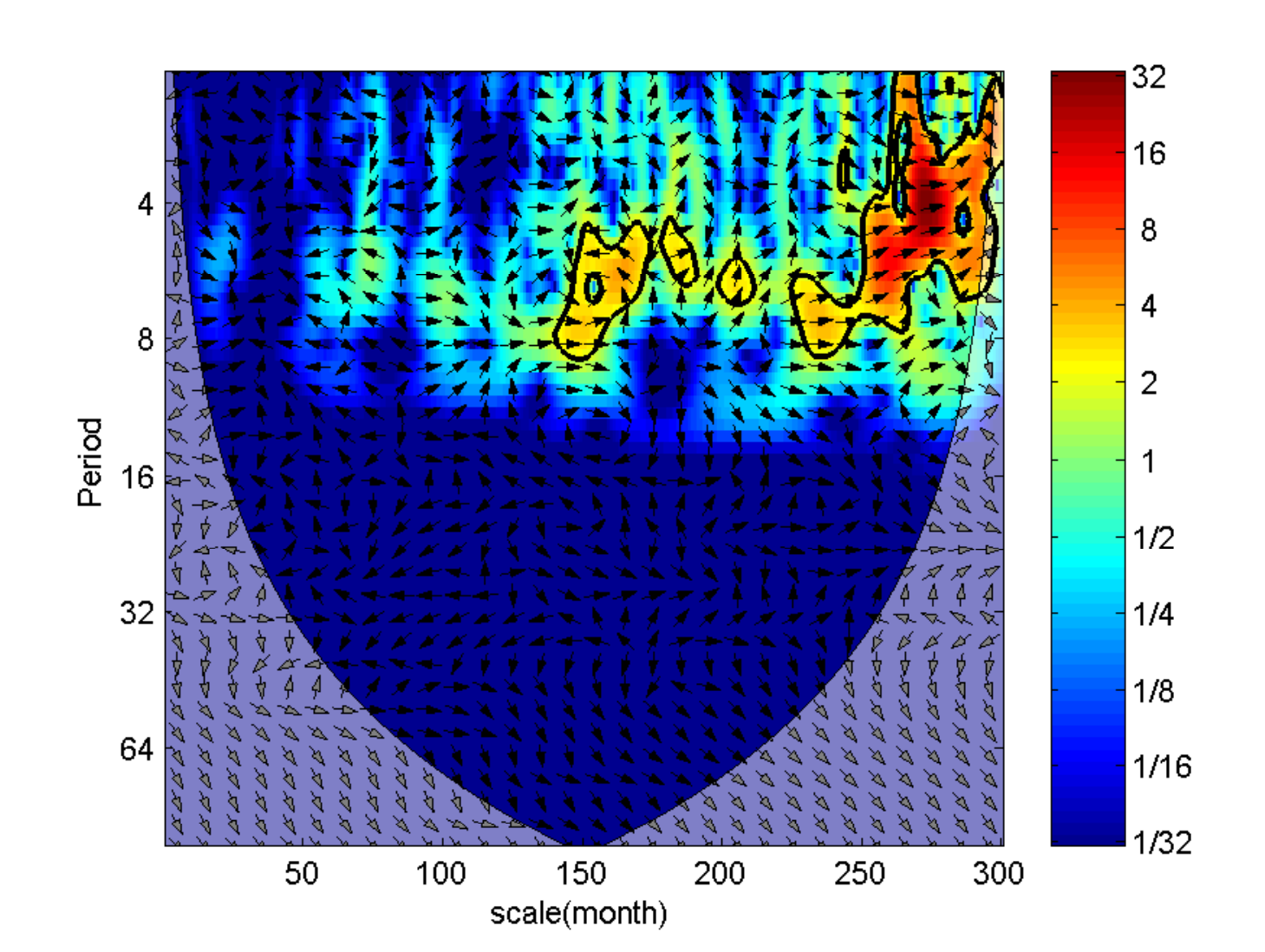}}
\hfill
\subfigure[level 3 ]
{\includegraphics[height=3cm,width=4cm]{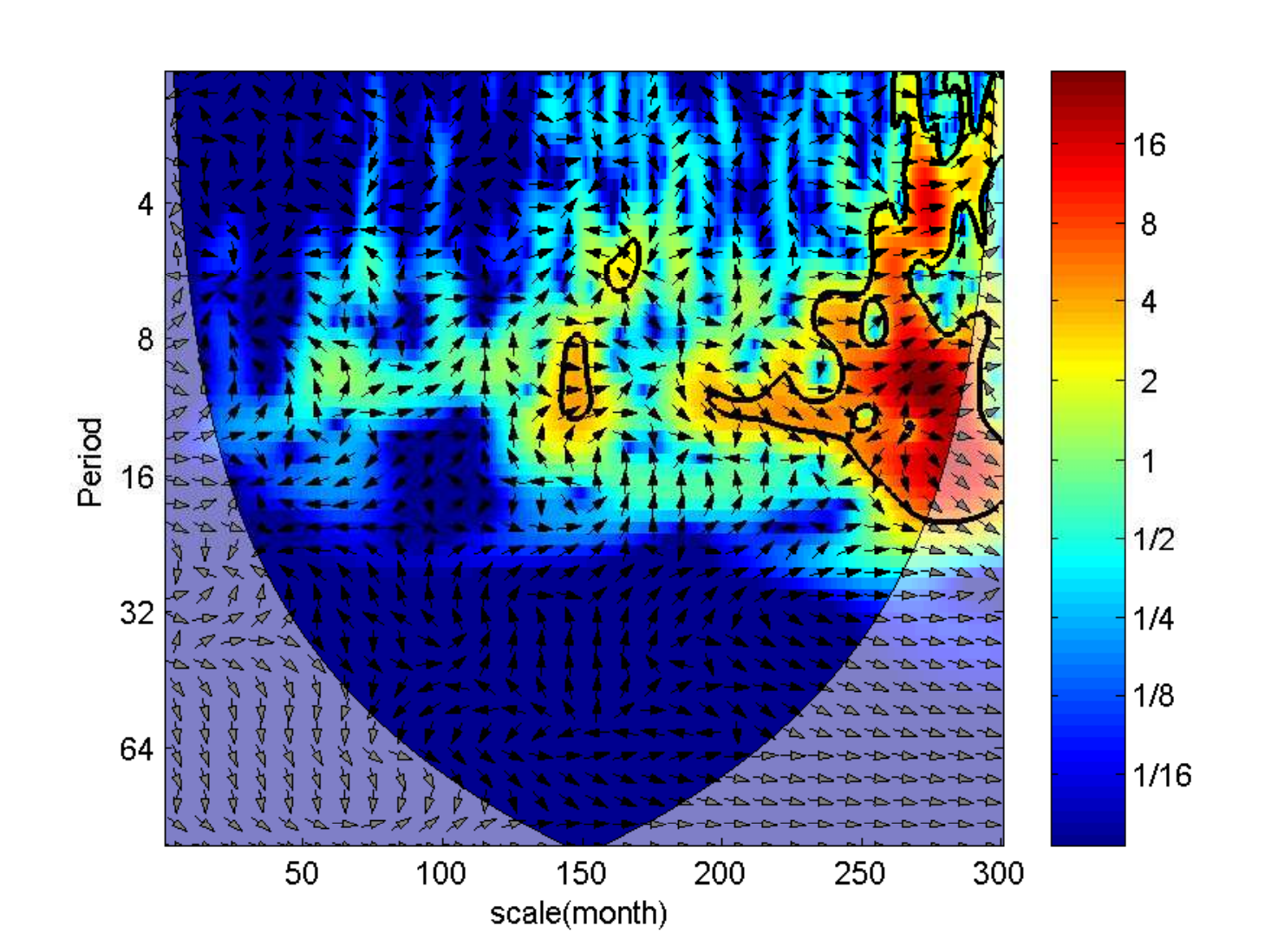}}
\hfill
\subfigure[level 4 ]
{\includegraphics[height=3cm,width=4cm]{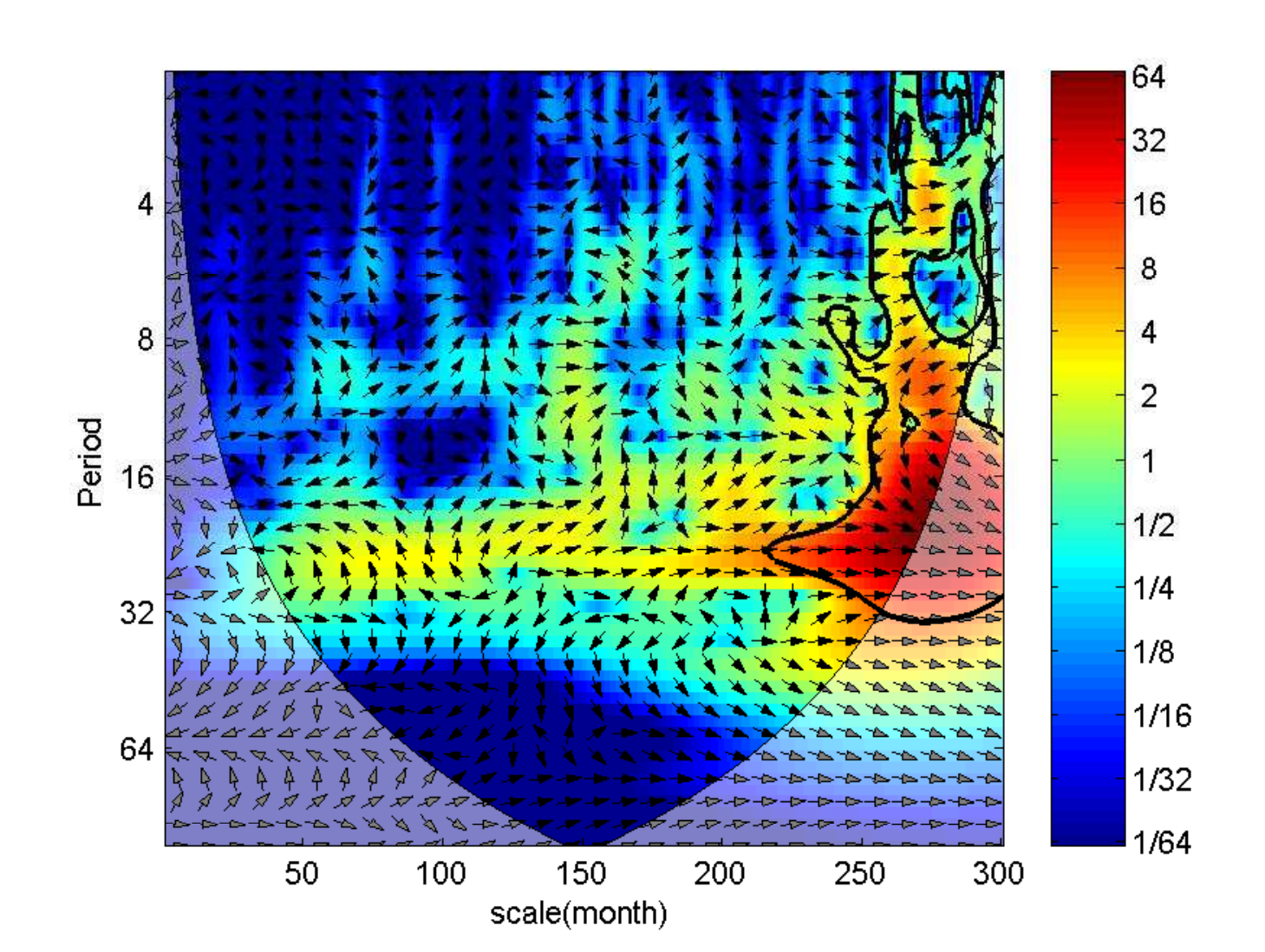}}
\hfill
\caption{Cross wavelet transform plots of fluctuations at 95\% confidence interval, with the cone of influence shown in lighter shade, indicating the region affected by edge affects. The colour code for power ranges from blue (low power) to red (high power). The nature of phase relationship between the two stock exchanges is represented by the direction of the arrows. They are directed toward right, right-upward, right-downward, left, left-upward, left-downward, respectively, showing that both the variables are in phase, NYSE is lagging, NYSE is leading, both are out of phase, NYSE is leading and NYSE is lagging. The share of power between the variables are plotted for   (a) level 1 (b) level 2 (c) level 3 and (d) level 4. It is more in the second half of the plot, at around time scale of 2 months for the level 1 and 6 months for level 2 and at around 12-16 months for level 3 }
\label{fig:h_xwt}
\end{figure}

\subsubsection{Wavelet Coherence  }\label{subsubsec:WTC}

Local correlation between two stock indexes, in time frequency space \cite{mallat1,grinsted1,torrence1,torrence2,book:percival_spect}, is now probed through  wavelet coherence. 
It helps in unveiling locally phase locked behaviour. Here, the significance level has been  determined using Monte Carlo simulation.\\ 

Sporadic blobs are  present at 4 and 8 months scale, as seen in Figs.\ref{fig:h_wtc}(a) and (b), respectively. With the advent of more structured variations, i.e,  a bridge like structure can be seen at  3 year scale, as shown in Fig.\ref{fig:h_wtc}.  Arrows are pointed in rightward direction, this indicates that the two exchanges are phase locked.
In case of average behaviour, see Fig.\ref{fig:l_wtc},  arrows are either aligned in rightward or right-downward direction,  this entails that both the exchanges are either in-phase or NYSE is driving BSE, respectively, the later is more ubiquitous at the scale of 3 years. 

\begin{figure}
\centering
\hfill
\subfigure[level 1 ]
{\includegraphics[height=3cm,width=4cm]{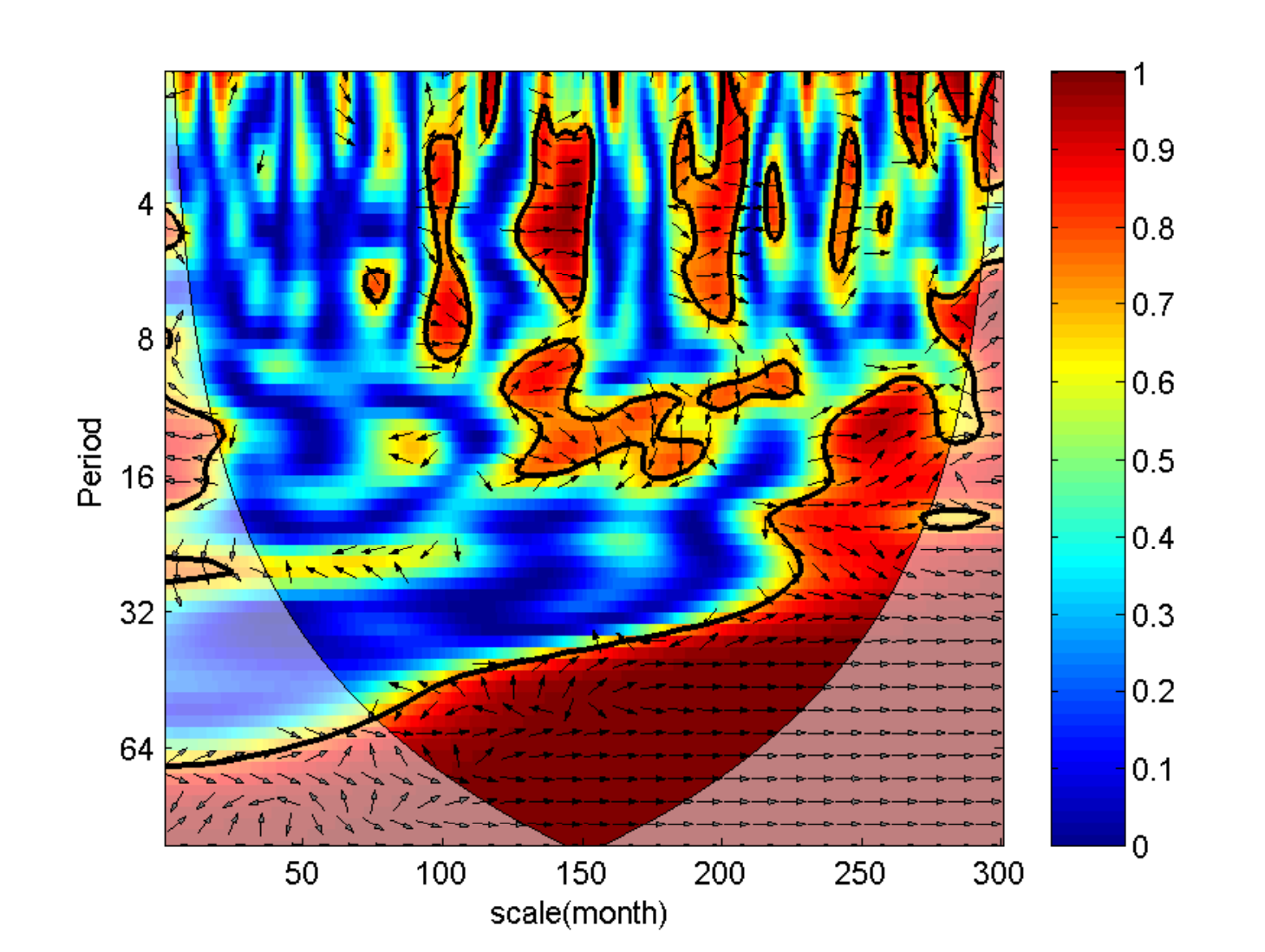}}
\hfill
\subfigure[level 2 ]
{\includegraphics[height=3cm,width=4cm]{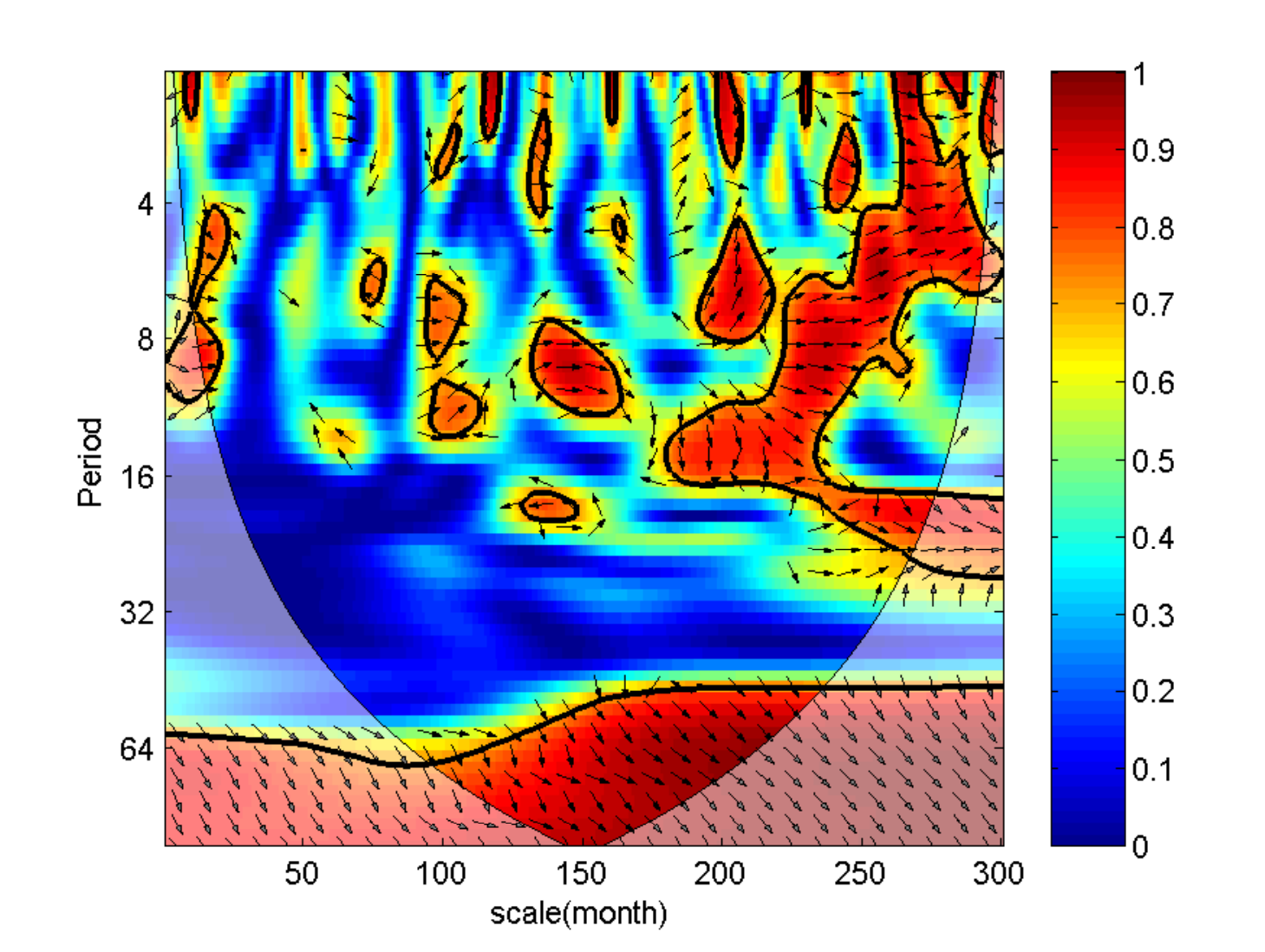}}
\hfill
\subfigure[level 3 ]
{\includegraphics[height=3cm,width=4cm]{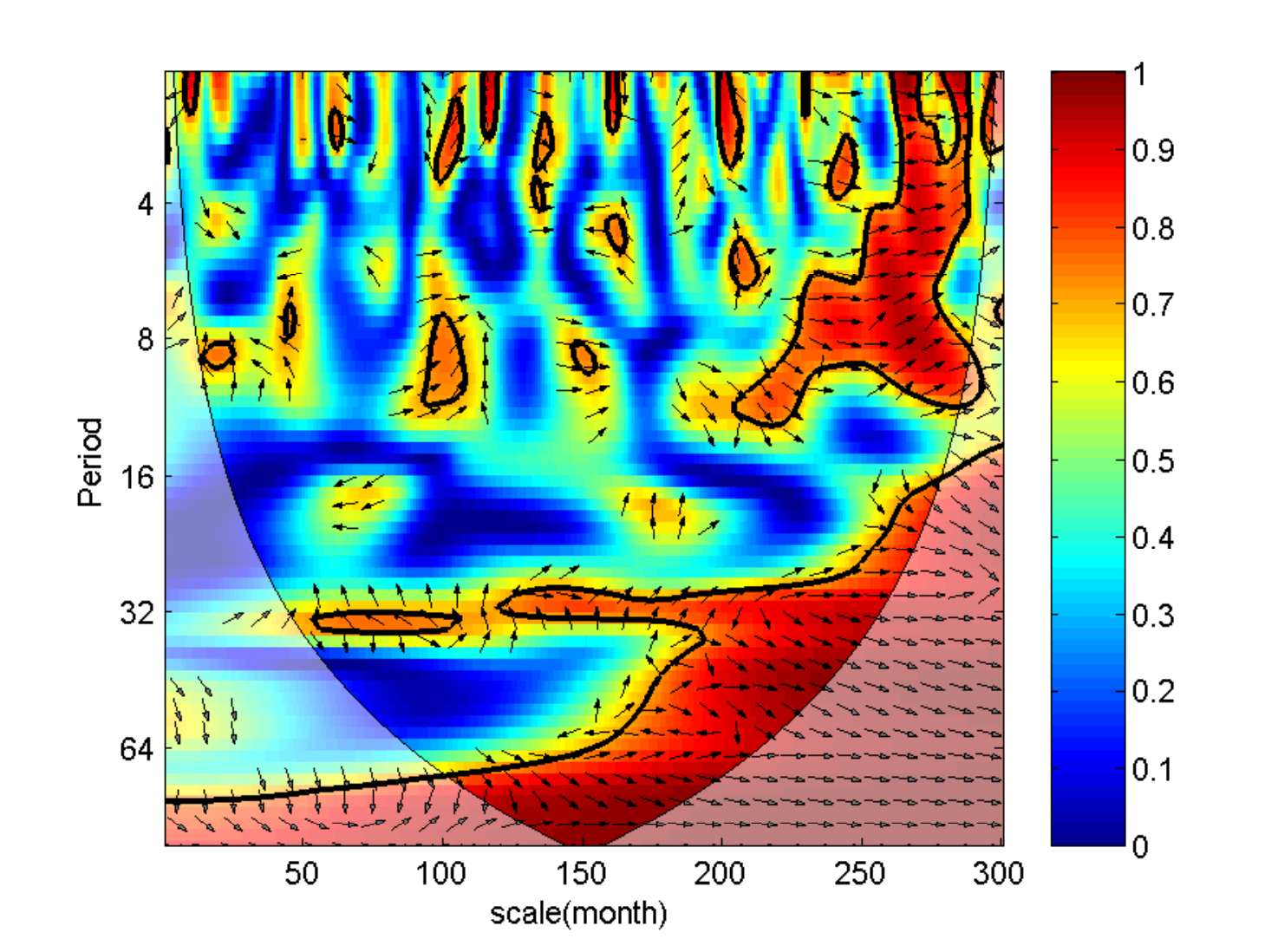}}
\hfill
\subfigure[level 4 ]
{\includegraphics[height=3cm,width=4cm]{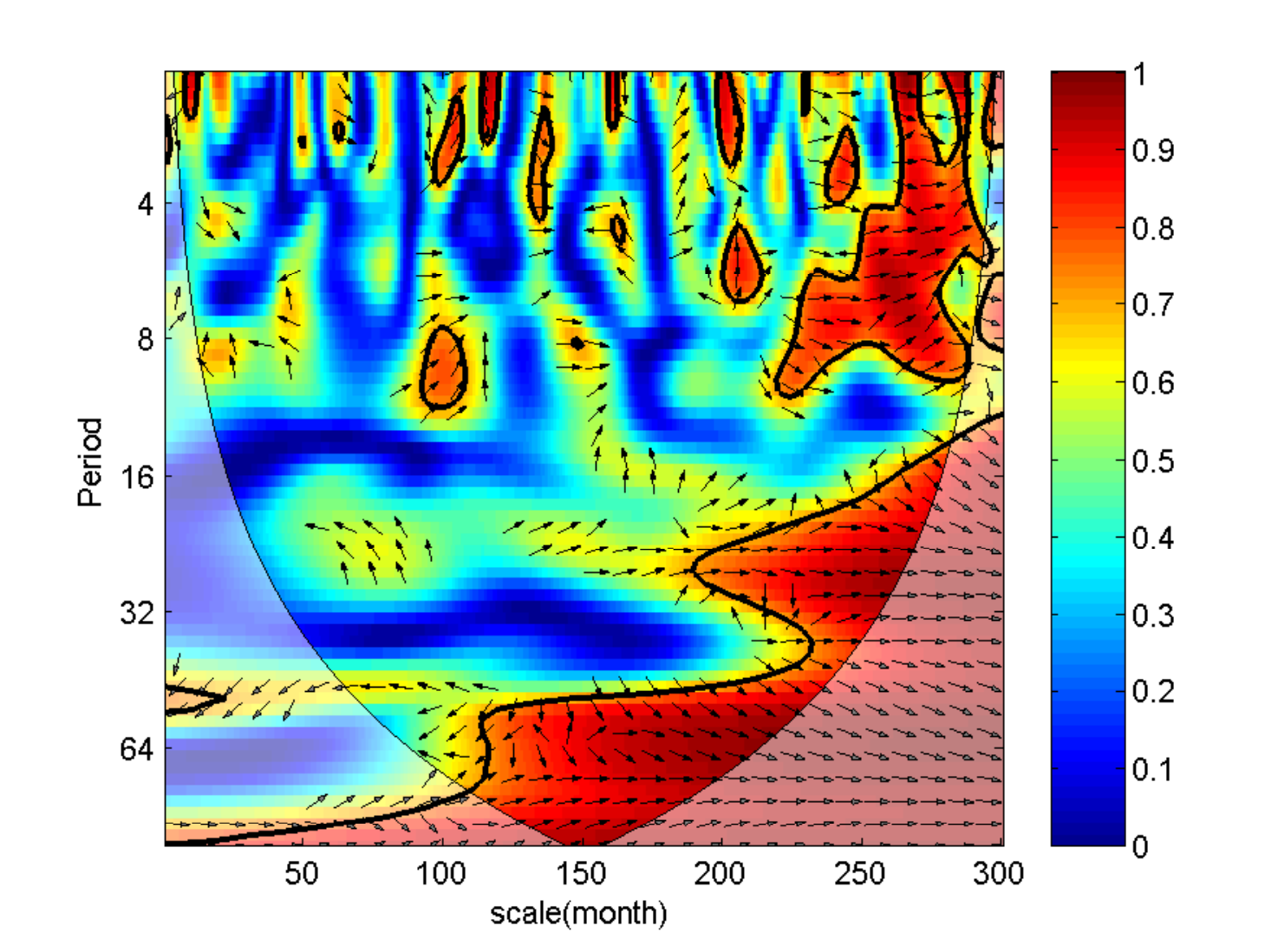}}
\hfill
\caption{Wavelet coherence plots of fluctuations  for  at (a) level 1 (b) level 2 (c) level 3 (d) level 4, of BSE and NYSE at 95\% confidence interval with the cone of influence shown in lighter shade. The colour code representing power ranges from blue (low power) to red (high power). The X-axis and Y-axis represent  the time (month) and scale, respectively.}
\label{fig:h_wtc}
\end{figure}
\begin{figure}[H]
\centering
\hfill
\subfigure[level 1 ]
{\includegraphics[height=3cm,width=4cm]{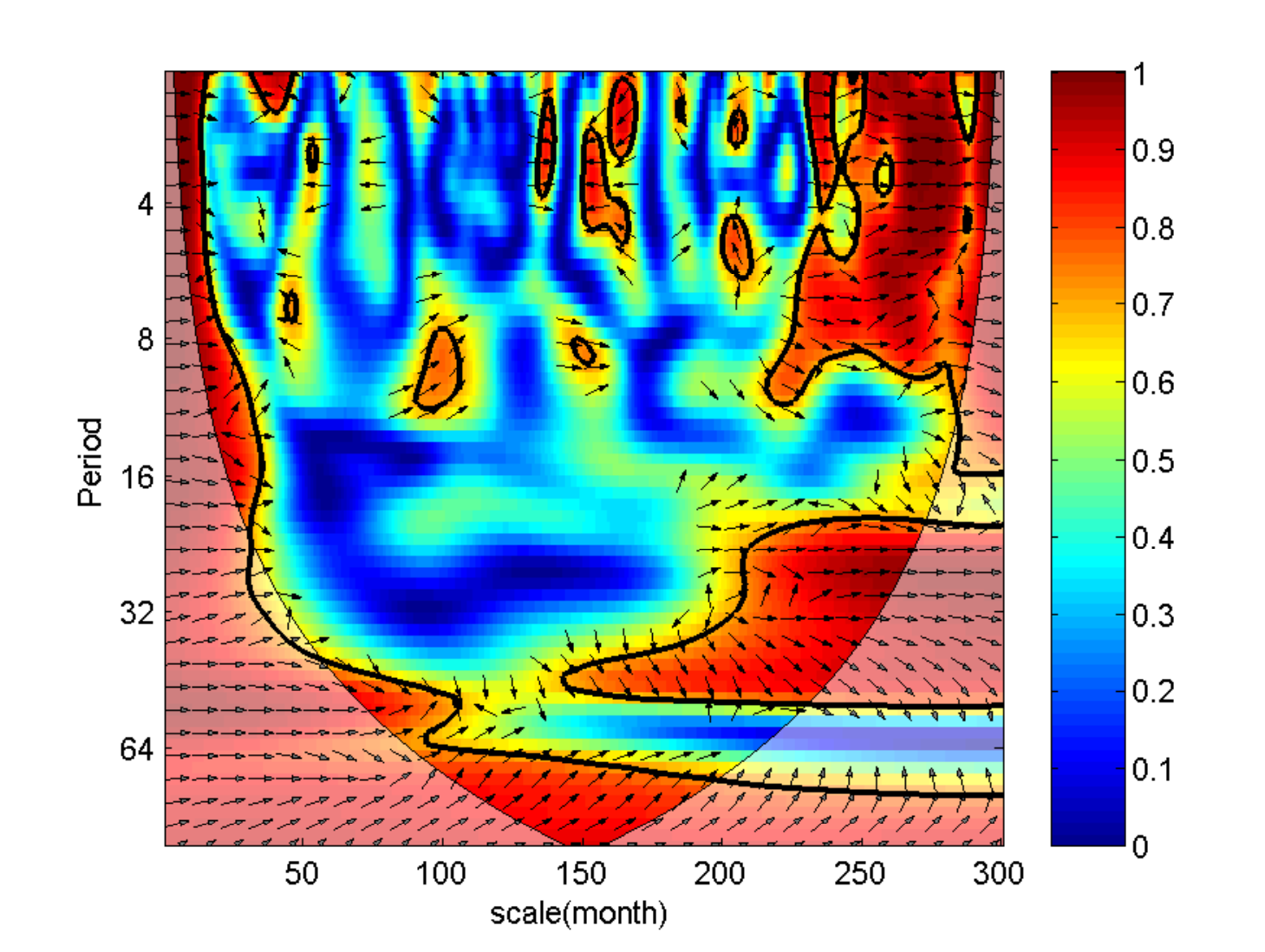}}
\hfill
\subfigure[level 2 ]
{\includegraphics[height=3cm,width=4cm]{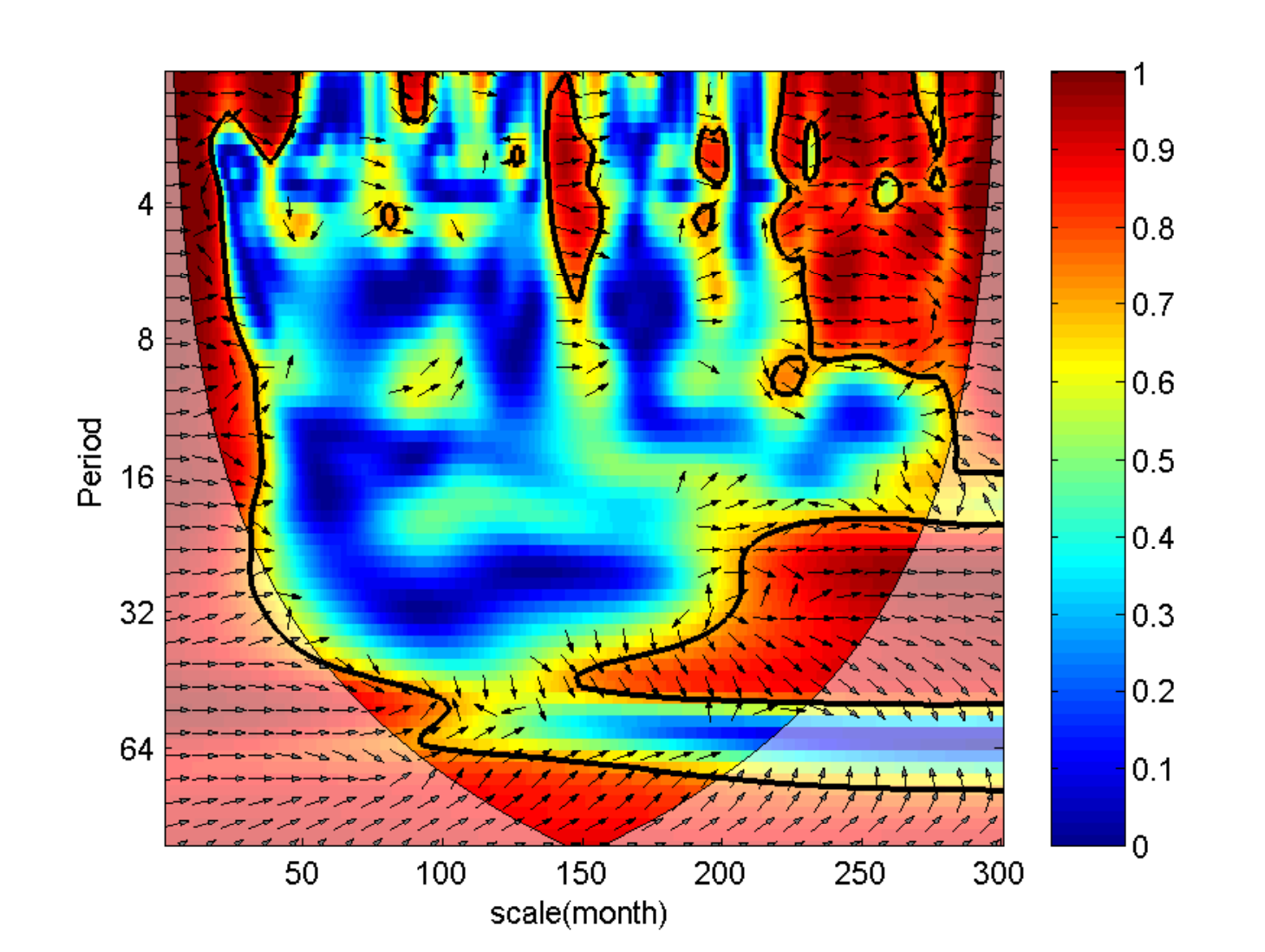}}
\hfill
\subfigure[level 3 ]
{\includegraphics[height=3cm,width=4cm]{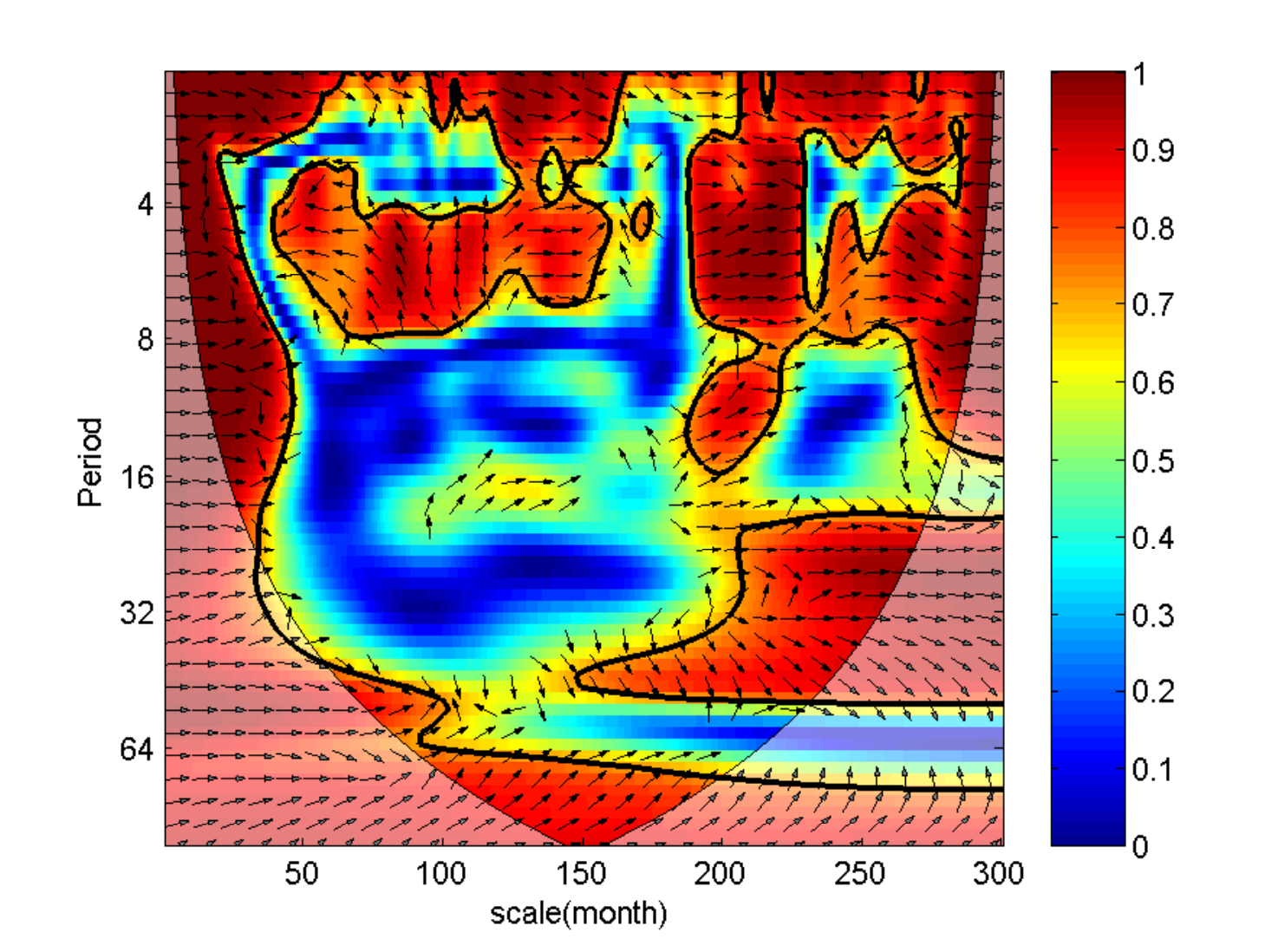}}
\hfill
\subfigure[level 4 ]
{\includegraphics[height=3cm,width=4cm]{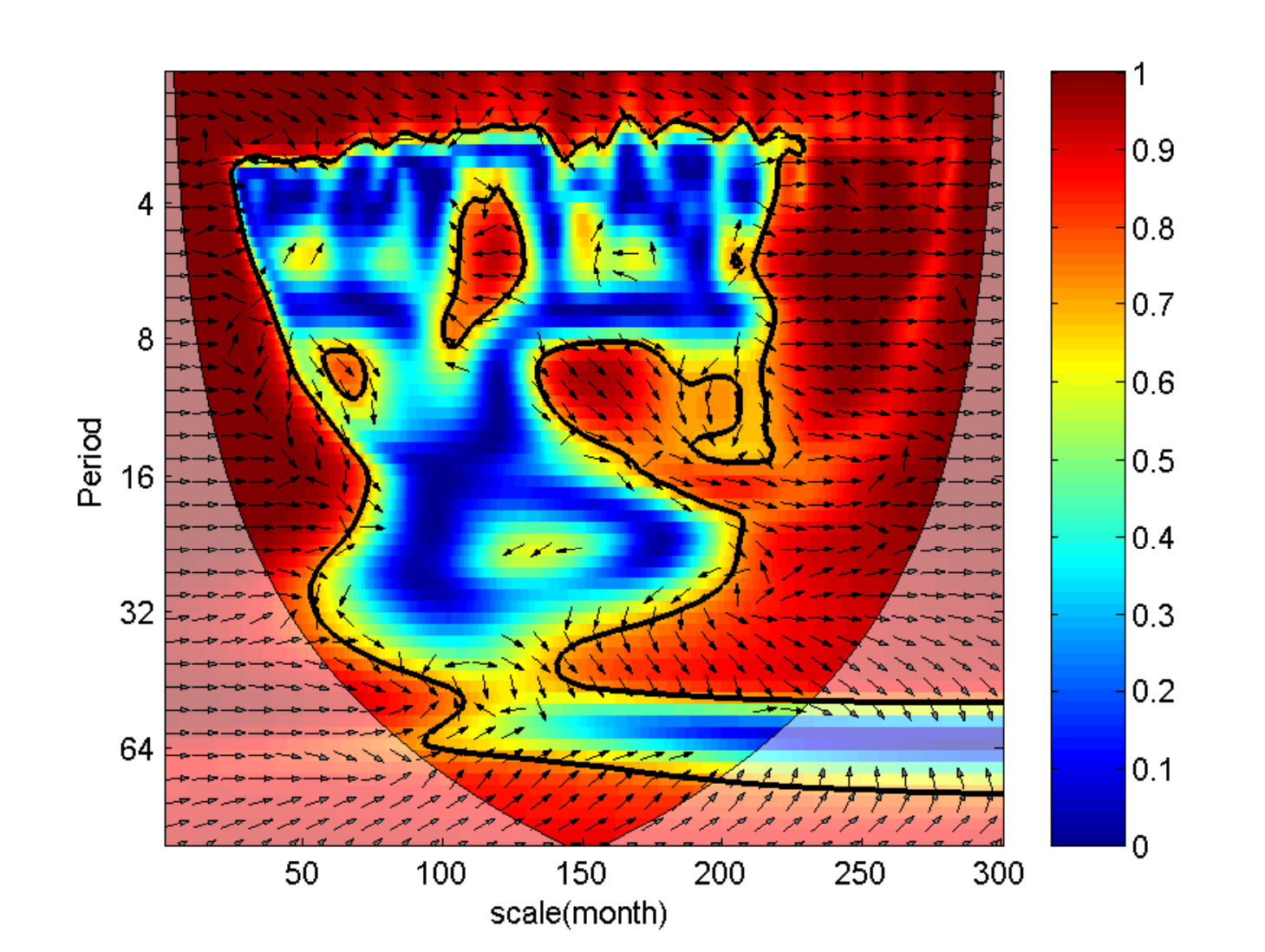}}
\hfill
\caption{Wavelet coherence plots of average behaviour for (a) level 1, (b) level 2, (c) level 3 and (d) level 4,  at 95\% confidence interval }
\label{fig:l_wtc}
\end{figure}

\section{Multi-scale Causality}\label{sec:causality}

To test the causal relationship between two  or multiple time series, the standard Granger causality test (GCT) is usually implemented. According to GCT, if X causes Y then the forecast of Y is better if the information in X is used than when it is not \cite{grang2}. The test relies on Vector Auto Regressive (VAR) model:
\begin{equation}\label{eq:caus1}
\begin{aligned}
Y_t= {} & \ a_0 + a_1Y_{t-1} + a_2Y_{t-2}+....+a_pY_{t-p} +\\
 & b_1X_{t-1} + b_2X_{t-2}+....+b_pX_{t-p} + u_t\
\end{aligned}
\end{equation}
 and 
\begin{equation}\label{eq:caus2}
\begin{aligned}
 X_t= {} & \ c_0 +c_1X_{t-1} + c_2X_{t-2}+....+c_pX_{t-p} +\\
& d_1Y_{t-1} + d_2Y_{t-2}+....+d_pY_{t-p} + v_t\
\end{aligned}
\end{equation}\\
In each case, rejection of the null hypothesis means there is  Granger causality. 
Here, we have used Toda-Yamamoto Granger causality (TYGC) instead of the traditional Granger causality, as  F-statistics used in the former, can lead to spurious causality, when one or both time series are non-stationary or when they are not co-integrated  \cite{maekawa,toda1,johan1,grang1,johan2}. 
This has been taken into account  in Toda-Yamamoto version of Granger causality. Hence, tests to check for co-integration need not  be conducted in TYGC.

Tests like Augmented Dickey Fuller (ADF) and Phillips and Perron are conducted to check whether the time series is integrated or not,  the power of these tests are low in comparison to the alternative hypothesis of (trend) stationarity. To overcome this shortcoming,  
\\Kwiatkowski–Phillips–Schmidt–Shin (KPSS) test is conducted, it has (trend) stationairty as null hypothesis.\\ 
Toda-Yamamoto   methodology  estimates an augmented VAR, when there is co-integration of different orders and with the VAR being stationary (about a trend), integrated or co-integrated of an arbitrary order ~\cite{toda1}.\\
Here, $Y_{t}$ and $X_{t}$ are the variables to be tested and $u_{t}$ and $v_{t}$ are the mutually uncorrelated white noise errors, t stands for time period and p is the time lag.
Testing $H_{0} : b_1=b_2= .... =b_{p}$=0, against the alternative, $H_{A}$ : not $H_0$, is a test for X does not Granger cause Y. Similarly, $H_0 : d_1=d_2= .... = d_{p}$=0, against $H_{A}$ : not $H_{0}$, is a test for Y does not Granger cause X.
Steps involved in the test are~\cite{TYGC} : 

\begin{enumerate}
\item The first step is to check for stationarity, hence one conducts ADF  and KPSS tests.  Null hypothesis of ADF test ($H_{0}$) : unit root exists, to check whether signal is stationary or not~\cite{dick1} (supplement material).
 Null hypothesis ($H_{0}$) of the KPSS test implies stationarity in the univariate time series, with alternative that, it is non-stationary unit root process ~\cite{kpss1} (supplement material).
If the time series is non-stationary, one  takes a difference of it and conducts ADF and KPSS tests  again on the newly obtained  time series, repeating this process until it  becomes stationary. This  unveils the order of integration, which is equal to the order of differencing. Let the value of integration be $I$, which is maximum of the integration values of the two time series.
\item We now set up a bi-variate VAR model with any arbitrary lag value, using the non-differenced data~\cite{watson,pfaff,pfaff_book}.
The execution of the bi-variate VAR model yields information about the appropriate lags to be considered for future tests. 
Here, finding the correct lag value plays a very crucial role, as Granger causality is   susceptible  to incorrect lag values.
To avoid it,   Akaike Information Criteria (AIC), Hannan-Quinn (HQ), Schwartz Criterion (SC) and Final Prediction Error (FPE) values are used  to find the appropriate lag~\cite{pfaff,pfaff_book,aic}. Let the lag value be $P$.

\begin{table}[!htbp] \centering 
  \caption{Values of information criteria, namely AIC,HQ,SC and FPE, obtained after conducting  the VAR model, for the first twenty lags.} 
  \label{tab:var.info} 
\begin{tabular}{@{\extracolsep{5pt}} ccccc} 
\\[-1.8ex]\hline 
\hline \\[-1.8ex] 
 Lag & AIC(n) & HQ(n) & SC(n) & FPE(n) \\ 
\hline \\[-1.8ex] 
1 & $23.470$ & $23.512$ & $23.574$ & $15,599,045,289.000$ \\ 
2 & $23.456$ & $23.518$ & $23.612$ & $15,372,509,925.000$ \\ 
3 & $23.465$ & $23.548$ & $23.673$ & $15,512,814,949.000$ \\ 
4 & $23.467$ & $23.571$ & $23.727$ & $15,551,191,607.000$ \\ 
5 & $23.427$ & $23.552$ & $23.738$ & $14,932,254,565.000$ \\ 
6 & $23.452$ & $23.598$ & $23.816$ & $15,321,920,251.000$ \\ 
7 & $23.466$ & $23.633$ & $23.881$ & $15,534,252,500.000$ \\ 
8 & $23.455$ & $23.643$ & $23.923$ & $15,367,698,794.000$ \\ 
9 & $23.410$ & $23.619$ & $23.930$ & $14,697,147,626.000$ \\ 
10 & $23.431$ & $23.660$ & $24.002$ & $15,008,563,873.000$ \\ 
11 & $23.446$ & $23.696$ & $24.069$ & $15,233,542,555.000$ \\ 
12 & $23.433$ & $23.704$ & $24.109$ & $15,048,551,288.000$ \\ 
13 & $23.447$ & $23.738$ & $24.174$ & $15,251,869,661.000$ \\ 
14 & $23.455$ & $23.767$ & $24.234$ & $15,385,124,110.000$ \\ 
15 & $23.431$ & $23.764$ & $24.262$ & $15,024,027,602.000$ \\ 
16 & $23.451$ & $23.805$ & $24.333$ & $15,331,280,705.000$ \\ 
17 & $23.462$ & $23.837$ & $24.397$ & $15,513,561,630.000$ \\ 
18 & $23.456$ & $23.852$ & $24.443$ & $15,428,608,855.000$ \\ 
19 & $23.450$ & $23.867$ & $24.489$ & $15,349,459,131.000$ \\ 
20 & $23.440$ & $23.877$ & $24.530$ & $15,198,053,943.000$ \\ 
\hline \\[-1.8ex] 
\end{tabular} 
\end{table}  
\begin{table}[!htbp] \centering 
  \caption{The optimum values of lag value considered on the basis of the information criteria embodied in the Table \ref{tab:var.info}} 
  \label{tab:var.lag} 
\begin{tabular}{@{\extracolsep{5pt}} cccc} 
\\[-1.8ex]\hline 
\hline \\[-1.8ex] 
AIC(n) & HQ(n) & SC(n) & FPE(n) \\ 
\hline \\[-1.8ex] 
$9$ & $1$ & $1$ & $9$ \\ 
\hline \\[-1.8ex] 
\end{tabular} 
\end{table}
\item Coefficients corresponding to the lag $P$ are put through Pormanteau Test (asymptotic), to check, if they are serially correlated (null hypothesis $H_{0}$ : coefficients are not serially correlated).
If the coefficients come out to be serially uncorrelated, then only, they are subjected to stability test. The coefficients  need to pass both the tests in order to be considered fit for further analysis. 
\item VAR model is again set up, this time with "$P+I$" as the lag value, here P is  lag value corresponding to which coefficients passed the tests mentioned in step 3. 
\item The coefficients of VAR ($P+I$) model, found in step 4, are then used for WALD test (null hypothesis $H_{0}$ : one time series, does not Granger cause the another) \cite{pfaff,pfaff_book,dolado}.
\end{enumerate}}
The test result for monthly averaged normalised stock value for BSE and NYSE are tabulated in Table.\ref{tab:norm_G}.

Toda-Yamamoto test was carried out for monthly averaged normalized data of {\small BSE} and {\small NYSE} as well as for the average behaviour  and fluctuations,  of the two stock exchanges. 

In case of monthly averaged normalized data of the two exchanges, the outcome is {\small NYSE} Granger causes {\small BSE} at lag of 9 months, there is no causality either way, at  one month lag.
\begin{table}
\caption{TYGC test results for monthly averaged normalised stock}
\label{tab:norm_G}
\centering
\begin{tabular}{|p{2cm}|p{0.001cm}|p{0.4cm}|p{1.7cm}|p{0.7cm}|p{1.2cm}|}
\hline 
Null hypothesis ($H_{0}$) &  & & Chi-squared test result & &   Result\tabularnewline
\hline 
\hline 
 &  & $X_{2}$ & df & p-value & \tabularnewline
\hline 
BSE  does not G-cause NYSE &  & 8.6 & 6 & 0.2 & accepted\tabularnewline
\hline 
NYSE does not G-cause BSE &  & 16.2 & 6 & 0.013 & rejected\tabularnewline
\hline 
\end{tabular}
\end{table}

\begin{table}
\caption{Results of the Toda-Yamamoto Granger causality test for the average behaviour and fluctuations of the two stock exchanges  }
\label{tab:granger}
\centering
\begin{tabular}{|p{0.6cm}|p{2cm}|p{1.2cm}|p{1.8cm}|p{0.4cm}|}
\hline 
 & Fluctuations & Lag (month) & Average behaviour  & Lag\tabularnewline
\hline 
\hline 
Level  &  &  &  & \tabularnewline
\hline 
1 & no causality &  & no causality & \tabularnewline
\hline 
2 & no causality & & no causality & \tabularnewline
\hline 
3 & BSE causing NYSE & 8 & no causality & \tabularnewline
\hline 
4 & BSE causing NYSE & 6 & no causality & \tabularnewline
\hline 
\end{tabular}
\end{table}  
Multi-scale causal relations have been established between the two stock exchanges, to understand their short and long term relationship~\cite{ramsey1,sato,cifter,zhang1}, as well as to   know about the developments driving the causality relations between the two economies. It may help in framing policies and taking measures at the time of economic slowdown. 
Causality results for the average behaviour and fluctuations are tabulated in the Table. \ref{tab:granger}.
In case of average behaviour, no causality is observed at any scale. In case of high frequency variations, BSE Granger causes NYSE,
the absence of NYSE Granger causing BSE at lower scales can be attributed to the insularity of the Indian economy. \\
The finding that BSE Granger causing NYSE at smaller scales may appear  spurious as, 
\begin{enumerate}
\item The {\small NYSE} is more stable than {\small BSE}.
\item {\small NYSE} represents a developed economy, while {\small BSE} is a developing one.
\item The market capitalisation of {\small NYSE} is significantly higher  than that of  {\small BSE},  {\small NYSE} is being the world's largest stock exchange  according to market capitalisation and trade value.
\end{enumerate}
The observed counter intuitive behaviour can be attributed to the fact that BSE being a smaller index is more prone to  fluctuations arising due to perturbations at a smaller time scale, as compared to much bigger NYSE. As is well known, the fluctuations represent the universal component of the financial time series and, hence, are also manifested in NYSE, albeit with a time lag.

\section{Conclusion}\label{sec:conc}
In conclusion, a systematic multi-scale analysis of  monthly averaged, normalised stock values of BSE and NYSE stock exchanges, over a span of 300 months, revealed a Janus faced relationship between the variations of the two. In a longer yearly time scale, the much bigger NYSE clearly drove the variations in the BSE index. Surprisingly, in the one year scale the two indexes showed completely similar response, in the  period of market turbulence, revealing crisis driven correlation, while the three year period, unveiled the correlation on a broader time scale.
In the region of high frequency fluctuations, NYSE behaviour resembled to those of BSE, with a time lag. This behaviour were clearly captured by the local wavelet transform, both DWT and CWT.
The monthly averaged normalised time series of NYSE Granger caused that of BSE, with a lag of nine months, but no linear Granger causality was found between the averaged behaviour of the two stock exchanges.
The large variations in BSE, driven by NYSE, Granger caused commensurate movement in BSE. However, the high frequency fluctuations of BSE  were found to be manifested in NYSE, with a time lag. BSE being a  smaller index is more prone to fluctuations arising due to perturbations at a smaller scale, as compared to much bigger NYSE. And as fluctuations being an universal component of financial time series, hence, they also get manifested in NYSE, albeit wiht a time lag. 
It is well understood that low frequency fluctuations invoke correlations among different sectors of an economy, while high frequency excites different sectors of it. Therefore, sectorial dynamics of a bigger economy can be simulated with the help of a smaller economy. 
Thus, BSE fluctuations can be used to simulate the NYSE variations, as it equilibrate in short span of time and this can be attributed to the universal nature of the high frequency fluctuations, which form an inalienable part of financial time series.

\newpage
\newpage
\bibliographystyle{elsarticle-num}

\bibliography{bib}
%
\newpage
\clearpage

\appendix
\section*{Appendices}
\addcontentsline{toc}{section}{Appendices}


\begin{figure}[H]
\hfill
\subfigure[BSE]{\includegraphics[height=4cm,width=4cm]{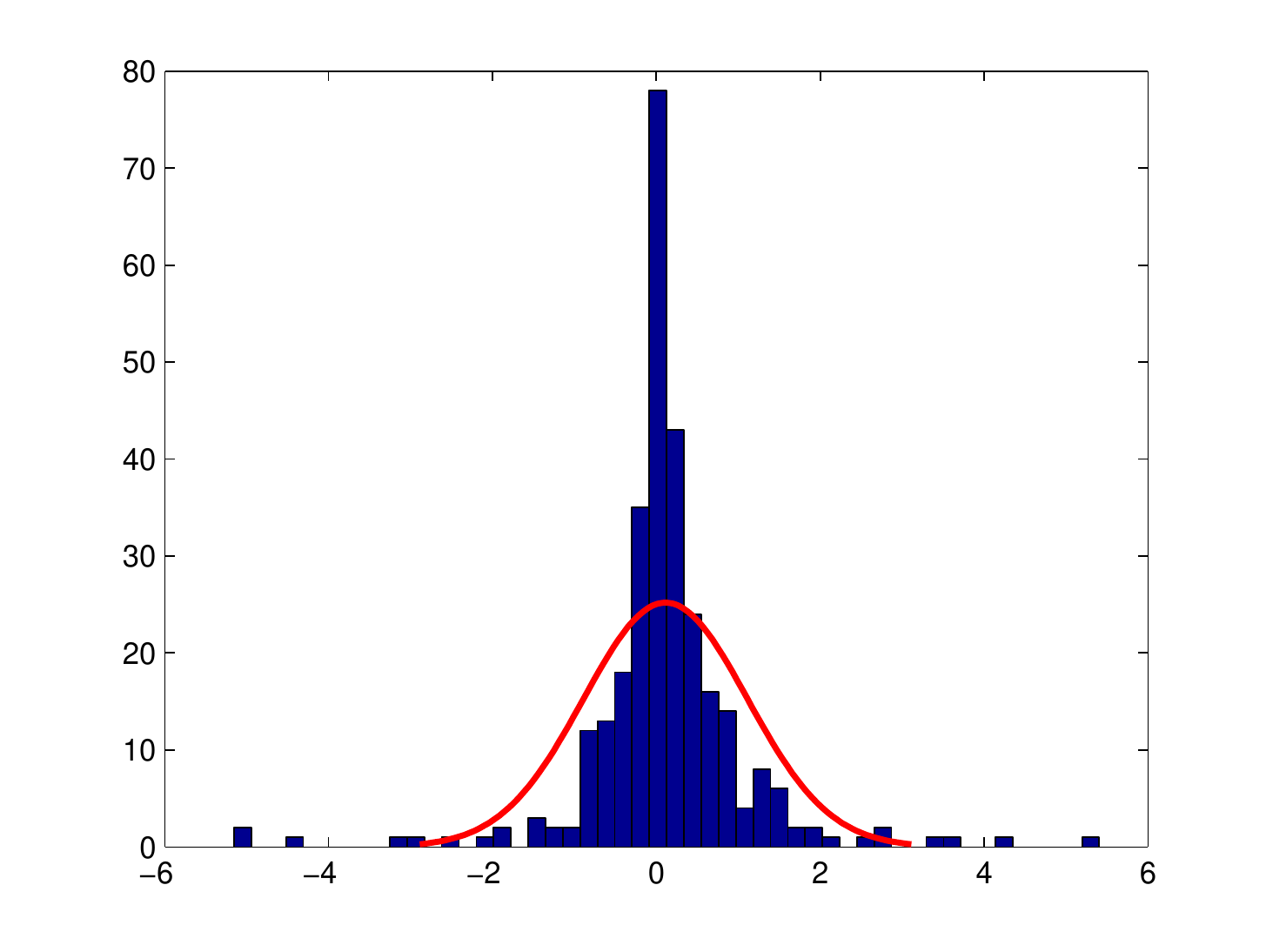}}
\hfill
\subfigure[NYSE]{\includegraphics[height=4cm , width= 4cm]{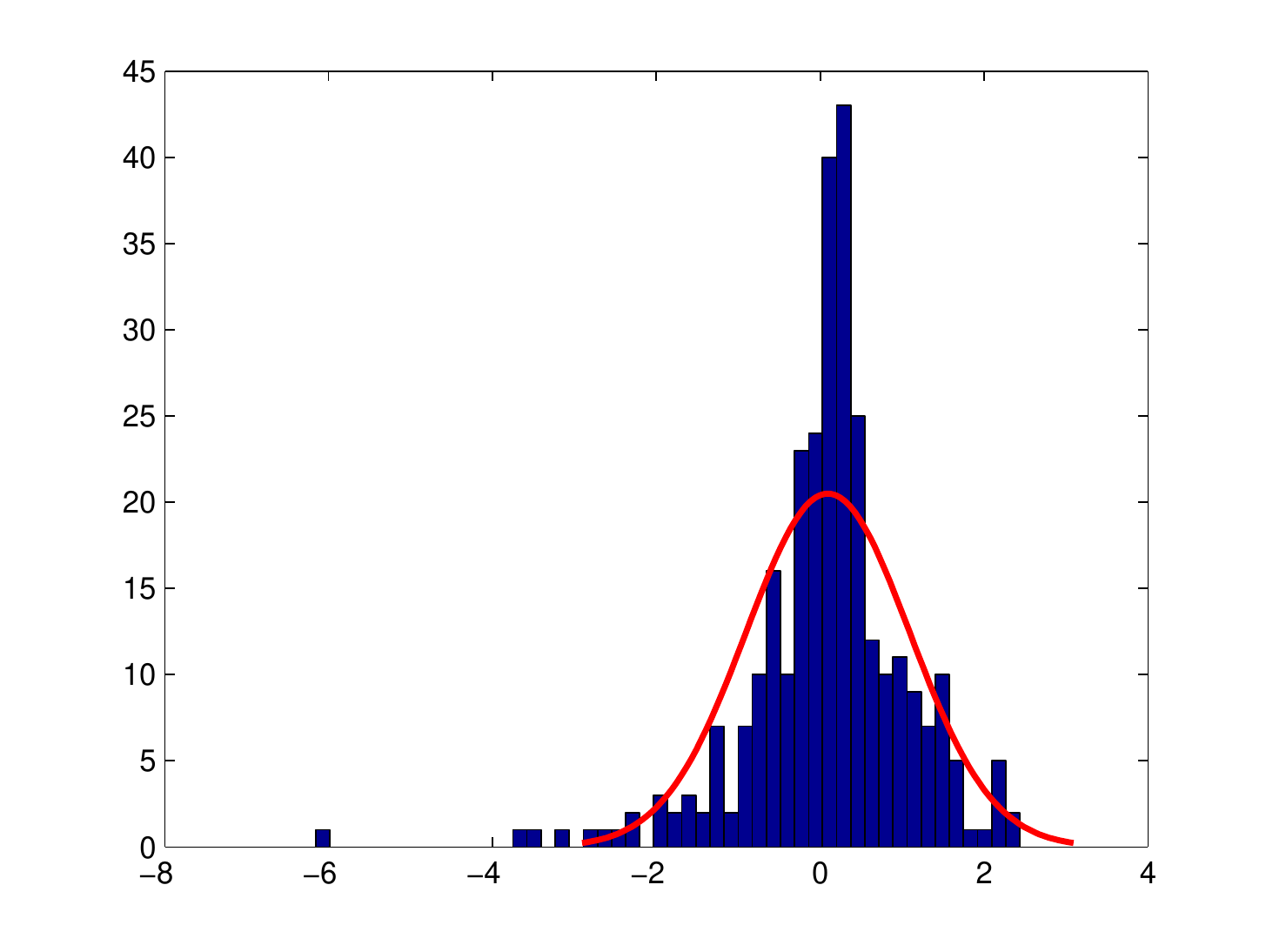}}
\hfill
\caption{Histograms of the normalised returns  for {\small BSE } and {\small NYSE }, show deviations from Gaussian behaviour for the former at the same time has been bestwoed with outliers too}
\label{fig:return_hist}
\end{figure}

\section{Mean Reversion and Stationarity}
\subsection{Augmented Dickey Fuller Test }
\begin{equation}\label{eq:adf1}
 \triangle Y_t= \alpha_1Y_{t-1} +  \sum_{j=1}^{P} \gamma_{j} \triangle Y_{t-j} + \epsilon_t
\end{equation}

\begin{equation}\label{eq:adf2}
 \triangle Y_t= \alpha_0 + \alpha_1Y_{t-1} +  \sum_{j=1}^{P}\gamma_{j}\triangle Y_{t-j}+ \epsilon_t
\end{equation}

\begin{equation}\label{eq:adf3}
 \triangle Y_t= \alpha_0 + \alpha_1Y_{t-1} + \alpha_2t +  \sum_{j=1}^{P}\gamma_{j}\triangle Y_{t-j}+ \epsilon_t
\end{equation}
$\epsilon_{t}$ is white noise. The test are based on null hypothesis ($H_{0}$):$Y_{t}$ is not I(0). If the calculated statistic are lower than the Fuller's statistics then the null hypothesis is accepted and the series is non-stationary.\\
\subsection{KPSS Test}
\begin{equation}\label{eq:kpss1}
\ Y_t = c_t + \delta t + u_{1t}
\end{equation}
\begin{equation}\label{eq:kpss2}
c_t=c_{t-1}+ u_{2t},
\end{equation}
Here, $\delta$ is a trend coefficient, $u_{1t} $ is a stationary process and $u_{2t}$ is an iid process with mean 0 and variance $\sigma^2$.
Unlike unit root tests, KPSS  provides a straightforward test for stationarity against the alternative of a unit root.
\begin{equation}
Y_{t}=\beta t+(r_{t}+\alpha ) + u_{t}
\end{equation}
Where,
\begin{equation}
\ r_{t}=r_{t-1}+ u_{t},
\end{equation}
is a random walk, the initial value $\ r_{0}$= $\alpha$ act as an intercept, \ $u_{t}$ is independent identical distribution with mean zero and a non-zero constant variance, and $'t'$ is the time. The model without the trend part is also used to test for stationarity, i.e $\beta$ = 0,  the null hypothesis, \\
$ H_{0} : Y_{t}$ is trend stationary, \\and alternative ;
$ H_{1} : Y_{t}$ has a unit root

\begin{table}[H]
\caption{p-values of ADF and KPSS tests corresponding to the logarithmic returns of BSE and  NYSE.}
\begin{tabular}{|c|c|c|c|c|c|}
\hline
 & ADF-test &  &  & KPSS-test & \tabularnewline
\hline
\hline 
lag & BSE  & NYSE & & BSE & NYSE \tabularnewline
\hline 
\hline 
1 & 0.001 & 0.001 & & 0.1 & 0.1 \tabularnewline
\hline 
2 & 0.001 & 0.001 & & 0.1 & 0.1 \tabularnewline
\hline 
3 & 0.001 & 0.001 & & 0.1 & 0.1 \tabularnewline
\hline 
4 & 0.001 & 0.001 & & 0.1 & 0.1 \tabularnewline
\hline 
5 & 0.001 & 0.001 &  & 0.1 & 0.1 \tabularnewline
\hline 
6 & 0.001 & 0.001 & & 0.1 & 0.1 \tabularnewline
\hline 
7 & 0.001 & 0.001 & & 0.1 & 0.1 \tabularnewline
\hline 
8 & 0.001 & 0.001 & & 0.1 & 0.1 \tabularnewline
\hline 
9 & 0.001 & 0.001 & & 0.1 & 0.1 \tabularnewline
\hline 
10 & 0.001 & 0.001 & & 0.1 & 0.1 \tabularnewline
\hline 
11 & 0.001 & 0.001 & & 0.1 & 0.1 \tabularnewline
\hline 
12 & 0.001 & 0.001 & & 0.1 & 0.1 \tabularnewline
\hline 
13 & 0.001 & 0.001 & & 0.1 & 0.1 \tabularnewline
\hline 
14 & 0.001 & 0.001 & & 0.1 & 0.1 \tabularnewline
\hline 
15 & 0.001 & 0.001 & & 0.1 & 0.1 \tabularnewline
\hline 
16 & 0.001 & 0.001 & & 0.1 & 0.1 \tabularnewline
\hline 
\end{tabular}
\label{table:ADF}
\end{table}
\begin{table}[H]
\centering
\caption{p-values of KPSS trend stationary test corresponding to logarithmic returns of {\small BSE} and {\small NYSE}}
\begin{tabular}{|c|c|c|}
\hline 
Lag & BSE & NYSE\tabularnewline
\hline 
\hline 
1 & 0.1 & 0.1\tabularnewline
\hline 
2 & 0.1 & 0.1\tabularnewline
\hline 
3 & 0.1 & 0.1\tabularnewline
\hline 
4 & 0.1 & 0.1\tabularnewline
\hline 
5 & 0.1 & 0.1\tabularnewline
\hline 
6 & 0.1 & 0.1\tabularnewline
\hline 
7 & 0.1 & 0.1\tabularnewline
\hline 
8 & 0.1 & 0.1\tabularnewline
\hline 
9 & 0.1 & 0.1\tabularnewline
\hline 
10 & 0.1 & 0.1\tabularnewline
\hline 
11 & 0.1 & 0.1\tabularnewline
\hline 
12 & 0.1 & 0.1\tabularnewline
\hline 
13 & 0.1 & 0.1\tabularnewline
\hline 
14 & 0.1 & 0.1\tabularnewline
\hline 
15 & 0.1 & 0.1\tabularnewline
\hline 
16 & 0.1 & 0.1\tabularnewline
\hline 
\end{tabular}
\label{table:KPSS_trend}
\end{table}
\section{Shapiro-Wilk  and KS test}
\label{sioioioec:KS}
\begin{table}[H]
\caption{p-values corresponding to average behaviour and fluctuation data of  BSE for {\small Shapiro-Wilk } and {\small Kolmogorov-Smirnov (KS) tests}}
\begin{tabular}{|p{0.5cm}|p{1.6cm}|p{1.6cm}|p{1.6cm}|p{1.6cm}|}
\hline 
 & SW-test &  & KS-test & \tabularnewline
\hline 
\hline 
 &  &  &  & \tabularnewline
\hline 
level & average behaviour & fluctuation & average behaviour & fluctuation\tabularnewline
\hline 
 &  &  &  & \tabularnewline
\hline 
1 & 1.79E-19 & 2.65E-15 & 1.11E-16 & 0\tabularnewline
\hline 
2 & 1.65E-19 & 3.51E-15 & 2.22E-16 & 0\tabularnewline
\hline 
3 & 1.55E-19 & 1.63E-14 & 3.33E-16 & 0\tabularnewline
\hline 
4 & 1.60E-19 & 2.62E-18 & 0 & 0\tabularnewline
\hline 
\hline 
\end{tabular}
\label{table:KS_sen}
\end{table}

The null hypothesis is rejected for both { {\small  Shapiro-wilk } and {\small KS-test } thus low pass as well as high-pass (variations) obtained at various levels do not belong to normal distribution.\\
\begin{table}[H]
\centering
\caption{p-values corresponding to low and high pass data of     NYSE for {\small Shapiro-Wilk } and  {\small Kolmogorov-Smirnov (KS) Tests}.}
\begin{tabular}{|p{0.5cm}|p{1.6cm}|p{1.45cm}|p{1.6cm}|p{1.45cm}|}
\hline 
 & SW-test &  & KS -test & \tabularnewline
\hline 
\hline 
 &  &  &  & \tabularnewline
\hline 
level  & average behaviour  & fluctuation & average behaviour & fluctuation\tabularnewline
\hline 
 &  &  &  & \tabularnewline
\hline 
1 & 3.06E-11 & 0.000124 & 1.11E-16 & 0\tabularnewline
\hline 
2 & 2.12E-11 & 4.82E-07 & 2.22E-16 & 0\tabularnewline
\hline 
3 & 1.46E-11 & 4.82E-09 & 3.33E-16 & 0\tabularnewline
\hline 
4 & 7.18E-12 & 1.27E-14 & 0 & 0\tabularnewline
\hline 
\end{tabular}
\label{table:KS_nys}
\end{table}
\section{Spearman and Pearson correlation coefficients}
\label{sec:SP}
Pearson correlation coefficients help in establishing linear relationship between two variables, if any. It is defined as the ratio of covariance of two variables to the product of their respective standard deviation.
Whereas, Spearman's correlation coefficients are rank  based version of  Pearson's correlation coefficients ( supplement material).
\begin{equation}\label{eq:spear}
 r_s= \frac {\sum_{i=1}^{n}((rank(x_{i})-\overline{rank(x)})(rank(y_{i})-\overline{rank(y)}))}{\sqrt{\sum_{i=1}^{n}((rank(x_{i})-\overline{rank(x)})^{2}}\sum_{i=1}^{n}((rank(y_{i})-\overline{rank(y)})^{2}}
\end{equation}
Here, $rank(x_{i})$ and  $rank(y_{i})$ are the ranks of the data points in the sample.

\begin{equation}\label{eq:pear1}
\ \rho = \frac{Cov(x,y)} {(\sigma_{x}\sigma_{y} )}\
\end{equation}

\begin{equation}\label{eq:pear2}
\begin{aligned}
\ r_{p}=\frac{\sum_{i=1}^{n}((x_{i}-\overline{x})(y_{i}-\overline{y})} {\sqrt{(\sum_{i=1}^{n}((x_{i}-\overline{x})^2)(\sum_{i=1}^{n}((y_{i}-\overline{y})^2)}}\
\end{aligned}
\end{equation}
where, $\overline{x}=\frac{\sum_{i=1}^{n}x_{i}}{n}$ and $\overline{y}= \frac{\sum_{i=1}^{n}y_{i}}{n}$\\

\section{ Stationarity check for  fluctuations and average behaviour}\label{sec:granger_stationarity}

\begin{table}[H]
\centering
\caption{p-values of the ADF and KPSS tests  for the fluctuations, for the first four levels, null hypothesis for ADF and KPSS tests are $H_0$: signal is non-stationary and signal is stationary, respectively.}
\begin{tabular}{|p{0.5cm}|c|p{0.8cm}|c|c|p{0.8cm}|c|}
\hline 
 &  & ADF-test  &  &  & KPSS-test  & \tabularnewline
\hline 
\hline 
Level & BSE & NYSE & $H_0$ & BSE & NYSE  & $H_0$\tabularnewline
\hline 
1 & 0.01 & 0.01 & reject  & 0.1 & 0.1 & accept\tabularnewline
\hline 
2 & 0.01 & 0.01 & reject  & 0.1 & 0.1 & accept\tabularnewline
\hline 
3 & 0.01 & 0.01 & reject  & 0.1 & 0.1 & accept\tabularnewline
\hline 
4 & 0.01 & 0.01 & reject  & 0.1 & 0.1 & accept\tabularnewline
\hline 
\end{tabular}

\label{tab:adf_kpss_granger}
\end{table}
%
\begin{table}[H]
\caption{p-values of the ADF and KPSS tests to check the stationarity of the  average behaviour of BSE and NYSE, for the first levels.}
\begin{tabular}{|c|c|p{0.8cm}|c|c|c|p{0.8cm}|c|}
\hline 
 &  & ADF-test  &  & & & KPSS-test  & \tabularnewline
\hline 
\hline 
Level & BSE & NYSE & $H_0$ & &BSE & NYSE & $H_0$\tabularnewline
\hline 
1 & 0.01 & 0.01 & reject &  & 0.1 & 0.1 & accept\tabularnewline
\hline 
2 & 0.01 & 0.01 & reject & & 0.1 & 0.1 & accept\tabularnewline
\hline 
3 & 0.01 & 0.01 & reject & & 0.1 & 0.1 & accept\tabularnewline
\hline 
4 & 0.01 & 0.01 & reject & & 0.1 & 0.1 & accept\tabularnewline
\hline 
\end{tabular}
\end{table}
%

\end{document}